\documentclass[twocolumn,aps,superscriptaddress,showpacs,nofootinbib,floatfix]{revtex4}
\usepackage{epsfig,bm,feynmf}
\usepackage{graphics}
\usepackage{graphicx}
\usepackage{amsmath}
\usepackage{dcolumn}
\usepackage{xcolor}
\usepackage[normalem]{ulem}  
\newcommand{\be}{\begin{equation}}
\newcommand{\ee}{\end{equation}}
\newcommand{\bea}{\begin{eqnarray}}
\newcommand{\eea}{\end{eqnarray}}

\renewcommand\sout{\bgroup \color{red} \ULdepth=-.5ex \ULset}

\begin{document}

\title{In-medium effects in $\phi$ meson production  in heavy-ion collisions from 
subthreshold to relativistic energies}


\author{Taesoo Song}\email{T.Song@gsi.de}
\affiliation{GSI Helmholtzzentrum f\"{u}r Schwerionenforschung GmbH, Planckstrasse 1, 64291 Darmstadt, Germany}

\author{Joerg Aichelin}\email{Aichelin@subatech.in2p3.fr}
\affiliation{SUBATECH UMR 6457 (IMT Atlantique, Universit\'{e} de Nantes, IN2P3/CNRS), 4 Rue Alfred Kastler, F-44307 Nantes, France}
\affiliation{Frankfurt Institute for Advanced Studies, Ruth-Moufang-Strasse 1, 60438 Frankfurt am Main, Germany}

\author{Elena Bratkovskaya}\email{E.Bratkovskaya@gsi.de}
\affiliation{GSI Helmholtzzentrum f\"{u}r Schwerionenforschung GmbH, Planckstrasse 1, 64291 Darmstadt, Germany}
\affiliation{Institute for Theoretical Physics, Johann Wolfgang Goethe Universit\"{a}t, Frankfurt am Main, Germany}
\affiliation{Helmholtz Research Academy Hesse for FAIR (HFHF), GSI Helmholtz Center for Heavy Ion Physics, Campus Frankfurt, 60438 Frankfurt, Germany}


\begin{abstract}
We investigate the hidden strange $\phi$ meson production in heavy-ion collisions from subthreshold ($E_{kin}\approx 1$ A GeV)  to relativistic ($E_{kin}\approx 21$ A TeV) energies as well as its coupling to the open strange mesons (kaons, antikaons) and their productions. 
Our study is based on the off-shell microscopic Parton-Hadron-String Dynamics (PHSD) transport approach which is applicable for  the dynamical description of strongly interacting hadronic and partonic degrees-of-freedom  created in heavy-ion collisions.  
Implementing  novel meson-baryon and meson-hyperon production channels for $\phi$ mesons, calculated within a T-matrix coupled channel approach based on the extended SU(6) chiral effective Lagrangian model, along with the collisional broadening of the $\phi$-meson in-medium spectral function,
we find a substantial enhancement of $\phi$ meson production in heavy-ion collisions, especially at sub- and  near-thresholds. This allows to describe the experimentally observed strong enhancement of the $\phi/K^-$ ratio at low energies without including hypothetical decays of heavy baryonic resonances to $\phi$ as in alternative approaches. Moreover, we show that in spite 
of a stronger contribution from enhanced $\phi$ to $K^-$ production, the majority of the experimental data for different A+A systems at low energies favour the scenario with in-medium modifications of the kaon and antikaon properties in the hot and dense environment. Moreover, we study the influence of the final state interactions of $K, \bar K$ mesons on the reconstruction of $\phi$'s by the  by invariant mass method.
\end{abstract}


\maketitle

\section{Introduction}

The understanding of  strangeness production in heavy-ion collisions since a few decades is one of the challenging topics in heavy-ion physics. Strangeness is a newly produced flavor, since initially the colliding nuclei are composed of 'up' and 'down' quarks only.
Due to strangeness neutrality of the initial system,  $s$ and $\bar s$ quarks must be produced in pairs and end up in (anti-)strange mesons and baryons or are combined in mesons with 'hidden' strangeness,  
as the $\phi$-meson.

An enhancement of strangeness production in A+A compared to 
p+p collisions has been observed experimentally from low (below $E_{kin}$ = 1 A GeV) 
to ultrarelativistic ($\sqrt{s} >$ 100 AGeV) energies. The reason for that is rather different at different energies.
In low energy A+A collisions strangeness 
can be produced well below the physical threshold for $p+p$ scattering.
The observed yield is much higher than expected from the enhances due to Fermi motion of the nucleons inside a nucleus. This
indicates an additional mechanisms for strangeness production in heavy-ion
collisions compared to elementary reactions \cite{Cassing:1999es,Hartnack:2011cn}.
A detailed analysis, based on microscopic transport approaches, shows that one of reasons is
 multi-step reactions, which are absent in elementary collisions \cite{Aichelin:1985rbt,Cassing:1996xx,Cassing:1999es,Hartnack:2001zs}. Indeed, in a primary collisions two nucleons produce a baryonic resonance, e.g. a $\Delta$ resonance, which subsequently collides with another nucleon. Since a $\Delta$ is heavier than a nucleon,  a $\Delta N$ collision can easier reach a sufficient $\sqrt{s}$ to produce strange hadrons than a $NN$ collision. 
Moreover, the newly produced mesons - dominantly pions, contribute to strangeness
production. This increases further the multiplicity of strange hadrons.  

Another reason is the in-medium modification of hadron properties 
in the dense and hot medium created in heavy-ion collisions (cf. the reviews 
\cite{Hartnack:2011cn,Tolos:2020aln,Bleicher:2022kcu} and references therein). 
Different chiral  models (staring from the early work of Kaplan and Nelson \cite{Kaplan:1986yq,Nelson:1987dg} and later continued by a variety of groups  \cite{Lutz:1997wt,Waas:1996xh,Waas:1996fy,SchaffnerBielich:1999cp,Lutz:2001yb,Ramos:1999ku,Mishra:2003tr})  
predict a modest enhancement of $K^+$ masses and a rather strong reduction of the $K^-$ mass with increasing baryon density. 
Later calculations within the coupled-channel unitarized G-matrix approach, using meson-exchange potentials \cite{Tolos:2000fj,Tolos:2002ud}, predicted a more
moderate in-medium change of antikaon masses (-40 to -60 MeV at density $\rho_0$), however, as well a visible broadening of the $K^-$ spectral function in the medium. 
Such a mass reduction of antikaons leads to a lowering of the production threshold for antikaons in the medium and, consequently, makes  the
subthershold $K^-$ production (even forbidden kinematically in the vacuum) possible and leads to
an enhancement of the $K^-$ yield in $A+A$ collisions. On the other hand, the yield of
$K^+$ is reduced in the medium due to the repulsive $KN$ potential which makes kaons heavier \cite{Korpa:2004ae,Cassing:1999es,Hartnack:2011cn}. 
The comparison of  experimental data of KaoS, FOPI, HADES collaborations for A+A collisions at beam energies of 1-2 A GeV 
\cite{Best:1997ua,Laue:1999yv,Menzel:2000vv,Sturm:2000dm,Stefanon:2017len,Zinyuk:2014zor,Agakishiev:2010rs,Adamczewski-Musch:2018xwg}  for different observables 
(as multiplicities, rapidity and $p_T$-spectra as well as directed ($v_1$) and elliptic flow ($v_2$)) with theoretical transport calculations
\cite{Aichelin:1986ss,Fang:1994wu,Li:1994cu,Li:1997zb,Li:1994vd,Li:1996aq,Pal:2001nk,Cassing:1996xx,Bratkovskaya:1997pj,Cassing:1999es,Cassing:2003vz,SchaffnerBielich:1999cp,Hartnack:2001zs,Fuchs:2000kp,Fuchs:2005zg,Mishra:2004te,Hartnack:2011cn,Kolomeitsev:2004np}) showed the traces of in-medium modifications of the properties of (anti-)kaons as well as a sensitivity to the equation-of-state  (EoS) of nuclear matter \cite{Fuchs:2000kp,Fuchs:2005zg,Hartnack:2011cn}.

At high energies strangeness enhancement in A+A versus p+p collisions 
has been suggested as one of the signatures for the formation of a quark-gluon plasma (QGP) \cite{Rafelski:1982pu,Stock:2002ww}, since  strangeness production is easier 
in a QGP than in a hadron gas due to the mass difference between the strange quark pair and the strange meson pair~\cite{Koch:1986hf} or due to the larger available energy in a QGP as compared to a hadron-hadron collision in the 'corona' region \cite{Werner:1992gn,Drescher:2001hp,Aichelin:2008mi}. 
Experimentally a sharp rise and drop in the excitation function of the $K^+/\pi^+$ ratio (so called "horn")  has been observed  \cite{NA49:2002pzu,NA49:2007stj,E917:2003gec,STAR:2004yym,NA49:2008goy},  
predicted in Ref. \cite{Gazdzicki:1998vd}, as an appearance 
of a QGP phase at a center-of-mass energy of $\sqrt{s_{NN}}\approx 7\,$GeV.
In Refs. \cite{Cassing:2015owa,Palmese:2016rtq} the 'horn' phenomenon has been explained from a microscopic point employing the PHSD (Parton-Hadron-String Dynamics) approach ~\cite{Cassing:2008sv,Cassing:2008nn,Cassing:2009vt,Bratkovskaya:2011wp,Linnyk:2015rco,Moreau:2019vhw}
as an interplay  between chiral symmetry restoration (CSR)
in the dense and hot hadronic medium (which modifies the Schwinger mechanism \cite{Schwinger:1951nm} for the hadronic particle production via the string formation and decay) and the formation of the QGP.

Strangeness enhancement is seen not only for 'open' strange mesons and baryons,
but also for 'hidden' strange states as the $\phi$ meson, which contains a $s$ and $\bar s$ quark. 
According to the Okubo-Zweig-Iizuka (OZI) rule~\cite{Okubo:1963fa,Zweig:1964jf,Iizuka:1966fk}, the $\phi$ meson
dominantly decays to the open strange $K, \bar K$ pair and not to light quark mesons. Thus, the $\phi$ dynamics
in heavy-ion collisions is directly coupled to the kaon and antikaon dynamics.

Since  $\phi$ meson production is strongly suppressed as compared to other light vector mesons ($\rho$ and $\omega$), it is a very challenging task to measure $\phi$'s experimentally.
One way is to measure the $\phi$ via dilepton decay $\phi \to e^+e^-$, which provides a clear signal, since  dileptons are not affected by the strong interaction. However, the small branching ratio of the $\phi$ meson to dileptons makes this measurement very complicated. Another way is to measure the $K^+K^-$ pairs from the $\phi$ meson decay. This process has a large branching ratio, however, there are other difficulties related to the final state interaction of the decay products.  $K^+$ and $K^-$ scatter elastically and inelastically in the hadronic medium. Especially the $K^-$ mesons with a low momentum have a large scattering cross section with nucleons. 
If the $K^+, K^-$ scatter or are being absorbed,  the $\phi$ cannot be reconstructed anymore by the invariant mass method; thus,  the experimentally 'measured' $\phi$ multiplicity is less than 
the real number of $\phi$ produced in the reactions and it's mass distribution is distorted by the
final state interaction.

Recently it has been reported by the HADES collaboration that also $\phi$ mesons are produced at relatively large quantity in Au+Au collisions at subthreshold energies and reach
$1\times 10^{-4}$ at 1.23 A GeV~\cite{HADES:2017jgz}. Besides the $\phi$ multiplicity also the ratio of hidden strangeness to open strangeness   has been measured and reaches values of $\approx$ 0.5~\cite{HADES:2017jgz}.
The same tendency of an enhanced $\phi$ production in Ni+Ni and Al+Al collisions at 1.93 A GeV  has  been reported earlier by the FOPI collaboration \cite{Herrmann:1996zg,FOPI:2016jgt}.
With increasing  beam energy the ratio decreases to 0.2 as has been measured recently by the STAR collaboration \cite{STAR:2021hyx} and becomes relatively small ($\approx 0.15$) already at 
$\sqrt{s_{NN}} \approx 4$ GeV; at high energies the dependence on the collision energy is mild \cite{NA49:2002pzu,E917:2003gec,STAR:2004yym,NA49:2007stj,NA49:2008goy}. 

One of the early theoretical attempts to explain these FOPI data was presented in Ref. \cite{Chung:1997mp} 
where in the RVUU transport model $\phi$ production by  scattering of $\Delta$ resonances with nucleons and pions as well as $\Delta\Delta$ 
scattering were included. The $BB, mB$ cross sections were defined within a chiral Lagrangian model. These studies came to the conclusion that these processes cannot fully account for the measured $\phi$ yield. 
Similar conclusions have been found in other transport calculations \cite{Schade:2009gg}.

The experimentally observed high $\phi$ multiplicity and the surprising behaviour of the $\phi/K^-$ ratio have triggered theoretical activities since both are higher than predicted by transport approaches. Both - along with the $\phi/\Xi^-$ ratio -  can find an explanation if one assumes that high lying nuclear resonances ($N^*$ with masses above 2 GeV) have a considerable branching ratio into $\phi$ through $N^*\rightarrow N+\phi$. Such a scenario has been proposed in a UrQMD study ~\cite{Steinheimer:2015sha} by accounting for 
$N^*(1990)$, $N^*(2080)$, $N^*(2190)$, $N^*(2220)$ and $N^*(2250)$ states decaying to $\phi$ mesons with a branching ratio $\approx 0.2$\%. According to \cite{Steinheimer:2015sha} the formation of such heavy states in A+A collisions is enhanced due to  Fermi motion and secondary meson-baryon collisions while their excitation in p+p collisions is small and doesn't contribute significantly to the near threshold data of $\phi$ production measured by the ANKE collaboration \cite{ANKE:2007dyc}.
This idea has been taken over later by the SMASH group in  Ref. \cite{Steinberg:2018jvv} applying, however, a much larger branching ratio of 0.5\%. 
The possibility of $N^*$ decay to other vector meson - $\omega$ -  has been discussed in Ref. \cite{Fabbietti:2015tpa}.
The decay of heavy baryonic resonances to $\phi$ mesons has, however, not been observed experimentally so far~\cite{ParticleDataGroup:2020ssz}.

The goal of this study is to show that the observed 'enhanced' $\phi$ multiplicity and $\phi/ K^-$ ratio close to threshold can be understood also in a more conventional approach,   \\
i) by incorporatingthe  dynamical modifications of the $\phi$ meson properties in the medium 
such as a collisional broadening of the $\phi$ meson spectral function; 
\\
ii) by accounting for additional multi-step meson-baryon and meson-hyperon reactions for 
$\phi$ meson production as predicted by the SU(6) extension of the meson-baryon chiral Lagrangian within a unitary coupled channel T-matrix approach. These reactions of produced mesons and/or baryonic non-strange strange resonance become possible in heavy-ion collision  and lead to an enhancement of the $\phi$ yield. 

As a theoretical laboratory for our study we use the microscopic PHSD transport approach 
~\cite{Cassing:2008sv,Linnyk:2015rco,Moreau:2019vhw} 
which is based on Kadanoff-Baym off-shell dynamics suited for the description of strongly interacting hadronic and partonic systems. The PHSD has been intensively applied for studying the 
in-medium effects and strangeness production in p+A and A+A collisions from low to ultrarelativistic energies 
\cite{Cassing:2003vz,Palmese:2016rtq,Linnyk:2015rco,Moreau:2019vhw,Song:2020clw}.

The consequences of the partial chiral symmetry restoration of the $\phi$ meson properties in a hot and/or dense matter are expected to be seen in proton-nucleus and heavy-ion collisions. 
In the QCD sum rule approach the decrease of the quark condensate leads to a mass shift 
of the vector meson~\cite{Brown:1991kk,Hatsuda:1991ez}.
Theoretical calculations based on chiral models predict a mass shift as well and a broadening of the $\phi$ meson spectral function
\cite{Klingl:1997tm,Oset:2000eg,Cabrera:2002hc,Gubler:2016itj,Kim:2022eku}.
There have been many efforts to measure experimentally the in-medium properties of vector mesons in p+A and A+A collisions, in particular by the dilepton decay channel, as well as to describe the data in theoretical approaches (cf. the  reviews \cite{Linnyk:2015rco,Bleicher:2022kcu}). These studies are, however, mostly concentrated on $\rho$ or $\omega$ mesons. The very low production rate and the rather small width  make it difficult to study the $\phi$ in-medium modification.
Recently experimental evidence for observation of the in-medium effects of $\phi$ mesons in p+A reactions has been reported by the KEK-PS E325 Collaboration  \cite{KEK-PS-E325:2005wbm} by measuring the dilepton spectra. 
On the other hand, the possibilities to observe  the in-medium modification of $\phi$ spectral function - using the strong decay mode $\phi \to K^+K^-$ - may be also explored  \cite{KEK:Sako2022}.

In this study we argue that the in-medium modification of the $\phi$ spectral function by collisional broadening during the propagation in the medium as well as the modification of the production cross sections in the medium play an important role for $\phi$ meson production as well as for the final reconstruction via the $K\bar K$ mode (similar to the $K^*$ reconstruction \cite{Ilner:2017tab}).

Further on, we argue that multi-step meson-baryon reactions with newly produced mesons and baryonic resonances lead to an enhancement of $\phi$ production in heavy-ion collisions compared to p+p reactions where such processes are not possible. In this study we substantially extend the traditionally considered $m+B$ channels involving pions, nucleons and $\Delta$'s
by accounting for s-wave scattering $m+B \to \phi + B$  reactions with $m=\eta, K, \rho, K^*$ mesons and $B=N, \Delta, \Lambda, \Sigma, \Sigma^*$ baryons for all possible $I=1/2$ and 3/2 channels. Since those cross sections are not known experimentally, we use T-matrix calculations based on the SU(6) extension of the meson-baryon chiral Lagrangian within a unitary coupled channel  approach \cite{Gamermann:2011mq}.
We note that the in-medium effects on vector and pseudoscalar strange mesons are included via the spectral functions also for these novel channels. The backward reactions are accounted for via detailed balance.

This paper is organized as follows: in Section~\ref{broadening} we describe the modification of $\phi$ meson spectral function in the medium; in Section \ref{off-shell} we recall how the off-shell particle propagates in a nuclear medium.
Then the $\phi$ meson production through elementary reactions in the nuclear medium is presented based on the coupled channel T-matrix approach in Section~\ref{elementary}.
Section~\ref{heavy-ion} is devoted to $\phi$ production in heavy-ion collisions and the results are compared with experimental data. 
The contribution from $\phi$ meson decay to the (anti-)kaon yield is investigated in Section~\ref{influence} and the strange particle ratios are presented in Section~\ref{ratios}.
Finally a summary is given in Section~\ref{summary}.

\section{In-medium modification of $\phi$ meson properties}\label{broadening}

In order to explore the influence of in-medium effects on the
vector-meson spectral function we introduce collisional broadening, $\Gamma_{coll}$,
following the approach of Refs. \cite{Cassing:1996xa,Bratkovskaya:1998pr,Bratkovskaya:2001ce,Bratkovskaya:2007jk} 
\begin{eqnarray}
\Gamma^*_V(M,|\vec p|,\rho)=\Gamma_V(M) + \Gamma_{coll}(M,|\vec
p|,\rho),
\label{gammas}
\end{eqnarray}
where $M$ is the mass and $\Gamma_V(M)$ the total width of the vacuum spectral function of the vector meson ($V=\rho,\omega, \phi$).
For the vacuum width of the $\phi$ meson we use 
\begin{eqnarray}
\Gamma_\phi(M) \simeq \Gamma_{\phi\to\rho\pi(3\pi)}^{exp}\frac{\Gamma_{\phi\to\rho\pi(3\pi)}(M)}{\Gamma_{\phi\to\rho\pi(3\pi)}(M_0)}~~~~~\nonumber\\ 
+\Gamma_{\phi\to K\bar{K}}^{exp}
\left(M_0\over M\right)^2 \left(q\over q_0\right)^3 \theta(M-2m_K)
\label{Widthrho},
\end{eqnarray}
where $M_0$ is the vacuum pole mass of the vector meson spectral function and
$\Gamma_{i}^{exp}$ is the experimental decay width of channel~$i$.
The momentum of a $\phi$ meson with masses $M$ and $M_0$ is given by
\begin{eqnarray}
q = {(M^2-4m_K^2)^{1/2}\over 2}, \nonumber\\
q_0 = {(M_0^2-4m_K^2)^{1/2}\over 2}. \nonumber
\end{eqnarray}

Here $\Gamma_{\phi\to\rho\pi(3\pi)}(M)$ is the decay width of $\phi\to\rho\pi(3\pi)$, calculated in the effective chiral lagrangian with parity anomalous terms~\cite{Kaymakcalan:1983qq}.
The collisional width in Eq.~(\ref{gammas}) is approximated as
\begin{eqnarray}
\Gamma_{coll}(M,|\vec p|,\rho) = \gamma \ \rho < v \
\sigma_{VN}^{tot} > \approx  \ \alpha_{coll} \
\frac{\rho}{\rho_0}, \label{dgamma}
\end{eqnarray}
where $v=|{\vec p}|/E$ is the velocity with ${\vec p}$ and $E$ being respectively the 3-momentum and energy of the vector meson in the rest frame of the nucleon current, and $\gamma^2=1/(1-v^2)$.
Furthermore, $\rho$ is the nuclear density,  $\rho_0 = 0.168 \ {\rm fm}^{-3}$ the normal
nuclear density and $\sigma_{VN}^{tot}$ the meson-nucleon total cross section in vacuum.
In order to simplify the calculations of $\Gamma_{coll}(\rho)$ we use the linear density approximation for the collisional width with coefficient $\alpha_{coll}$  \cite{Bratkovskaya:2007jk}.
Experimental and theoretical results constraint $\alpha_{coll}$ for $\phi$ mesons: 
The first is the KEK E325 experiment, where the $\phi$ width has been measured to be 15.3 MeV at normal nuclear matter  density~\cite{KEK-PS-E325:2005wbm}.
The second are calculations based on hadronic models, which tend to get a larger broadening {(up to 40 MeV width)}~\cite{Cabrera:2016rnc,Gubler:2016itj,Kim:2022eku}. 
The value of 25 MeV, which we use in this study, is roughly the average value. 

We note that the CBELSA-TAPS collaboration has shown (for the $\omega$ meson) that the pole mass of vector mesons changes at finite baryon density \cite{CBELSATAPS:2005iwc}. We model this change  following the Hatsuda/Lee \cite{Hatsuda:1991ez} or Brown/Rho scaling
\cite{Brown:1991kk} as
\begin{eqnarray}
\label{Brown}
M_0^*(\rho)= \frac{M_0} {\left(1 + \alpha {\rho / \rho_0}\right)},
\end{eqnarray}
where $\rho$ is the nuclear density at the $\phi$ location point.
$\alpha \simeq 0.12$ for the $\omega$ meson \cite{Metag:2007hq}, and $\alpha \simeq 0.035$ for the $\phi$ meson. We note that the parametrization of Eq.~(\ref{Brown}) may also be employed at much higher collision
energies without the need to introduce a
cut-off density in order to avoid negative pole masses. Eq.~(\ref{Brown}) is uniquely fixed by the standard expression
$M_0^*(\rho) \approx M_0 (1 - \alpha \rho/\rho_0)$ in the low
density regime.
In this study, however, we consider only the broadening of the width of the $\phi$ mesons, neglecting the (small) mass shift in the medium.

\begin{figure}[hbt!]
\includegraphics[width=8.6 cm]{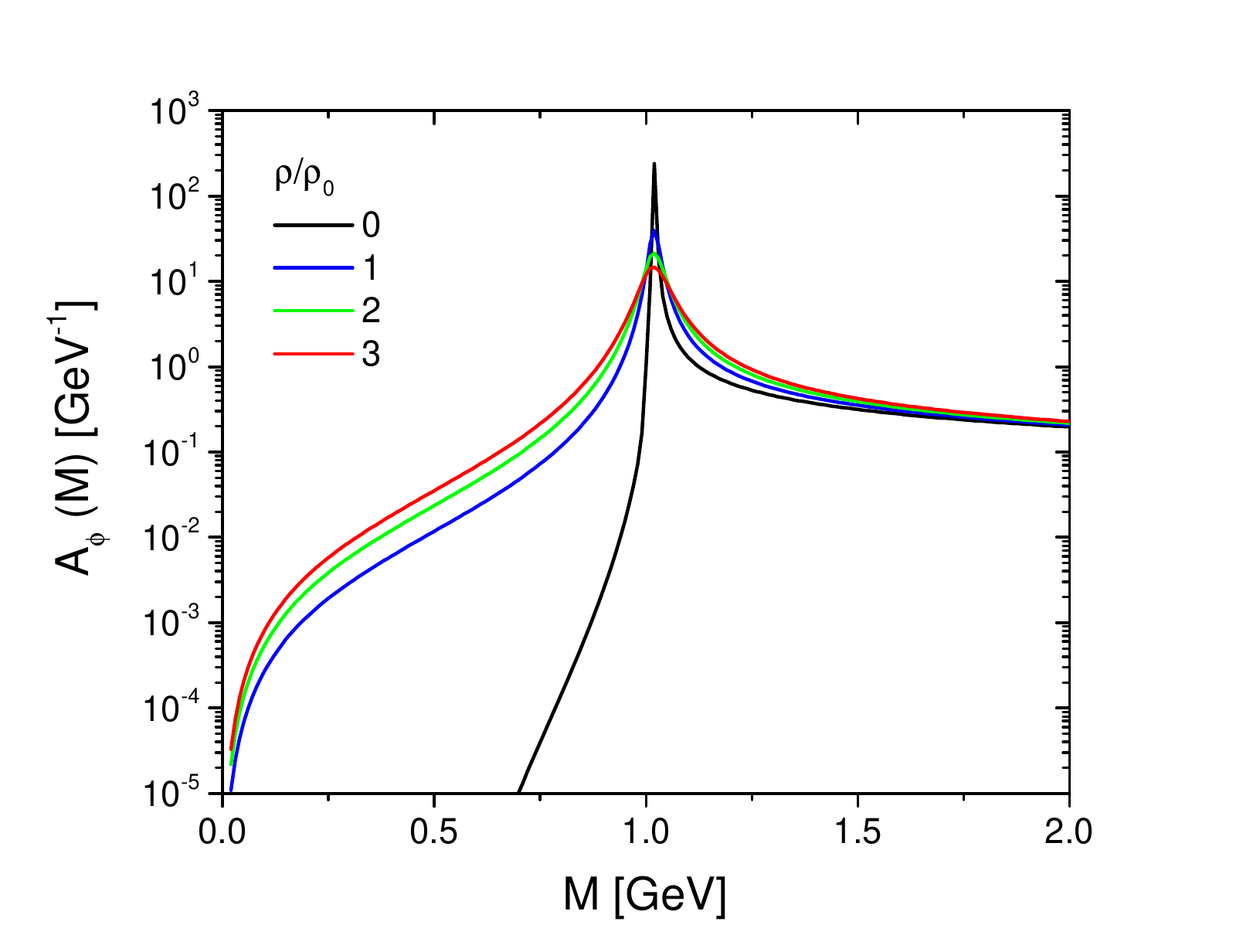}
\caption{ The spectral functions of $\phi$ mesons in the 'collisional broadening' scenario for nuclear densities of 0,1,2,3$\times\rho_0$}.
\label{spectra}
\end{figure}

For the spectral function of a vector meson of mass $M$ at baryon density $\rho$ we take a Breit-Wigner form:
\begin{eqnarray}
A_V(M,\rho) =~~~~~~~~~~~~~~~~~~~~~~~~~~~~~~~~~~~~~~~\nonumber\\
C_1\cdot {2\over \pi} \ {M^2 \Gamma_V^*(M,\rho)
\over (M^2-M_{0}^{*^2}(\rho))^2 + (M {\Gamma_V^*(M,\rho)})^2}\, ,
\label{spfunV}
\end{eqnarray}
where $C_1$ is fixed by the normalization condition for arbitrary $\rho$:
\begin{eqnarray}
\int_{M_{min}}^{M_{lim}} A_V(M,\rho) dM =1,
\label{SFnorma}\end{eqnarray} 
with $M_{lim}=2$~GeV being chosen as
an upper limit for the numerical integration. 
The lower limit for
the vacuum spectral function is given by the two-pion decay, $M_{min}=2 m_\pi$ for a  $\rho$ meson and the three-pion decay, $M_{min}=3 m_\pi$ for a $\phi$ meson decay, whereas for the in-medium collisional broadening case by $M_{min}=2 m_e \to 0$ with $m_e$ denoting the electron mass.
$M_0^*$ is the pole mass of the vector meson spectral function with $M_0^*(\rho=0)=M_0$ in vacuum. It may be  shifted in the medium for the dropping mass scenario according to Eq.  (\ref{Brown}), which is, however, not considered here for $\phi$ mesons.

The resulting spectral function of $\phi$ mesons
is displayed in Fig. \ref{spectra} for the case of the 'collisional broadening' scenario for densities of 0,1,2, and 3 times $\rho_0$.  Note that the hadronic width of $\phi$ mesons vanishes below the three-pion mass in vacuum.
With increasing nuclear density the elastic and inelastic interactions of the vector mesons shift strength towards low invariant masses.

\section{Off-shell PHSD approach}\label{off-shell}

The Parton-Hadron-String Dynamics (PHSD) 
~\cite{Cassing:2008sv,Cassing:2008nn,Cassing:2009vt,Bratkovskaya:2011wp,Linnyk:2015rco,Moreau:2019vhw}
is a microscopic covariant  transport approach for the dynamical  description of strongly interacting 
hadronic and partonic matter created in heavy-ion collisions. 
It is based on a solution of the Cassing-Juchem generalised off-shell transport equations for test particles \cite{Cassing:1999wx,Cassing:1999mh} derived from the first-order gradient expansion of  Kadanoff-Baym equations \cite{KadanoffBaym}, see Refs. \cite{Juchem:2004cs,Cassing:2008nn}. 
PHSD differs from the semi-classical BUU model, since it  propagates Green's functions
(in phase-space representation) which contain information not only on the occupation probability
(in terms of the phase-space distribution functions), but also on the properties of hadronic and  partonic degrees-of-freedom via their spectral functions.
The PHSD approach allows to describe the full evolution of a relativistic heavy-ion collision
from the initial hard nucleon-nucleon scatterings with string formation and decay, to the formation
of the strongly-interacting quark-gluon plasma (QGP) as well as hadronization and the subsequent 
interactions in the expanding hadronic phase extending the early realization of the Hadron-String-Dynamics (HSD) transport approach \cite{Cassing:1999es} by including the in-medium effects such as a collisional broadening of the vector meson spectral functions \cite{Bratkovskaya:2007jk} and the modification of strange degrees-of-freedom in line with G-matrix calculations \cite{Cassing:2003vz,Song:2020clw}. 
Moreover, the implementation of detailed balance on the level of  $2\leftrightarrow 3$ 
reactions is realized for the main channels of strangeness production/absorption 
by baryons ($B=N, \Delta, Y$) and pions \cite{Song:2020clw}, as well as for the multi-meson fusion reactions $2\leftrightarrow n$ for the formation of $B +\bar B$  pairs  \cite{Cassing:2001ds,Seifert:2017oyb}.

Since the spectral function of the vector mesons (i.e. real and imaginary part of their self-energies) 
changes with density by collisional broadening, their propagation through the hot and dense medium
can only be treated in off-shell transport approaches in a consistent way (see example in Fig. 3 of \cite{Bratkovskaya:2007jk}).

Here we briefly describe the implementation of the off-shell transport dynamics (OSTD) in the PHSD
approach (for the details of off-shell transport theory we refer the reader to  the book \cite{Cassing:2021book}).
A vector meson $V$ - created at some space-time point $x$ at density $\rho$ - is distributed in mass according to its
spectral function $A_V(M,\rho)$ (in the kinematically allowed region which is not excluded by energy and momentum conservation). 
While propagating through the nuclear medium the self-energy and total width of
the vector meson $\Gamma_V^* (M,\rho)$, Eq.~ (\ref{gammas}), changes
dynamically. This modifies the spectral function, Eq.~(\ref{spfunV}), depending on the real part of the vector meson self energy $Re \Sigma^{ret}$ as well as on the imaginary part ($\Gamma_V^*\simeq -Im \Sigma^{ret}/M$)
\begin{eqnarray}
A_V(M,\rho) =~~~~~~~~~~~~~~~~~~~~~~~~~~~~~~~~~~~~~~~\nonumber\\
C_1\cdot {2\over \pi} \ {M^2 \Gamma_V^*
\over (M^2-M_0^2- Re \Sigma^{ret})^2 + (M {\Gamma_V^*})^2},
\label{spfun}
\end{eqnarray}
which is the in-medium form for a  boson spectral function.
The spectral function merges to the vacuum spectral function when the $\phi$ leaves the medium.

In the OSTD the general off-shell Cassing-Juchem equations of motion  for test
particles  with three-momentum $\vec P_i$, energy $\varepsilon_i$ at position
$\vec X_i$ read
\cite{Cassing:1999wx,Cassing:1999mh}
\begin{eqnarray}
\frac{d {\vec X}_i}{dt} \! & = & \frac{1}{1-C_{(i)}}
\  \frac{1}{2 \varepsilon_i} \: \bigg[ \, 2  {\vec P}_i  +  {\vec
\nabla}_{P_i} \, Re \Sigma^{ret}_{(i)}\nonumber\\
&&+  \frac{ \varepsilon_i^2
- {\vec P}_i^2 - M_0^2 - Re \Sigma^{ret}_{(i)}}{{\tilde
\Gamma}_{(i)}} \: {\vec \nabla}_{P_i} \, {\tilde \Gamma}_{(i)} \:\bigg],
\label{eomr} 
\\
\frac{d {\vec P}_i}{d t} \! & = & -
\frac{1}{1-C_{(i)}}\ \frac{1}{2 \varepsilon_{i}} \: \bigg[ {\vec
\nabla}_{X_i} \, Re \Sigma^{ret}_i \: \nonumber\\
&&+ \: \frac{\varepsilon_i^2 -
{\vec P}_i^2 - M_0^{2} - Re \Sigma^{ret}_{(i)}}{{\tilde
\Gamma}_{(i)}} \: {\vec \nabla}_{X_i} \, {\tilde \Gamma}_{(i)} \:
\bigg],
\label{eomp}\\
\frac{d \varepsilon_i}{d t} & = &
\frac{1}{1-C_{(i)}}\ \frac{1}{2 \varepsilon_i} \: \bigg[
\frac{\partial Re \Sigma^{ret}_{(i)}}{\partial t} \: \nonumber\\
&&+ \:
\frac{\varepsilon_i^2 - {\vec P}_i^2 - M_0^{2} - Re
\Sigma^{ret}_{(i)}}{{\tilde \Gamma}_{(i)}} \: \frac{\partial
{\tilde \Gamma}_{(i)}}{\partial t} \bigg],
\label{eome}
\end{eqnarray}
where the notation $F_{(i)}$ implies that the function is taken at
the coordinates of the test particle, i.e. $F_{(i)} \equiv
F(t,\vec{X}_{i}(t),\vec{P}_{i}(t),\varepsilon_{i}(t))$. In Eqs.
(\ref{eomr})-(\ref{eome}) $Re \Sigma^{ret}$ is a short-hand notation of the real part of the retarded self energy while ${\tilde \Gamma} = -Im
\Sigma^{ret} $ stands for the (negative) imaginary part. 
Apart from the propagation in the real potential $\approx Re
\Sigma^{ret}/2\varepsilon$ the equations (\ref{eomr}) --
(\ref{eome}) include the dynamical changes due to the imaginary
part of the self energy $Im \Sigma^{ret} \approx - M \Gamma_{V}^*$
with $\Gamma_V^*$ from Eq.~(\ref{gammas}).

In the above equations a common factor $(1-C_{(i)})^{-1}$
appears, which includes the energy derivatives of the retarded
selfenergy,
\begin{eqnarray} \label{pref} C_{(i)} & = &
\frac{1}{2 \varepsilon_i} \: \bigg[ \frac{\partial Re
\Sigma^{ret}_{(i)}}{\partial \epsilon_i}\: \nonumber\\
&&+ \:
\frac{\varepsilon_i^2 - {\vec P}_i^2 - M_0^{2} - Re
\Sigma^{ret}_{(i)}}{{\tilde \Gamma}_{(i)}} \: \frac{\partial
{\tilde \Gamma}_{(i)}}{\partial \epsilon_i} \bigg].
\end{eqnarray}
This common factor leads to a rescaling of the 'eigentime' of
particle $i$ but does not change the trajectories as demonstrated
in Fig. 3 of Ref.~\cite{Cassing:1999mh}. According to the model
studies in \cite{Cassing:1999mh} this prefactor is negative for
invariant masses $M$ below the pole mass $M_0$ and positive above.
This implies that the factor $1/(1-C_i)$ is less than 1 for
$M<M_0$ and larger than 1 for $M>M_0$ (as stated in
\cite{Cassing:1999mh}). This yields a time dilatation for masses $M <
M_0$ in their phase-space propagation.

The interpretation of Eqs.~ (\ref{eomr}) --
(\ref{eome}) becomes particularly transparent if $\tilde{\Gamma}$
is independent on the 3-momentum $\vec{P}$. 
Then, using $M^{2} =P^2 - Re \Sigma^{ret}$ as an independent variable instead of the energy $P_0 \equiv \varepsilon$, Eq.~(\ref{eome}) turns to
\begin{eqnarray}
\frac{dM_i^2}{dt} \; = \; \frac{M_i^2 - M_0^2}{{\tilde
\Gamma}_{(i)}} \; \frac{d {\tilde \Gamma}_{(i)}}{dt}. \label{eomm}
\end{eqnarray}
This is the equation for the time evolution of the invariant
mass squared of the test-particle $i$~\cite{Cassing:1999wx,Cassing:1999mh}. Now
the deviation from the pole mass, $\Delta M^2 = M^2 -M_0^2$, follows the equation
\begin{eqnarray}
{d\over dt}\Delta M^2 = {\Delta M^2\over \tilde{\Gamma}} \ {d\over
dt} \tilde{\Gamma}~~~~~~~~~~~~~\nonumber\\
 \leftrightarrow \ \frac{d}{dt}
\bigg[\ln\left(\frac{\Delta M^2}{\tilde{\Gamma}}\right) \bigg]
=0, \label{dm2}\end{eqnarray} which expresses the fact that the
off-shellness of the particle is proportional to the total width.



\section{$\phi$ meson production in elementary reactions in the nuclear medium}\label{elementary}

\subsection{$\phi$ meson production in baryon-baryon reactions}

\begin{figure}[hbt!]
\centerline{
\includegraphics[width=8.6 cm]{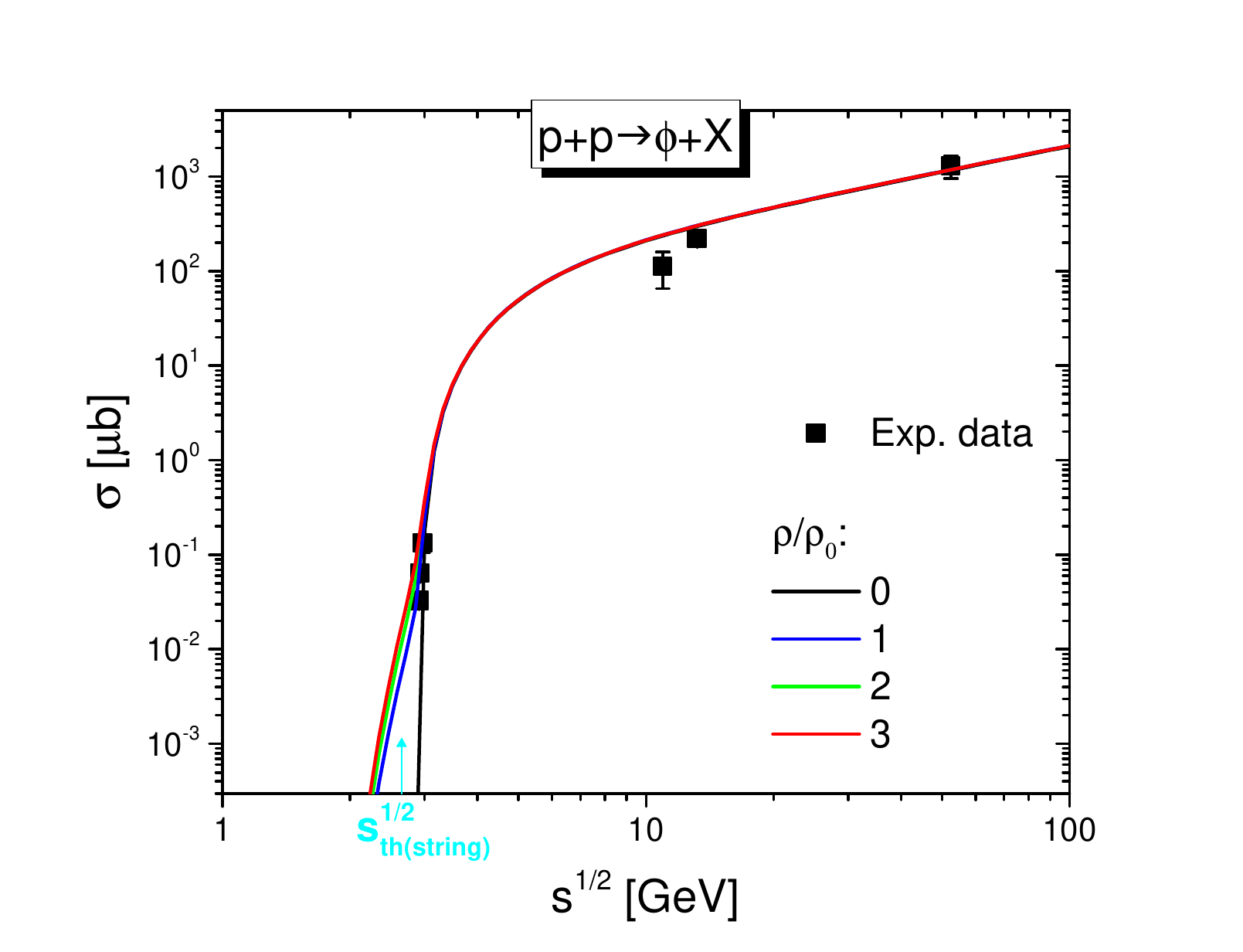}}
\caption{Inclusive cross section for $\sigma_{pp\rightarrow \phi X}$ as a function of the invariant energy $\sqrt{s}$ , calculated at $\rho/\rho_0=0,1,2,3$,
in comparison to the experimental data from Refs.
~\cite{ANKE:2007dyc,Schopper:1988hwx,DISTO:2000dfs,DISTO:2002pvr,Moskal:2002jm,AnnecyLAPP-CERN-CollegedeFrance-Dortmund-Heidelberg-Warsaw:1981whv,ACCMOR:1981wfy,Antipov:1981ze,Blobel:1975sb,DISTO:1999ujs,NA49:2000jee}.
The arrow indicates the threshold energy for string formation in baryon-baryon scattering.
}
\label{Anke}
\end{figure}

Since the vector mesons are produced in the hot and dense medium, which is created in heavy-ion collisions, their production cross section has to be also modified compared to the vacuum cross section. However, the theoretical calculation of such in-medium cross sections is not trivial and can presently not be achieved from first principle QCD calculations. Also a variety of effective chiral models and $T$-matrix approaches  provide 
only in-medium spectral functions (cf. \cite{Rapp:1999ej}) but no cross sections.  
We note that the situation in the strangeness channels is more advanced in this respect and  G-matrix calculations for the in-medium cross sections are available \cite{Cassing:2003vz,Cabrera:2014lca}.

In the traditional on-shell approaches, in-medium mass shifts and/or width broadening of particles have been incorporated by parametrizing the vacuum cross sections as a function of the invariant energy $\sqrt{s}$ and the threshold $\sqrt{s_0}$ (cf. \cite{Bratkovskaya:1996qe,Cassing:1999es,Sibirtsev:1998vz}). 
In-medium modifications of cross sections then in first order have been incorporated by shifting the
threshold energy
$\sqrt{s_0^*} = \sqrt{s_0} + M _0^* - M_0$ due to a lack of microscopic calculations for the corresponding in-medium transition matrix elements.

In order to account for the in-medium effects in production cross sections, we model the vector meson production cross sections in $NN$ and $\pi N$ reactions in the following phenomenological way:\\
i) for low energy $BB \to \phi BB$ reactions  (here $B=N,\Delta$)
the total cross section for $NN$ (or $BB$) scattering 
$\sigma_{NN \rightarrow VNN}(s,\rho)$  is given by
\be
\sigma_{NN \rightarrow VNN}(s,\rho) =\int\limits_{M_{min}}^{M_{max}}
dM\
\frac{d \sigma_{NN \rightarrow VNN}(s,M,\rho)}{d M} .
\label{xs_NNVVtot}
\ee
We note in advance that a similar strategy is used for modeling $\sigma_{\pi N \rightarrow VN}(s,\rho)$.\\
The mass differential cross sections are approximated by
\begin{eqnarray}
\frac{d \sigma_{NN \rightarrow VNN}(s,M,\rho)}{d
M}  =  \sigma_{NN \rightarrow VNN}^{0}(s,M,\rho)\nonumber\\
\times
A_V(M,\rho) {1
\over \int\limits_{M_{min}}^{M_{lim}} A_V(M,\rho) dM , }
\label{xs_NNVV}
\end{eqnarray} 
where $A_V(M,\rho)$ denotes the meson spectral
function, Eq.~(\ref{spfunV}), for a given total width $\Gamma_V^*$, Eq.~(\ref{gammas}); $M_{max}=\sqrt{s}-2 m_N \le M_{lim}$ is the 
kinematically allowed maximum invariant mass of the vector meson $V$ and $m_N$ is the nucleon mass. 
In Eq. (\ref{xs_NNVV}) $\sigma_{NN \rightarrow VNN}^{0}(s,M,\rho)$ is the vacuum cross section for vector mesons with mass $M$ (see Section 3.4 of Ref. \cite{Cassing:1999es} for the explicit parametrizations).
Note, that the physical threshold for the $\phi$ production in $NN$ reactions is defined as $\sqrt s_{th} = 2 m_N + M_{min}$, i.e. for the broadening scenario $\sqrt s_{th} \to 2 m_N +2m_e$. For the definition of $M_{lim}$ and $M_{min}$ see Eq. (\ref {SFnorma}).

Thus, Eq.~(\ref{xs_NNVV}) is used to model the vector meson production in $NN$ reactions in the vacuum and in the medium.
Substituting Eq.~(\ref{xs_NNVV}) into Eq.~(\ref{xs_NNVVtot}), we get
\be
\sigma_{NN \rightarrow VNN}(s,\rho) = {\int\limits_{M_{min}}^{M_{max}}
A_V(M,\rho)\sigma_{NN \rightarrow VNN}^{0}(s,M,\rho)dM
\over \int\limits_{M_{min}}^{M_{lim}} A_V(M,\rho) dM .}
\nonumber
\ee

ii) The implementation of the in-medium effects for $\phi$ production from $BB$ as well as from $mB$ and $mm$ reactions at high energies follows the general strategy in PHSD:
the treatment of energetic $BB \to X$ collisions, leading to multiparticle production 
is realized in the PHSD by the excitation of the strings based on string LUND model \cite{Nilsson-Almqvist:1986ast} (see Ref. \cite{Kireyeu:2020wou} for the details of 
the treatment of elementary collisions in the PHSD within the 'tuned' LUND string model.)
The threshold for the string formation is fixed to $s_{the(string)}^{1/2}(BB)=2.65$ GeV in PHSD.
The in-medium spectral functions for mesonic and baryonic resonances are incorporated in the PHSD 'tune' of the Lund model by including their in-medium self-energy and in-medium width
depending on the local baryon density and temperature \cite{Bratkovskaya:2007jk} .
It allows to study in-medium effects  such as a collisional broadening of spectral 
functions of vector mesons ($\rho, \omega, \phi, a_1$) \cite{Bratkovskaya:2007jk,Linnyk:2015rco}.

The total cross section for the channel $pp \rightarrow \phi X$ is displayed
in Fig. \ref{Anke} as a function of the invariant energy
$\sqrt{s}$ and for densities of $\rho/\rho_0 = 0, 1, 2, 3$. 
The arrow shows the threshold for string formation ($\sqrt{s}=2.65$ GeV).
The squares show the experimental data \cite{Schopper:1988hwx,DISTO:2000dfs,Moskal:2002jm} for the inclusive vector meson production $pp \rightarrow V X$, where $X$ stands for all final particles including two baryons and further mesons.
As seen from Fig. \ref{Anke}, the threshold for $\phi$ production shifts to lower $\sqrt{s}$ with
increasing baryon density $\rho$, which leads to the enhancement of their production in a medium compared to the vacuum (black line).

\subsection{$\phi$ meson production in meson-baryon reactions}

\begin{figure}[hbt!]
\centerline{
\includegraphics[width=8.6 cm]{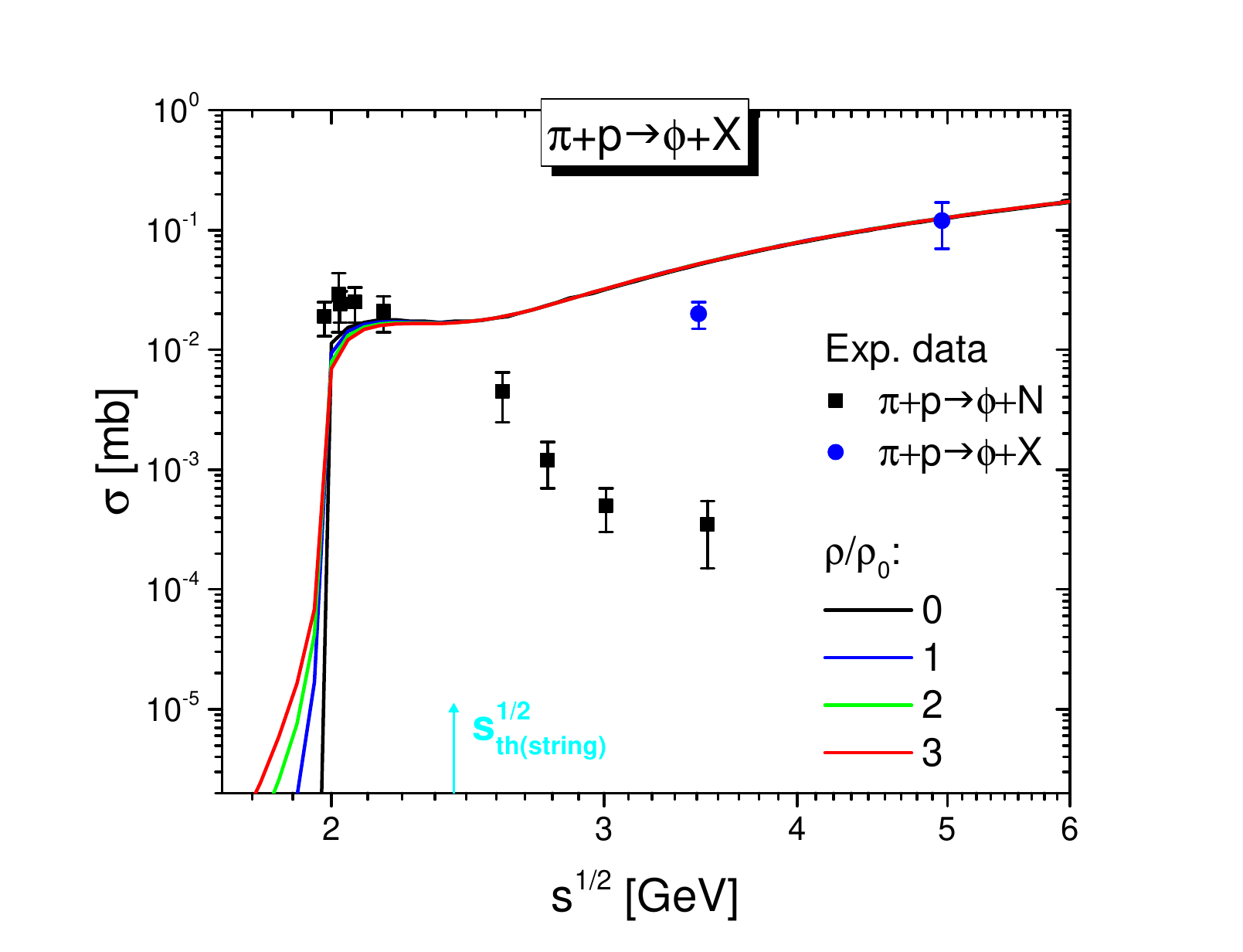}}
\caption{Inclusive cross section for $\pi+p\rightarrow \phi+X$ as a function of the invariant energy $\sqrt{s}$  for $\rho/\rho_0=$0, 1, 2, and 3 in comparison with inclusive and exclusive experimental data on $\phi$ production from \cite{Schopper:1988hwx,Moskal:2002jm}.
The vertical arrow indicates the threshold energy for string formation in meson-baryon collisions.}
\label{cs-piNphiN}
\end{figure}

The $\phi$ production by $mB$ collisions is (similar to $BB$ reactions) split into two regimes : the low energy binary collisions $mB \to \phi B$ and high energy $mB \to \phi +X$,  which are realized via string formation if the $\sqrt{s}$ is above the $mB$ string threshold $s_{th(string)}^{1/2} (mB)=2.4$ GeV. The implementation of in-medium effects is similar to the $BB$ case discussed in the previous section.
The actual parametrizations for binary $mB \to \phi B$ reactions for $m=\pi$ and $B=N, \Delta$ are taken in Breit-Wigner form - the same way as in the HSD code (see Section 3.5 of Ref.~\cite{Cassing:1999es}). 

In  Fig.~\ref{cs-piNphiN} we show the  cross section for the $\pi N \rightarrow \phi X$ reaction 
versus the invariant energy $\sqrt{s}$ for baryon densities $\rho/\rho_0 = 0, 1, 2, 3$.
The black line represents the  $\pi N \rightarrow \phi X$ cross sections in free space while the colored lines stand for the in-medium cross sections within the 'collisional broadening' scenario. The vertical arrow denotes the threshold for string formation in meson-baryon collisions.

As seen from Fig. \ref{cs-piNphiN}, similar to the $NN$ case, the $\pi N$ cross section is shifted towards the low $\sqrt{s}$ region due to the in-medium collisional broadening of the $\phi$ spectral function which allows for $\phi$ production in the medium  below the vacuum threshold. Consequently, the $\phi$ production is enhanced in heavy-ion collisions.

We note that the in-medium modifications are also accounted for the $\phi$
absorption and elastic scattering cross sections.
The total cross section for $\phi$ absorption by a nucleon $\phi +N \to X$ is parameterized by~\cite{Golubeva:1997na}: 
\begin{eqnarray}
\sigma_{\phi N}^{tot}=5 ~{\rm mb}+\frac{4.5~{\rm mb GeV/c}}{p_{lab}}
\end{eqnarray}
where $p_{lab}$ is the $\phi$ meson momentum (GeV/c) in the nucleon-rest-frame.

The elastic scattering of a $\phi$ mesons on nucleon is taken as in Ref. \cite{Golubeva:1997na} :
\begin{eqnarray}
\sigma_{\phi N}^{el}=\frac{10~{\rm mb}}{1+p_{lab}/{\rm (GeV/c)}}.
\end{eqnarray}


\subsection{Novel $mB\to \phi B$ production channels within a SU(6) based T-matrix approach}

In heavy-ion collisions even at relatively low energies additionally to pions, nucleons and $\Delta$ resonances a variety of mesons and baryonic non-strange and strange resonances are produced. Even if their abandunces are small at SIS energies, they can participate in multi-step  scattering
which leads also to $\phi$ production. The AMPT transport model \cite{Pal:2002aw,Lin:2004en} 
includes $mB\to \phi B$ reactions with $m=\pi,\rho$, $B=N,\Delta$ as well as reactions with 
initial strange mesons and hyperons - $K\Lambda \to \phi N$ \cite{Ko:1991kw}. However, such 
reactions were found to be not sufficient to explain the experimental data on 
$\phi$ production at SIS energies \cite{Chung:1997mp}. 
In this study we consider additional $mB$ channels, predicted by a meson-baryon chiral effective Lagrangian, and explore their influence on $\phi$ production.

The  consistent SU(6) extension of the meson-baryon chiral Lagrangian within a coupled
channel unitary approach   allows to include in the T-matrix  a lot of other  meson-baryon channels which produce a $\phi$ meson~\cite{Gamermann:2011mq}.
The s-wave scattering amplitude from the SU(6) chiral effective Lagrangian is written as 
\begin{eqnarray}
V_{ij}^{SIJ}=\varepsilon_{ij}^{SIJ}\frac{2\sqrt{s}-M_i-M_j}{4f_if_j}\sqrt{\frac{E_i+M_i}{2M_i}}\sqrt{\frac{E_j+M_j}{2M_j}},
\end{eqnarray}
where $i(j)$ indicates the initial (final) meson-baryon scattering states, $M_{i(j)}$ and $E_{i(j)}$ are, respectively, mass and center-of-mass energy of the baryon, $f_{i(j)}$ the decay constant of the meson to the $i(j)$ state, and $\varepsilon_{ij}^{SIJ}$ the degeneracy coefficient, corresponding to the scattering channel with $S,I, {\rm and}~ J$ being total strangeness, isospin and angular momentum of the collision, respectively~\cite{Oset:1997it,Cabrera:2014lca}.

A T-matrix approach can be formulated on the basis of the Born scattering amplitude $V_{ik}^{SIJ}$,
\begin{eqnarray}
T_{ij}^{SIJ}=V_{ij}^{SIJ}+V_{ik}^{SIJ}G_{kk}^{SIJ}T_{kj}^{SIJ},
\label{tmatrix1}
\end{eqnarray}
where $k$ is the intermediate meson-baryon state and the sum is performed over all possible states.  $G_{kk}^{SIJ}$ is the product of the meson and baryon propagators of the state $k$~\cite{Nieves:2001wt}, which is renormalized such that 
\begin{eqnarray}
G_{kk}^{SIJ}(s=m_N^2+m_\pi^2)=0
\end{eqnarray}
with $m_N$ and $m_\pi$ being nucleon and pion massess, respectively.

Since the second term of the right hand side of Eq.~(\ref{tmatrix1}) is factorizable in the S-wave channel, means that the loop momentum and $\sqrt{s}$ are separable, the T-matrix can be expressed as
\begin{eqnarray}
T_{ij}^{SIJ}=(1-V^{SIJ}G^{SIJ})_{ik}^{-1}V_{kj}^{SIJ},
\label{tmatrix2}
\end{eqnarray}
where $1_{jk}=\delta_{jk}$.

\begin{figure*}[t!]
\centerline{
\includegraphics[width=8.6 cm]{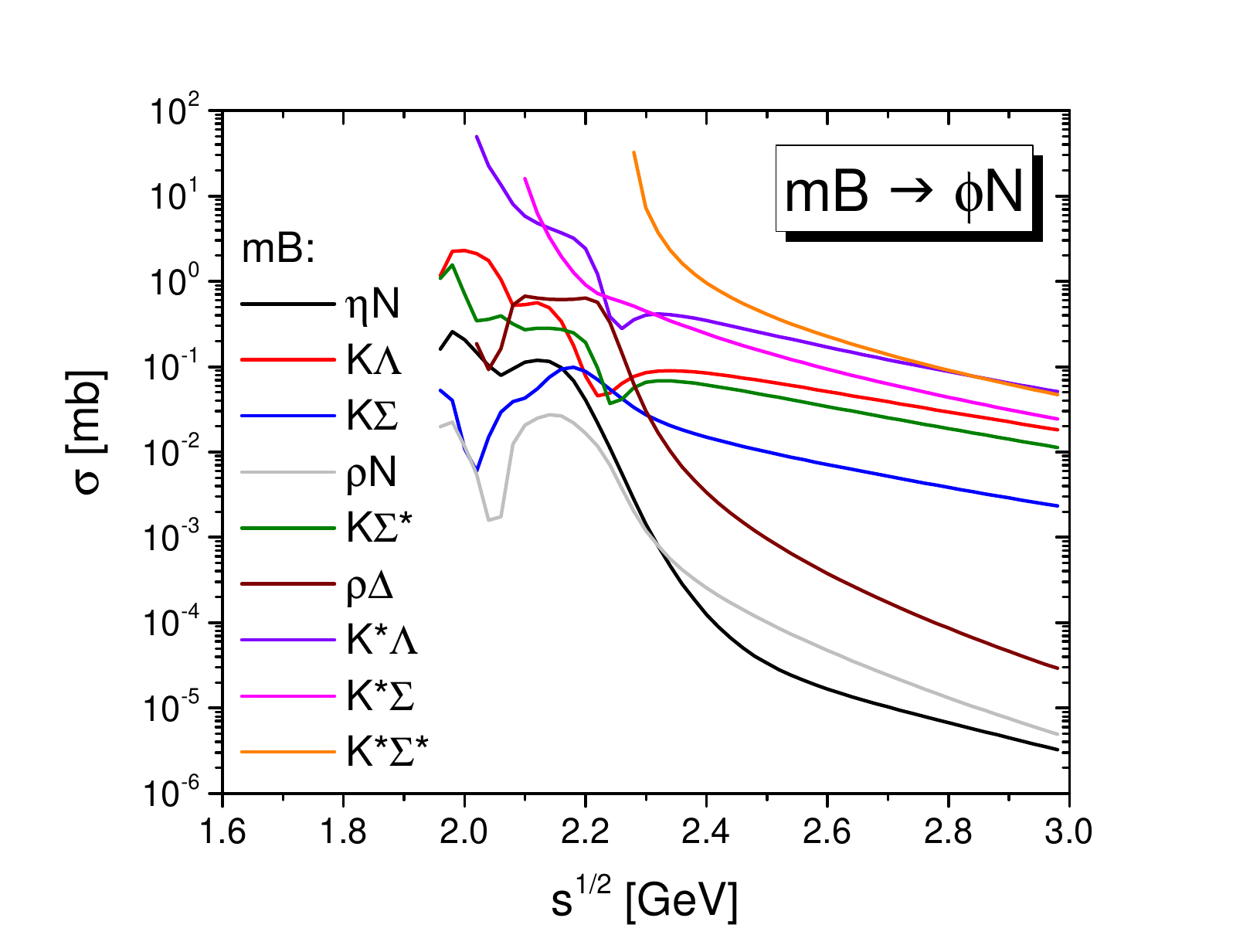}
\includegraphics[width=8.6 cm]{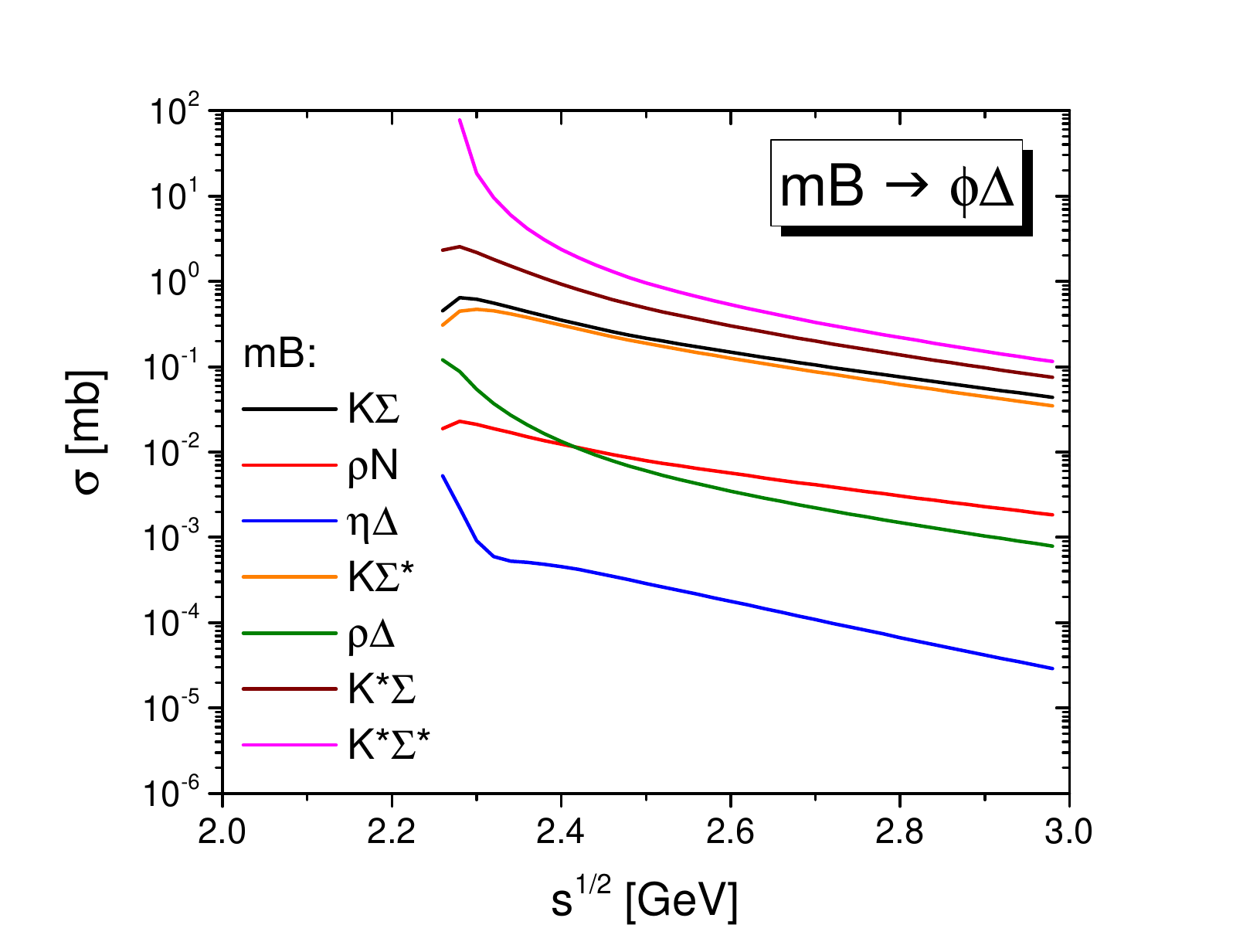}}
\caption{Cross sections of the $mB$ channels for $\phi N$ production  (left) and for $\phi\Delta$ production (right), calculated in the T-matrix approach. All resonances here are taken with their pole mass.}
\label{csSU6}
\end{figure*}

For $\phi$ production from meson-baryon $S-$wave scattering we explore reactions that invole amplitudes in the following sectors,
\begin{eqnarray}
(I,J)&=&(1/2,1/2),~(3/2,1/2),~(1/2,3/2),\nonumber\\
&~&(3/2,2/2),~(3/2,5/2),    
\end{eqnarray}
and $S=0$.
The channels considered in this study for $\phi$ meson production are 
\begin{eqnarray}
\eta N,~K\Lambda,~K\Sigma,~\rho N,~K\Sigma^*,~\rho\Delta, \nonumber\\
K^*\Lambda,~K^* \Sigma,~K^*\Sigma^* \rightarrow \phi N, 
\end{eqnarray}
for $I=1/2$ and 
\begin{eqnarray}
K\Sigma,~\rho N,~\eta\Delta,~K\Sigma^*,~\rho\Delta, \nonumber\\
K^* \Sigma,~K^*\Sigma^* \rightarrow \phi \Delta
\end{eqnarray}
for $I=3/2$.
For example, $\rho \Delta\rightarrow \phi\Delta$ has contributions from $(I,J)=(3/2,1/2)$, $(3/2,3/2)$ and $(3/2,5/2)$.

The $\phi$ production cross sections along with a nucleon and a $\Delta$ from the T-matrix  is defined as 
\begin{eqnarray}
\label{cross-sec}
\sigma_{ij}(\sqrt{s}) &=& \frac{1}{4 \pi} \frac{M_i M_j}{s} \frac{q_j}{q_i} \frac{\sum_J (2J+1)| T_{ij}^J|^2}{\sum_J (2J+1)}.
\end{eqnarray}

Figure~\ref{csSU6} shows the $\phi$ production cross sections of the $mB$ channels for $\phi N$ (left) and for $\phi\Delta$ (right), calculated in the T-matrix approach. All resonances here are taken with their pole mass for simplicity.
One can see that the cross sections show a very strong flavour dependence. The channels with initial strange
mesons and hyperons are the largest due to the OZI-rule. However, in heavy-ion collisions the relative contribution of the different $mB$  channels is defined by the abundances of the secondary hadrons involved in the $\phi$ production. The $\eta$ mesons are more abundant then $K^*$, thus their scattering with nucleons are more probable then $K^*$ scattering with strange baryons $\Lambda, \Sigma,\Sigma^*$ even if the $\eta N$ cross section is smaller then that for $K^*Y$. The latter will be demonstrated in Fig. \ref{dndy} in the Section \ref{heavy-ion}.

\subsection{$\phi$ meson production by meson-meson reactions}

Since $\phi$ mesons decay either into $K+\bar{K}$ or into $\rho +\pi$, they can be  
 formed in the medium by the inverse processes : $K\bar K \to \phi$, $\rho +\pi \to \phi$. 
Additionally we take into account the low energy channels $K+\bar{K}\rightarrow \phi+m$ and $\eta+m\rightarrow \phi+m$ with $m$ being a meson, though their contributions are very small.
We note that the $\phi$ production by the $mm$ string formation is also accounted for in the PHSD,
where the $mm$ string threshold is taken as $s_{th(string)}^{1/2} (mm)=1.3$ GeV.

\section{In medium strangeness production in the PHSD}

Before  showing the PHSD  results on the influence of the novel channels for $\phi$ meson production on the (anti-)kaon dynamics, we briefly recall the theoretical basis for including of the in-medium effects on (anti-)kaon production in heavy-ion collisions as realized in our previous work~\cite{Song:2020clw}.

The transition amplitudes of antikaon baryon scattering as well as the self-energy of the antikaon are calculated in the G-matrix approach \cite{Cabrera:2014lca}. It is based on the self-consistent unitarization of the transition amplitude from an effective Chiral lagrangian, and the summation of the Matsubara frequencies to obtain finite temperature results.
As a result, the transition amplitudes decrease in the nuclear medium due to the Pauli blocking and the self-energy of the antikaon, which both depend on the nuclear density, the temperature and the antikaon momentum in nuclear matter~\cite{Cabrera:2014lca}.
Both the real and the imaginary part of the self-energy decreases with nuclear density, while the temperature dependence is mild.
Furthermore, with increasing momentum the real part of the self-energy decreases rapidly, whereas the imaginary part of the self-energy changes only little.
As a result, the spectral function of antikaons - whose pole mass is shifted by the real part of the self-energy and whose width is generated by the imaginary part of the self-energy shows a width broadening and at low momenta, $p<300$ MeV, also a mass shift.
Since both, mass shift and width broadening, lower the threshold energy for antikaon production,  the antikaon yield in heavy-ion collisions is enhanced close to threshold. The produced off-shell antikaons propagate in the medium according to Eqs.~(\ref{eomr}), (\ref{eomp}) and (\ref{eomm}).
According to these equations, antikaons, which are lighter than the pole mass, loose momentum and the ${\rm p_T}$ spectrum softens~\cite{Song:2020clw}.

For the $K^+$ we simply assume that in the medium it is subject to a repulsive potential, which increases linearly with the nuclear density \cite{Korpa:2004ae}. This  increases effectively the $K^+$ mass. Therefore $K^+$ production is suppressed in heavy-ion collisions and the ${\rm p_T}$ spectrum becomes harder~\cite{Song:2020clw}.

We also note that in the present study the realization of the interaction of a $K^-$ with a $\Delta$ baryon has been improved; consequently, the antikaon yield is slightly suppressed compared with our previous results in Ref.~\cite{Song:2020clw}. 
However, this is compensated by the decay of $\phi$ mesons, whose yield is enhanced by including the novel meson-baryon channels through the T-matrix approach as well as by the
collisional broadening of  $\phi$ widths.

In order to pin down the influence of medium effects, we consider different in-medium scenarios for the $\phi$ meson and for strange $K, \bar K$ mesons:\\
I) with and without including collisional broadening of $\phi$ mesons as discussed in Section II ;\\
II) with and without including the in-medium effects for the $K$ and $\bar K$
as described in details in Ref. \cite{Song:2020clw}, i.e.
\begin{itemize}
\item  {\it with in-medium effects for the $K$ and $\bar K$ }: \\
i) for kaons ($K^+, K^0$)  a repulsive density-dependent 
potential is included which increases the kaon mass with increasing density. This yields a shift of the production threshold and modification of production and interaction 
cross sections.\\
ii) for antikaons ($K^-, \bar K^0$) the full density and temperature dependent complex self-energy within a G-matrix approach is accounted: 
the modification of the antikaon masses in line with the in-medium spectral function, off-shell dispersion relations including complex self-energies, in-medium cross sections for the production and interactions as well as the $\bar K$ off-shell propagation.
\item {\it without medium effects for the $K$ and $\bar K$ }: \\
i) for kaons ($K^+, K^0$)  on-shell dynamics without any potential is employed and  free cross sections are used.\\
ii) for antikaons ($K^-, \bar K^0$)  on-shell dynamics using the free production cross sections in the coupled channels is used.
\end{itemize}

\section{$\phi$ meson production in heavy-ion collisions}\label{heavy-ion}

In this section we study $\phi$ meson production close to the threshold by investigating Au+Au collisions at $E_{kin}=$ 1.23 A GeV as measured by the HADES Collaboration as well as $\phi$ production above threshold, at $\sqrt{s_{\rm NN}}=$ 3 GeV, as measured by the STAR collaboration. 
The PHSD calculations include the $\phi$ meson production channels discussed in the last section.  
Here we investigate the dominant channels for the $\phi$ meson production near and above the threshold, the time evolution of the $\phi$ abundances and the rapidity distributions. 
Since the $\phi$ meson is an unstable resonance, any observables related
to the $\phi$ mesons can not be measured directly in experiment, rather the
$\phi$ mesons are reconstructed from the final $K^+, K^-$ mesons
by the invariant mass method.
The difficulty is that the kaons or antikaons can scatter elastically in the medium and the antikaon can be absorbed by inelastic collisions. Moreover, they can change their properties in the medium \cite{Song:2020clw} and thus the 'true' number of produced $\phi$ mesons and their distribution differs from the reconstructed one.

\subsection{Channel decomposition of produced $\phi$ meson versus time}

\begin{figure*}[t!]
\centerline{
\includegraphics[width=7.6 cm]{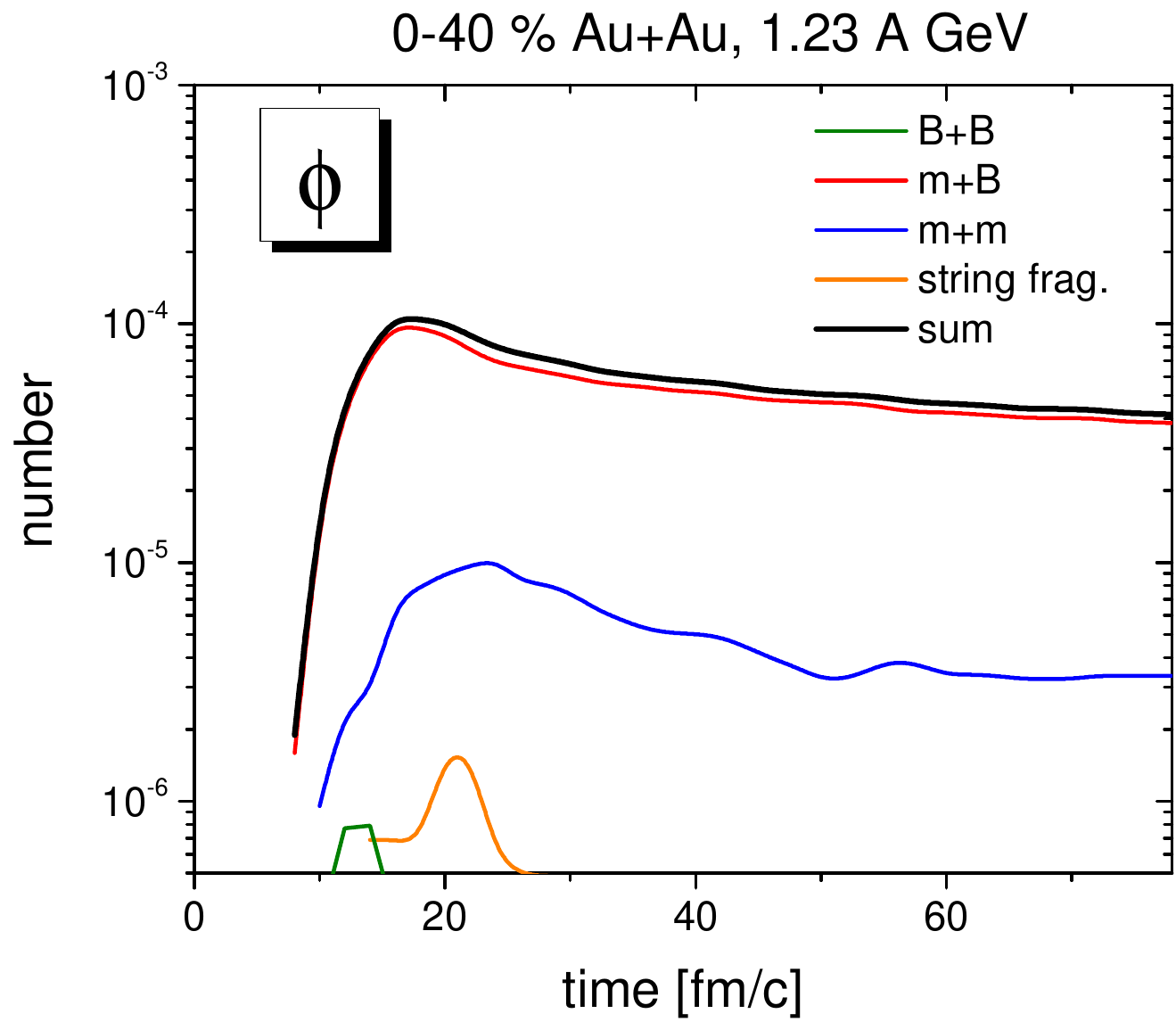} \hspace*{5mm}
\includegraphics[width=7.6 cm]{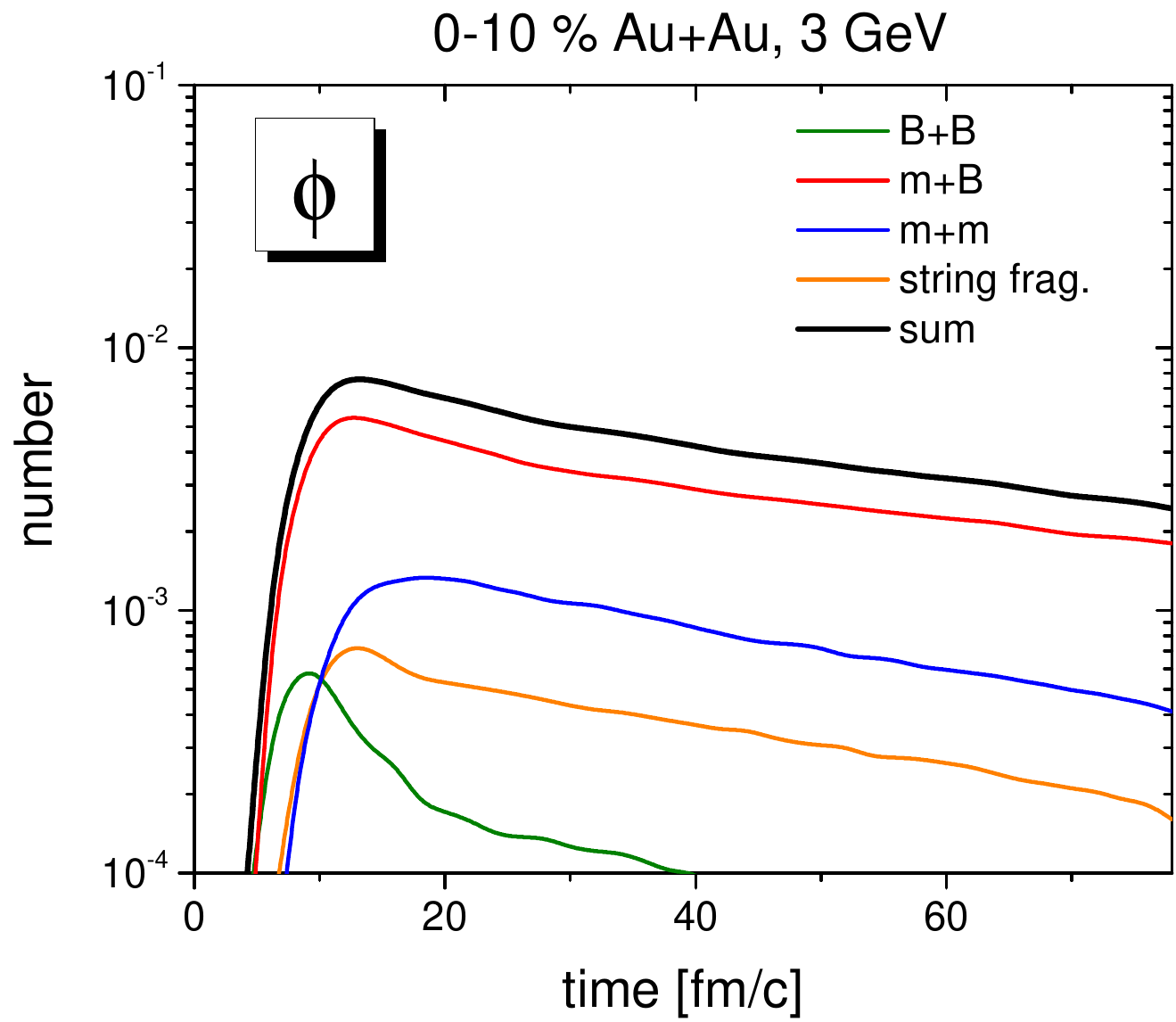}}
\caption{Channel decomposition of the numbers of produced $\phi$ mesons as a function of time in Au+Au collisions, for 0-40\% centrality at $E_{kin}=$ 1.23 A GeV (left) and 
for 0-10\% centrality at $\sqrt{s_{\rm NN}}=$ 3 GeV (right).
The green, red and blue lines show the number of $\phi$ mesons produced in low energy $BB, mB, mm$ reactions, respectively, while the orange line indicates the $\phi$'s produced by string fragmentation,
the black line shows the sum of all contributions.}
\label{production1}
\end{figure*}

Fig.~\ref{production1} shows the number of $\phi$ mesons in the system as a function of time for central Au+Au collisions of the HADES experiment at $E_{kin}$ = 1.23 A GeV (left) and of the STAR  experiment at $\sqrt{s}$ = 3 GeV (right).
The PHSD calculations have been performed including the collisional broadening
of $\phi$ mesons and taking into account the in-medium modifications of the $K$ and $\bar K$ mesons. 

In both cases meson-baryon reactions are the dominant channel for $\phi$ production. The meson-meson reactions, mainly $K +\bar K \to \phi$, are also very important and their contribution grows with energy due to the increasing number of strange mesons.
The $\phi$ production in baryon-baryon collisions and by string fragmentation is sub-dominant. We note that  $\sqrt{s}$ = 3 GeV is slightly above the threshold energy for $N+N\rightarrow N+N+\phi$. Therefore the contribution to the $\phi$ production from baryon-baryon scattering at $\sqrt{s}$ = 3 GeV is much larger than at $E_{kin}$ = 1.23 A GeV, but not dominant due to the small cross sections near threshold, as shown in Fig.~\ref{Anke}. 
Since the $\phi$ decays into a kaon-antikaon pair or three pions, the number of $\phi$ decreases with time after the stage of initial production. This decay is stronger in a nuclear medium because of the broadened width.

\subsection{Mass distribution of $\phi$ mesons}

\begin{figure*}[th!]
\centerline{
\includegraphics[width=8.6 cm]{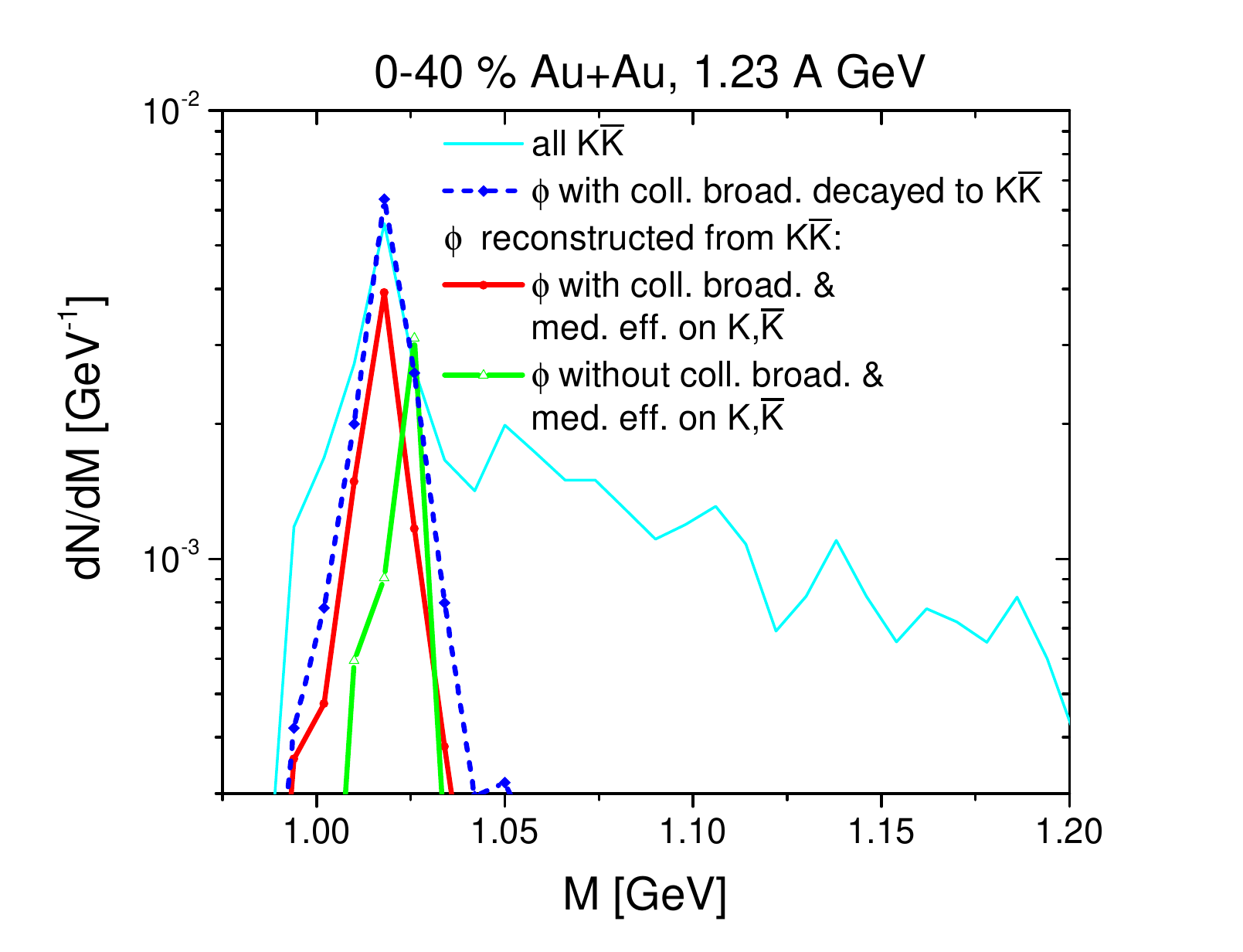}
\includegraphics[width=8.6 cm]{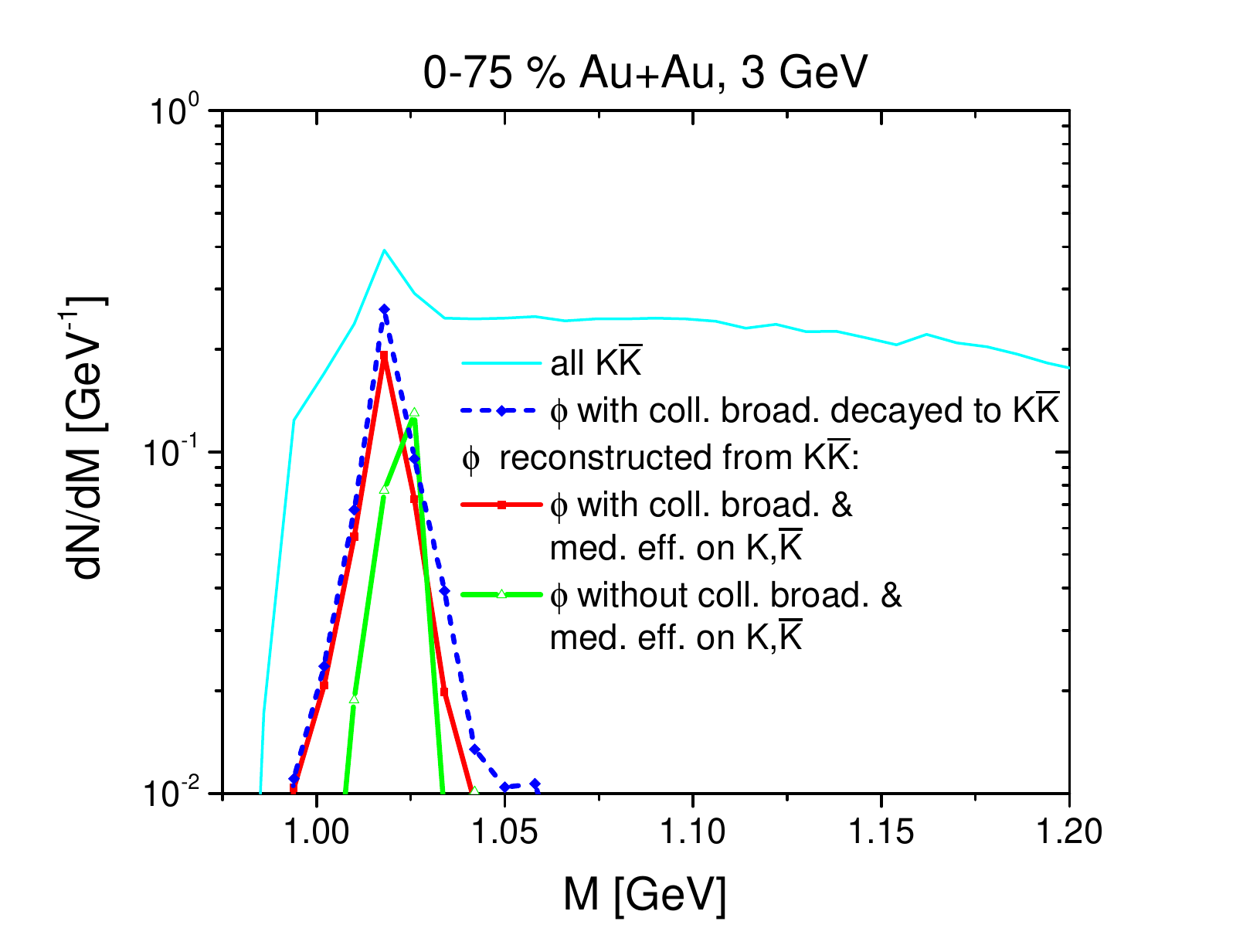}}
\caption{ Invariant mass distribution of $\phi$ mesons created in Au+Au collisions at $E_{kin}$ = 1.23 A GeV (left) and  at $\sqrt{s_{\rm NN}}=$ 3 GeV (right). 
The light blue solid lines indicate all possible combinations of $K\bar K$  pairs 
corrected by branching ratio ($K\bar{K} \equiv (K^+K^-)/Br(\phi\to K^+K^-)$) 
per event at the end of the PHSD calculations with the in-medium effects for $\phi$ and $K, \bar K$. 
The blue dashed lines shows, at the space-time point of the decay, all $\phi$ mesons (with in-medium effects for $\phi$ and $K, \bar K$) which decayed into $K\bar K$  pairs. 
The red solid lines depict the $\phi$'s reconstructed from the final $K\bar K$  pairs 
for the case of collisional broadening of $\phi$ and in-medium effects for $K, \bar K$ mesons, while the green lines show the reconstructed $\phi$'s for the calculation without medium effects for the $\phi, K, \bar K$ mesons.
The number of reconstructed $\phi$ from $K^+K^-$ pairs, divided  by the branching ratio $Br(\phi\to K^+K^-)$.
}
\label{production2}
\end{figure*}

Since the $\phi$ mesons are unstable particles they can not be measured directly and have to be reconstructed by their decay product using the invariant mass method.
$\phi$ mesons  decay dominantly to kaon-antikaon pairs  ($\phi \to K^0\bar K^0; \ K^+K^-$), which makes their reconstruction experimentally difficult  since the kaons and antikaons suffer from final-state interactions with other mesons and baryons in the expanding hadronic matter, i.e. they can rescatter or be absorbed by surrounding hadrons.  The consequences are shown in Fig.~\ref{production2} which presents the invariant mass distribution of $\phi$ mesons, reconstructed from $K^+K^-$ pairs and divided  by the branching ratio $Br(\phi\to K^+K^-)$,
in Au+Au collisions at $E_{kin}$ = 1.23 A GeV (left) and  at $\sqrt{s_{\rm NN}}=$ 3 GeV (right). 
The light blue solid lines indicate all possible combinations of $K\bar K$  pairs 
corrected by branching ratio ($K\bar{K} \equiv (K^+K^-)/Br(\phi\to K^+K^-)$) 
per event at the end of the PHSD calculations employing the in-medium effects for $\phi$ and $K, \bar K$. This means that each $K^-$ has been correlated with all $K^+$ from the same event. This provides an estimate for the combinatorial background. The blue dashed lines shows all $\phi$ mesons which decayed to $K^+K^-$ pairs at the  space-time point of their decay, i.e. the $\phi$ mass distribution. It can be larger than the light blue line because after the decay the (anti-)kaons can rescatter. 

The red solid lines depict the $\phi$'s reconstructed from the final $K\bar K$  pairs for the case of collisional broadening of $\phi$ and in-medium effects for the $K, \bar K$ mesons, while the green lines show the reconstructed $\phi$'s for the case without medium effects for the $\phi, K, \bar K$ mesons. At each energy the number of decayed $\phi$'s (blue dashed lines) is larger than the number of $\phi$'s reconstructed from $K\bar K$ (red solid lines). This reduction of the reconstructed $\phi$ mesons compared to the 'true' decayed $\phi$'s is related to the final state interactions of (anti-)kaons by inelastic reactions - absorption or charge exchange reactions.  We note that in our theoretically 'reconstructed' $\phi$'s we have accounted all final $K^+ K^-$ pairs coming from 'true' $\phi$ decays,  even if one (or both) of the decay products re-scatter elastically. Because elastic collisions - with a usually small momentum transfer - do not change substantially the invariant mass of the $K^+ K^-$ pairs, this theoretical $\phi$ reconstruction is the closest  to the experimental procedure to identify $\phi$'s.

The light blue line and the red line differ due to the background ((anti-)kaons which do not come from a $\phi$ decay) which contributes only to the light blue line.
One can see from Fig.~\ref{production2} that the total number of  $K\bar K$  pairs (light blue line) is much larger in Au+Au collisions at $\sqrt{s}=3$  GeV than at $E_{kin }$ = 1.23 A GeV due to the opening of multiple channels for the $K, \bar K$ production with increasing energy. On the other hand, the number of $\phi$'s ('true' decayed and reconstructed from the final $K\bar K$ pairs)  also grows with energy.

One sees also the large effect of the collisional broadening of $\phi$ mesons, green lines versus red lines, which leads to the widening of $dN/dM$ distribution if  medium effects are included compared to the case of no in-medium effects.

\subsection{Channel decomposition of the rapidity distribution of reconstructed $\phi$ mesons}

\begin{figure*}[th!]
\centerline{
\includegraphics[width=8.2 cm]{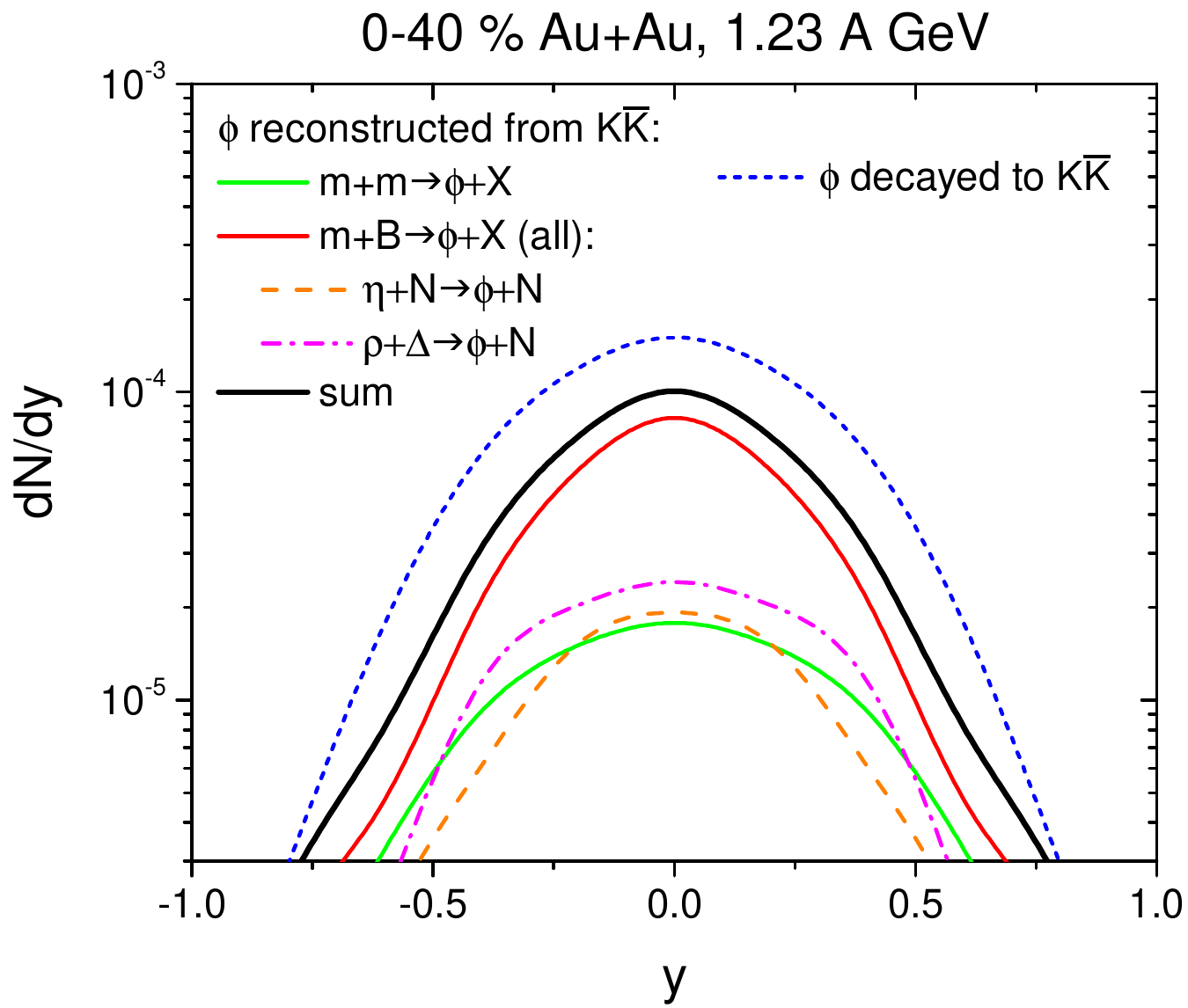} \hspace*{0.5cm}
\includegraphics[width=8.1 cm]{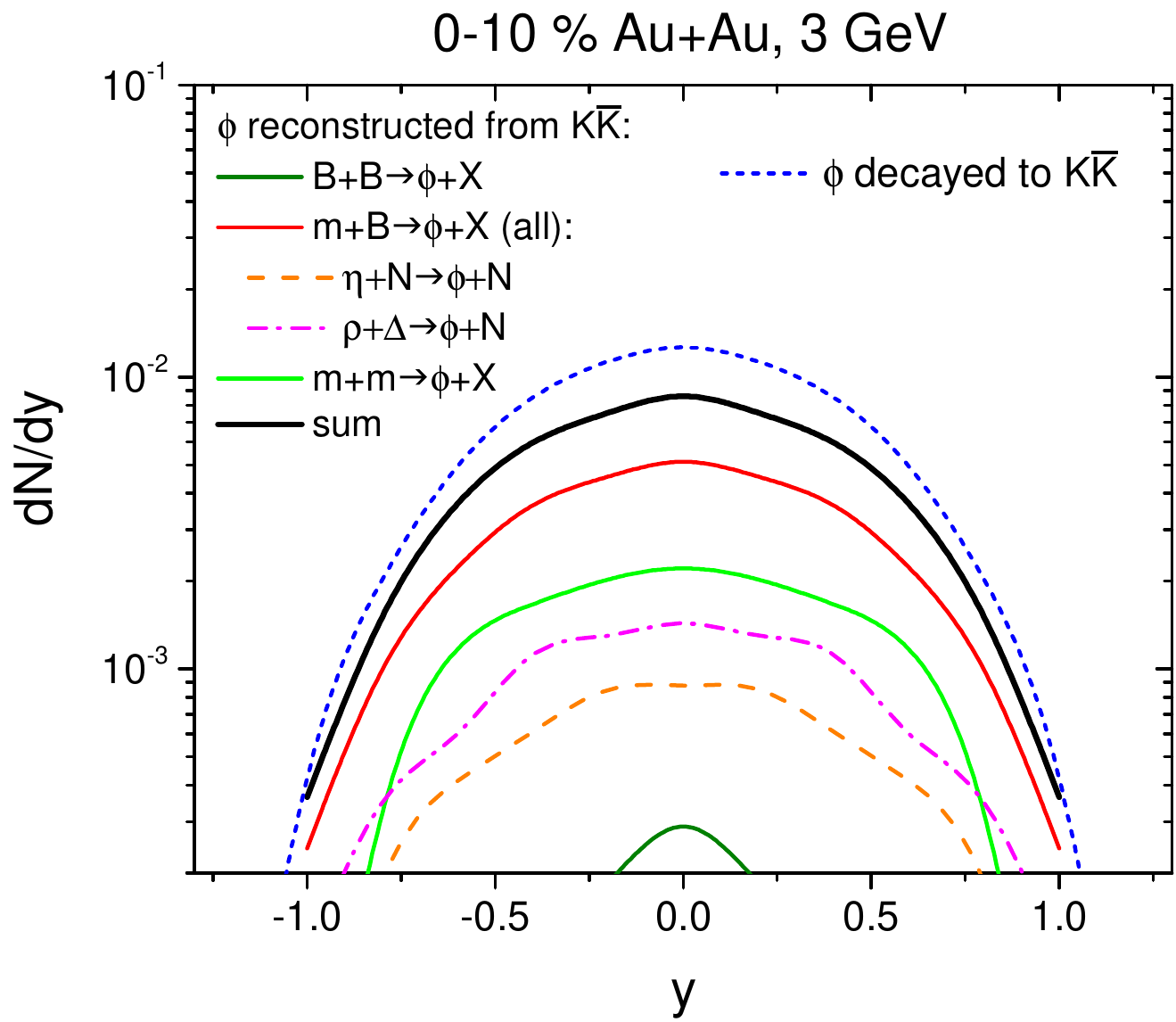}}
\caption{ Channel decomposition of the rapidity distribution of reconstructed $\phi$ mesons  in Au+Au collisions 
for 0-40\% centrality at $E_{kin}=$ 1.23 A GeV (left) and 
for 0-10\% centrality at $\sqrt{s_{\rm NN}}=$ 3 GeV (right).
The olive, red and green lines show the number of $\phi$ mesons produced in low energy $BB, mB, mm$ reactions, respectively, while the black line shows the sum of all contributions. The dominant $mB$ contributions are specified explicitly:
the dashed red line shows the contribution from $\eta+N\rightarrow \phi+N$ 
and the dashed-dotted magenta lines indicate  $\rho+\Delta \rightarrow \phi+N$ contributions.
The blue dotted lines display $dN/dy$ of $\phi$'s at their space-time decay points.
All PHSD calculations are performed including collisional broadening of $\phi$ mesons and in-medium effects for $K, \bar K$ mesons.
The number of reconstructed $\phi$ from $K^+K^-$ pairs, divided  by the branching ratio $Br(\phi\to K^+K^-)$.
}
\label{production3}
\end{figure*}

The rapidity distribution of the reconstructed $\phi$ mesons, separated by the production channels,  are presented in Fig.~\ref{production3} for Au+Au collisions for 0-40\% centrality at $E_{kin}=$ 1.23 A GeV (left) and for 0-10\% centrality at $\sqrt{s_{\rm NN}}=$ 3 GeV (right). All PHSD calculations are performed including collisional broadening of $\phi$ and in-medium effects for $K, \bar K$ mesons.
The olive, red and green lines show the number of $\phi$ mesons produced in low energy $BB, mB, mm$ reactions, respectively, while the black line shows the sum of all contributions. 
The dominant contribution for the $\phi$ production comes from meson-baryon scattering. 
We find that $\eta+N\rightarrow \phi+N$ (indicated by the dashed red lines) and $\rho+\Delta \rightarrow \phi+N$ (the dash-dotted magenta lines), both calculated by the T-matrix approach (section IV), are most important at the energies 
considered here.

The blue dotted lines show the $dN/dy$ of $\phi$ mesons at space-time point of their decay. They include those $\phi$ mesons, which could not be reconstructed experimentally (as discussed in the previous section VI.B) due to the final state interaction of one of the decay products with the hadronic medium.
Comparing the blue dotted and red solid lines, we find that about 60\% of the $\phi$ mesons can be reconstructed in 0-40\% central Au+Au collisions at $E_{kin}$= 1.23 A GeV and about 70\% in 0-75\% central Au+Au collisions at $\sqrt{s_{\rm NN}}=$ 3 GeV.

\begin{figure*}[th!]
\centerline{
\includegraphics[width=8.6 cm]{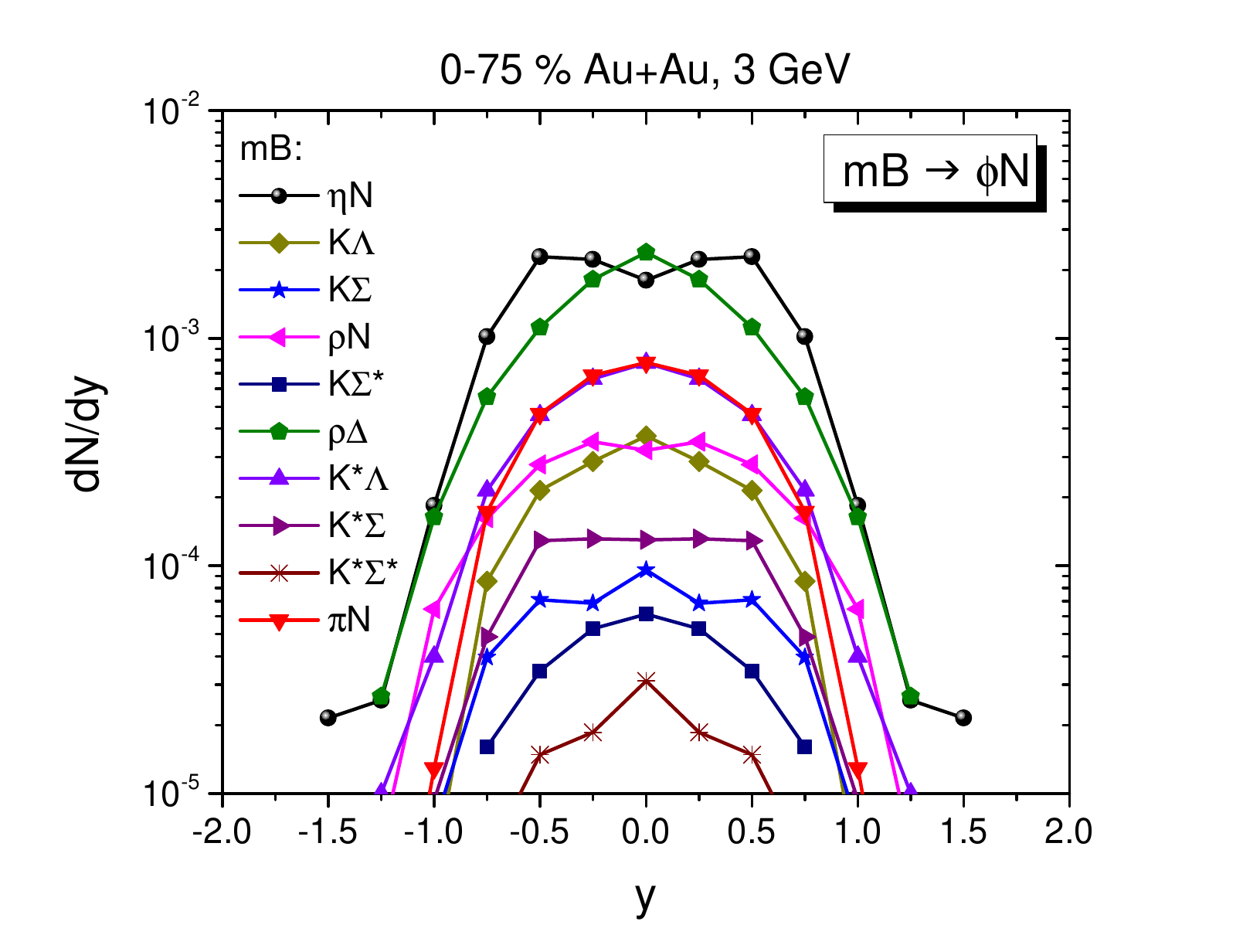}
\includegraphics[width=8.6 cm]{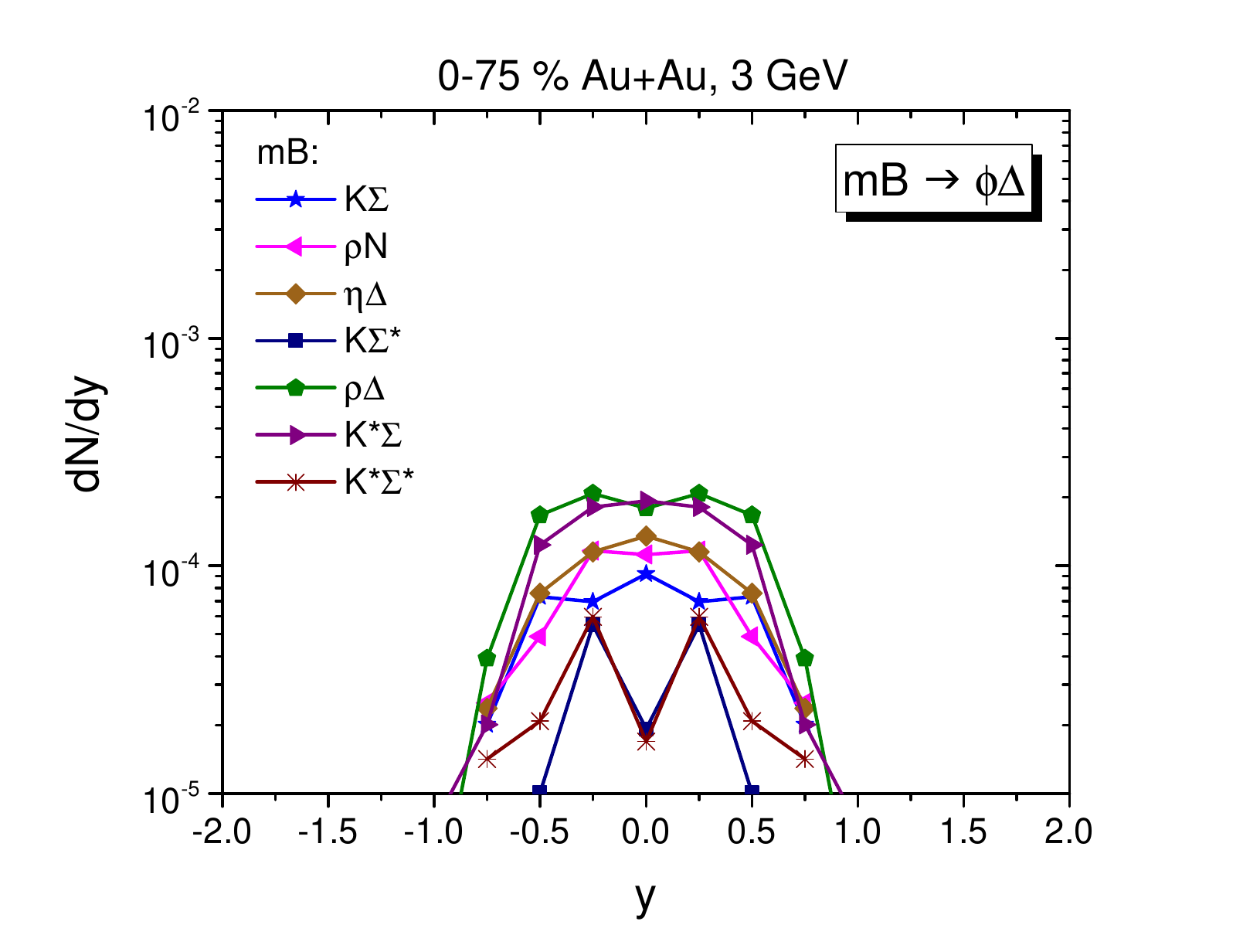}}
\centerline{
\includegraphics[width=8.6 cm]{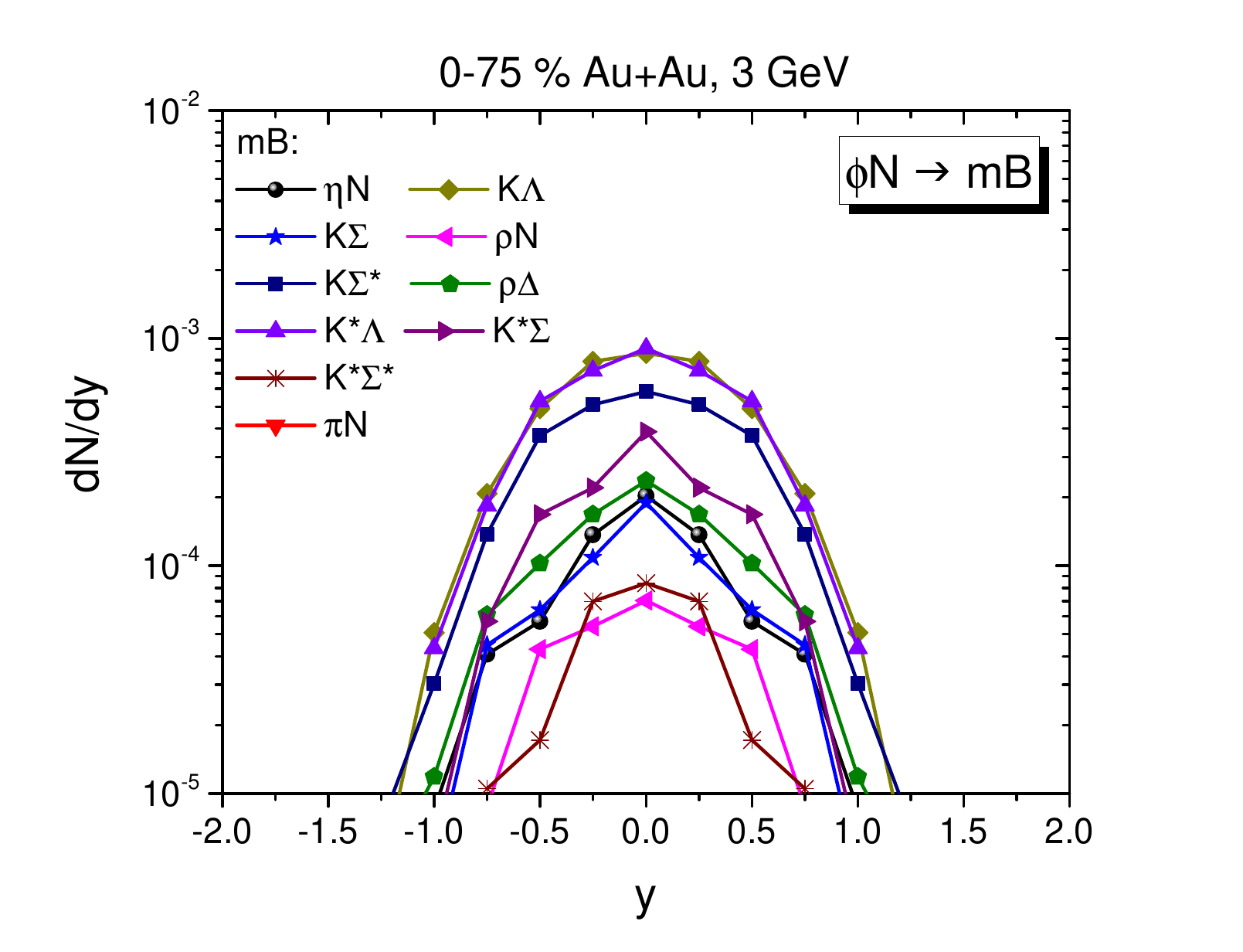}
\includegraphics[width=8.6 cm]{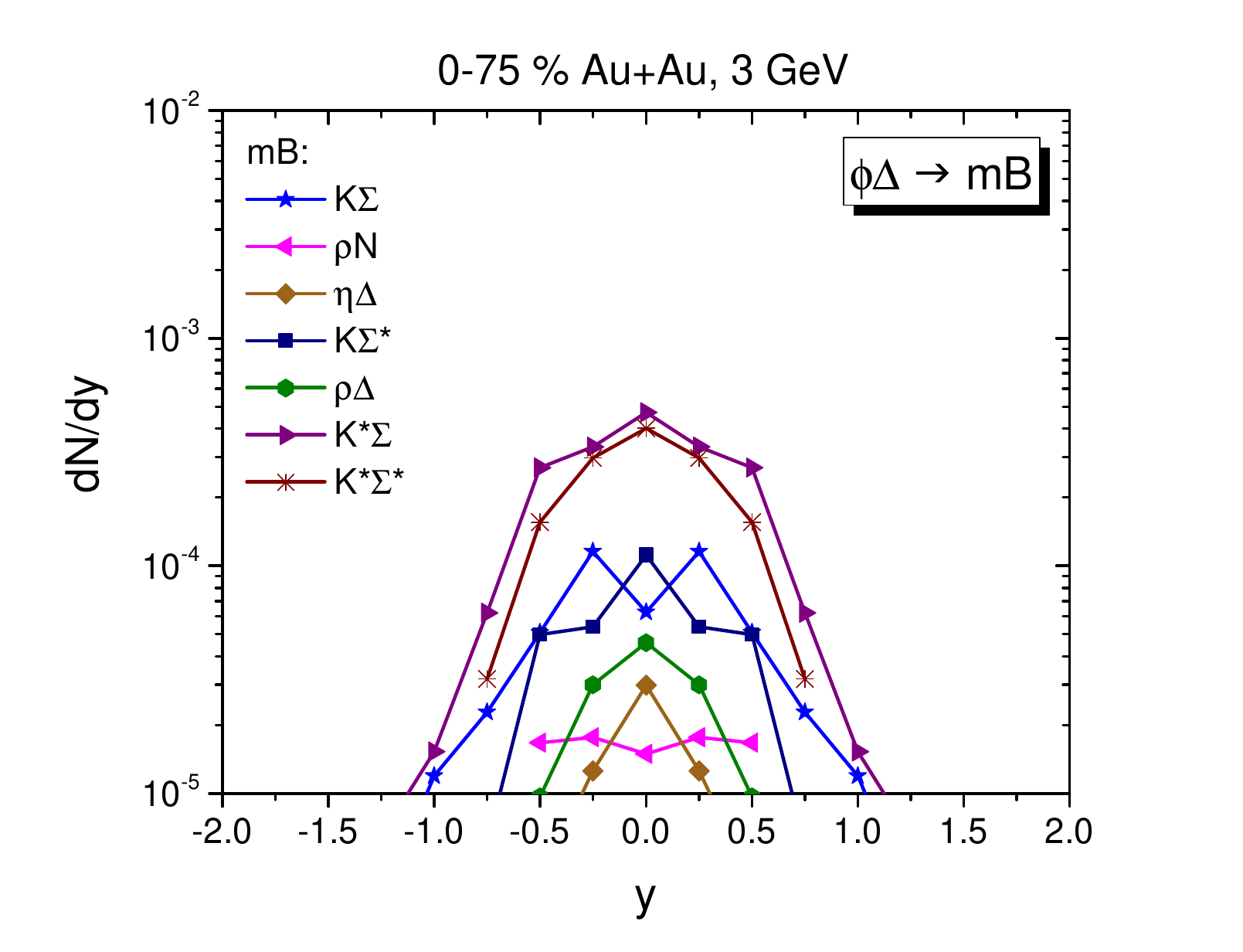}}
\caption{Top: channel decomposition of $\phi$ mesons production in 
 $mB\to \phi N$  (top left) and in $mB\to \phi \Delta$ reactions (top right) in 0-75 \% central Au+Au collisions at $\sqrt{s_{\rm NN}}=$ 3 GeV. 
Bottom:  channel decomposition of $\phi$ meson absorption  by $\phi N \to mB$ (bottom left)
and by $\phi \Delta \to mB$ reactions (bottom right). 
The individual channels, depicted in the legend, show the PHSD results  for the novel $mB \leftrightarrow \phi B$ channels from  the T-matrix approach.
All PHSD calculations are performed for the case of collisional broadening of $\phi$'s and in-medium effects for $K, \bar K$ mesons. 
}
\label{dndy}
\end{figure*}

Supplementary to Fig.~\ref{production3}, in Fig.~\ref{dndy} we display the 
accumulated rapidity distribution  for the 
$\phi$ production (top) and absorption (bottom) by nucleons (left) and by $\Delta$ resonances (right), separated for each meson-baryon channel from our $T$-matrix approach, in 0-75 \% central Au+Au collisions at $\sqrt{s_{\rm NN}}=$ 3 GeV. 
Since hyperons are rare compared to nucleons or $\Delta$'s, the contribution from $K+Y\rightarrow \phi+N$ is less relevant in spite of the larger cross section (see Fig. \ref{csSU6}). 
The most dominant channels are therefore $\eta+N\rightarrow \phi+N$ and $\rho+\Delta \rightarrow \phi+N$.
Though the number of nucleons is larger than that of $\Delta$s, $\rho+\Delta \rightarrow \phi+N$ is more dominant than $\rho+N \rightarrow \phi+N$, since the cross section for the latter is smaller and kinetically suppressed due to the smaller mass of  nucleon as compared to that of the $\Delta$. 
On the other hand, $\phi+N \rightarrow K+Y$ is the dominant channel for $\phi$ absorption due to its large cross section.

\subsection{Comparison of $\phi$ meson production with experimental data}\label{experiments}

\begin{figure*}[th!]
\centerline{
\includegraphics[width=8.6 cm]{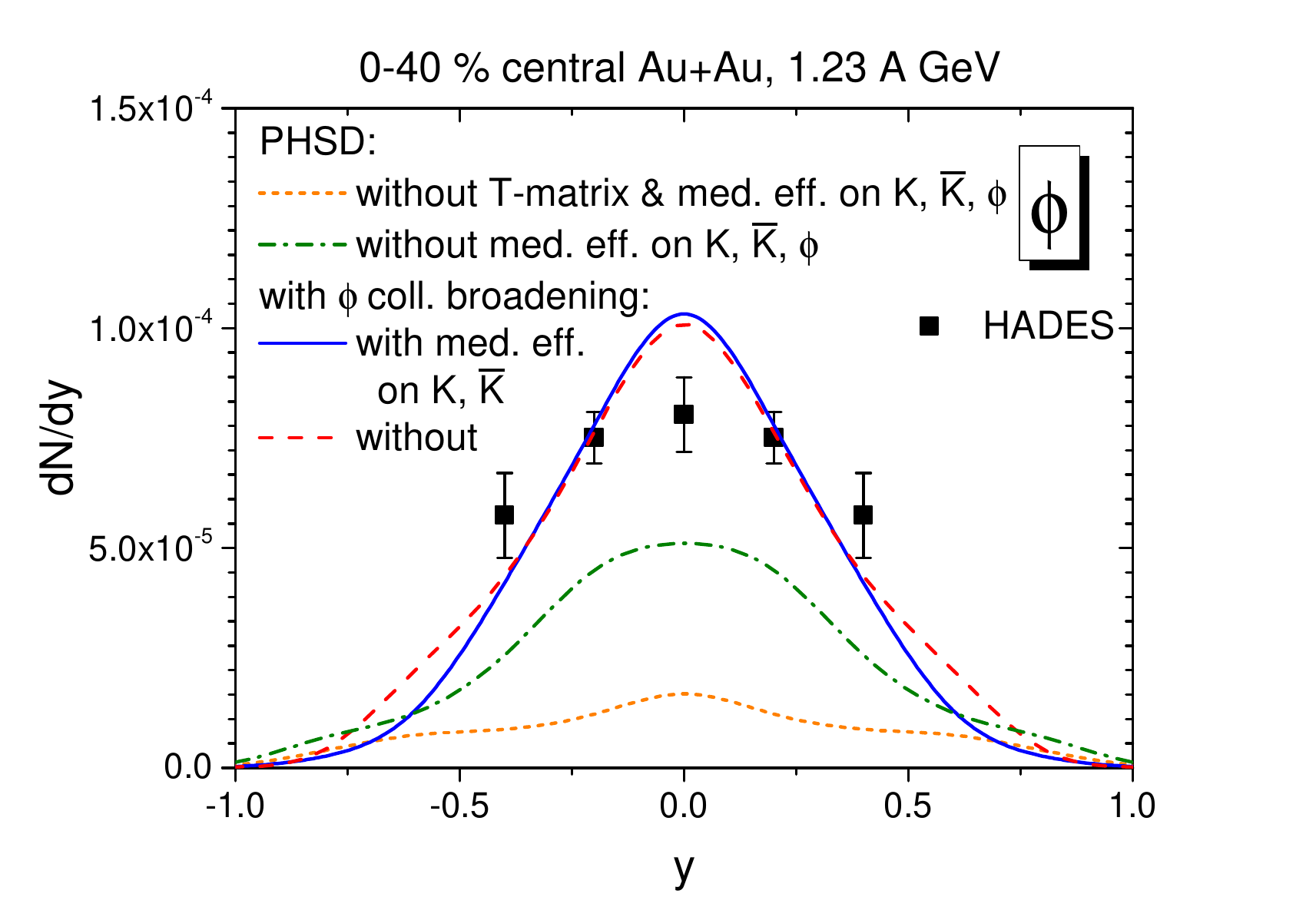} 
\includegraphics[width=8. cm]{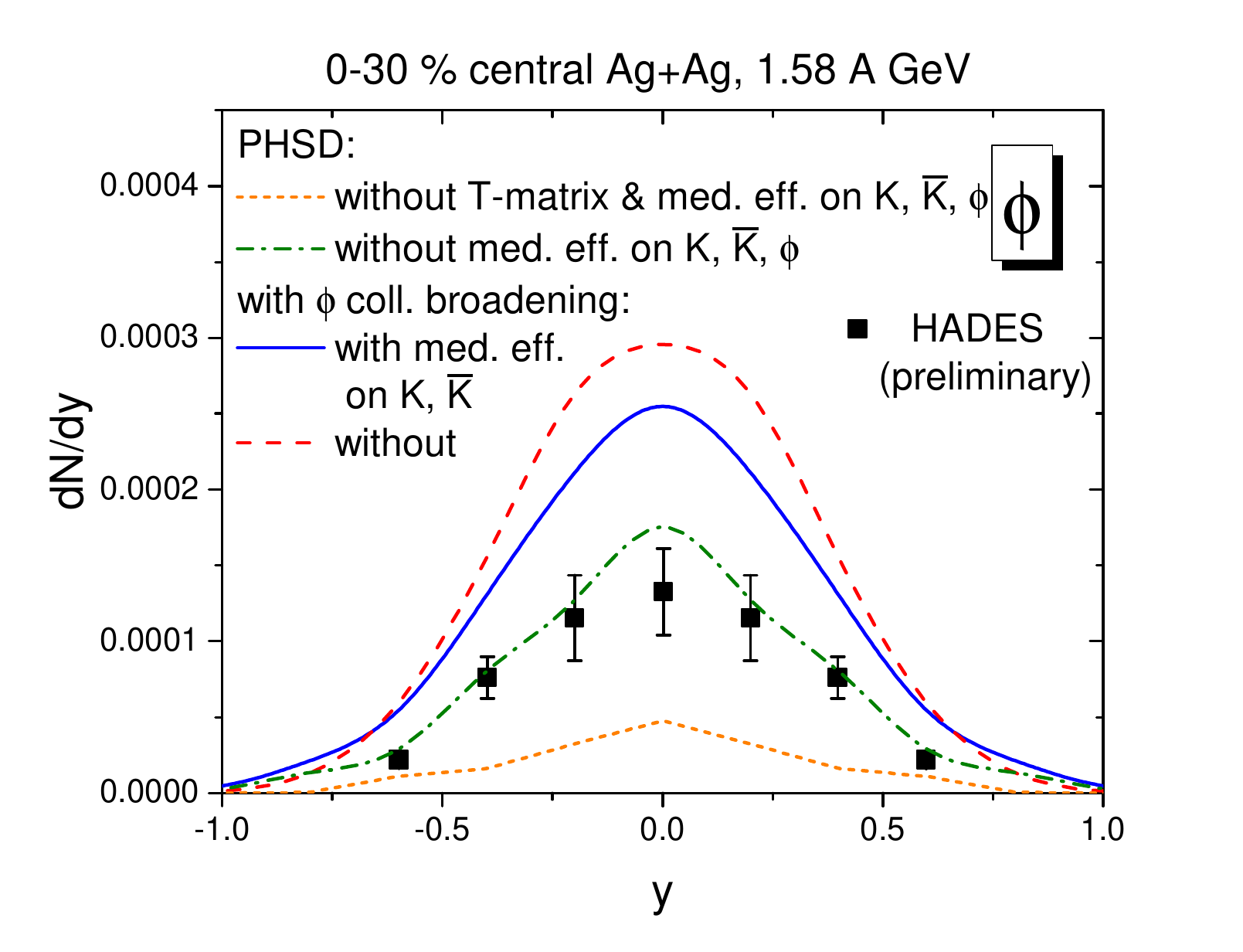} }
\caption{Rapidity distribution of reconstructed $\phi$ mesons 
in 0-40 \% central Au+Au collisions at $E_{kin}=$ 1.23 A GeV in comparison with experimental data from the HADES collaboration~\cite{HADES:2017jgz} and in 0-30 \% central Ag+Ag collisions at $E_{kin}$ = 1.58 A GeV in comparison with preliminary HADES data ~\cite{HADES-Ag}.
The short dashed orange lines  show the PHSD results  without including the novel $mB$ channels from the T-matrix approach and without any in-medium modifications of $\phi$ and $K, \bar K$ mesons.
The dash-dotted green lines show the results without in-medium modifications of $\phi$ and $K, \bar K$ mesons.
The dashed red lines indicate the $\phi$ rapidity distributions with $\phi$ collisional broadening, but without in-medium effects for $K, \bar K$ mesons. The solid blue lines show the results with collisional broadening  for $\phi$ mesons and with in-medium modifications of  $K, \bar K$ mesons.The number of $\phi$ mesons, reconstructed from $K^+K^-$ pairs, is divided  by the branching ratio $Br(\phi\to K^+K^-)$.}
\label{dndym}
\end{figure*}

\begin{figure*}[th!]
\centerline{
\includegraphics[width=8.6 cm]{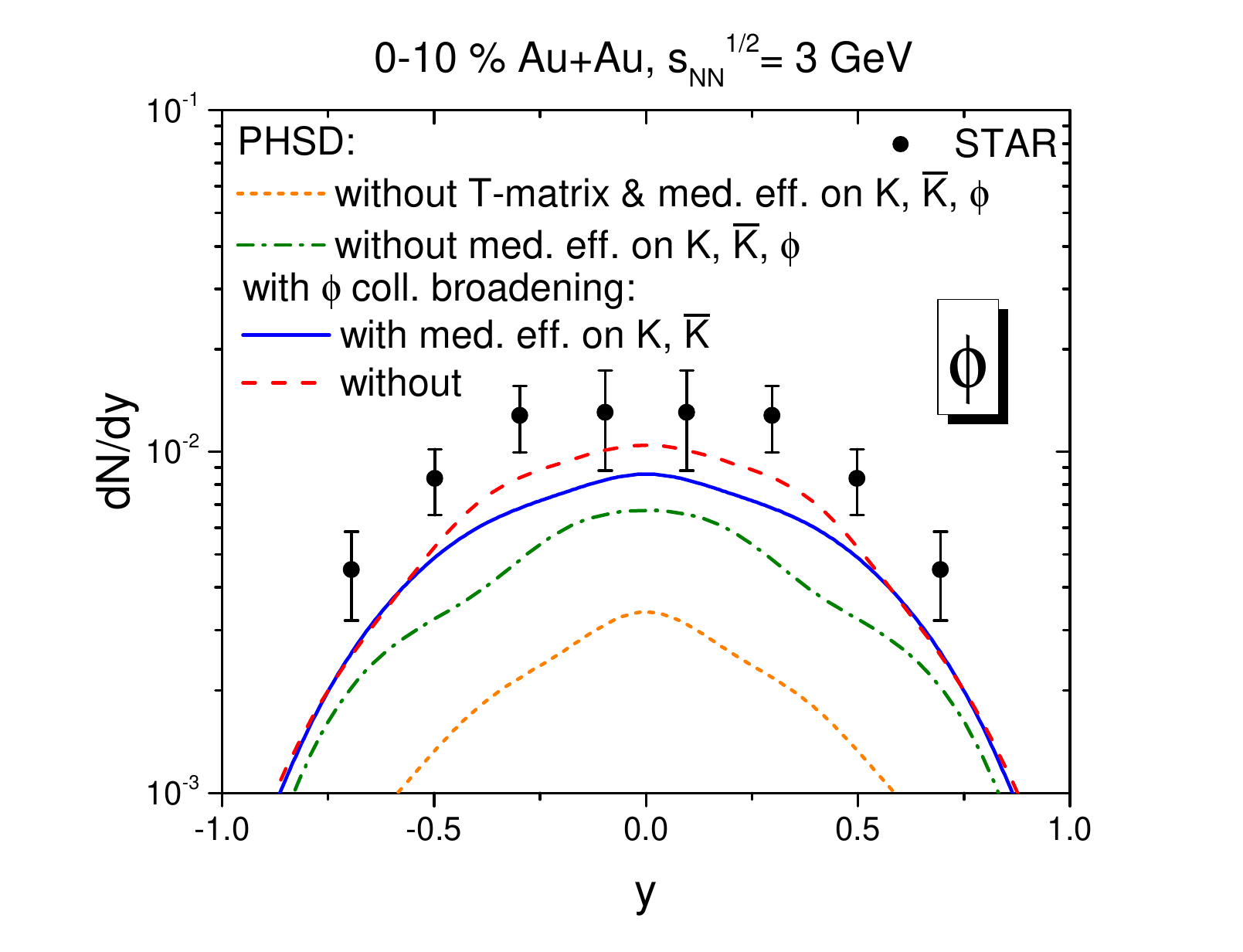} 
\hspace*{5mm}
\includegraphics[width=8.6 cm]{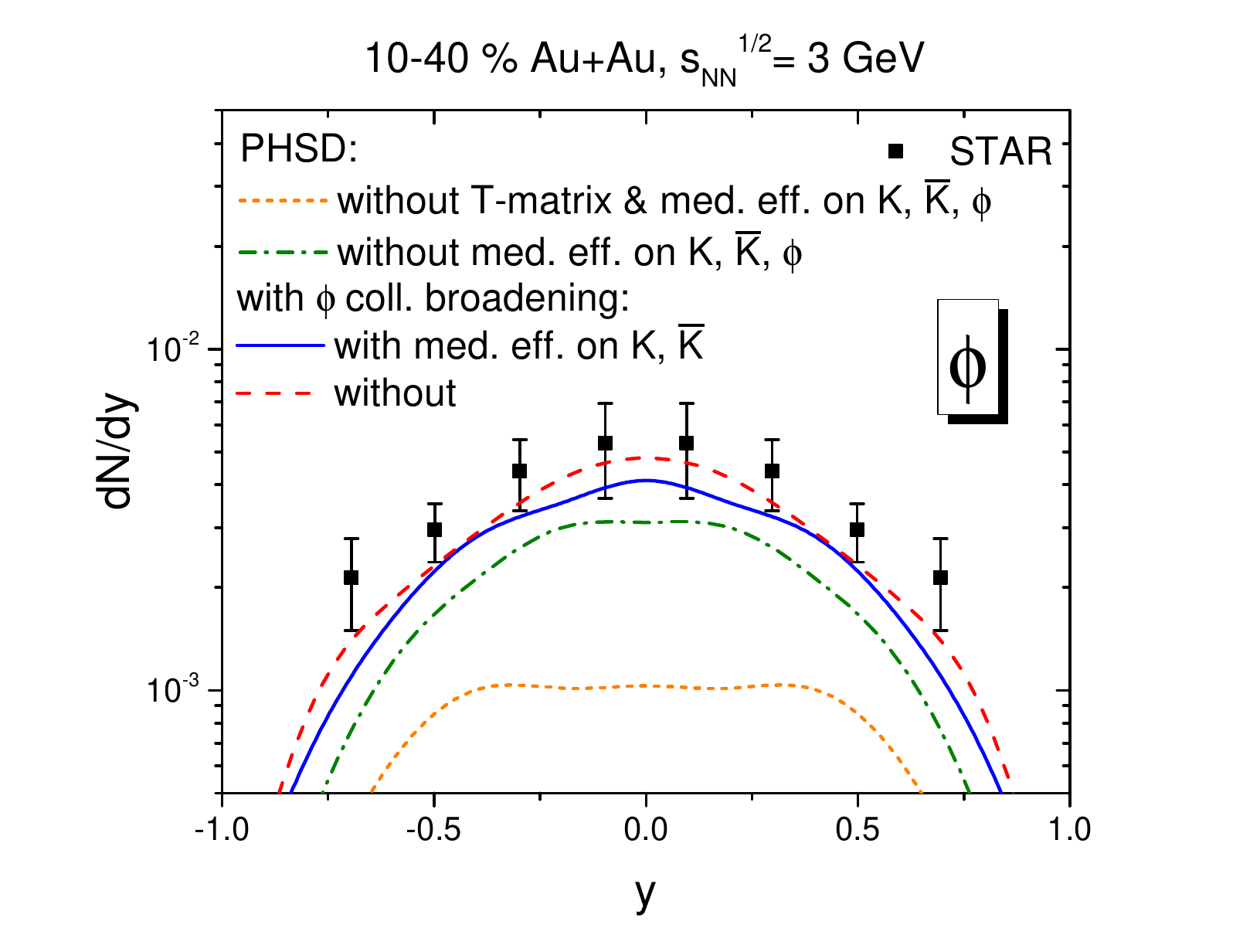}}
\caption{Rapidity distribution of reconstructed $\phi$ mesons in
0-10 \% (left)  and 10-40 \% (right) central Au+Au collisions at $\sqrt{s_{\rm NN}}=$ 3 GeV in comparison with experimental data from the STAR collaboration~\cite{STAR:2021hyx}.
The short dashed orange lines  show the PHSD results  without including the novel $mB$ channels from our T-matrix approach and without any in-medium modification of $\phi$ and $K, \bar K$ mesons.
The dash-dotted green lines show the results without in-medium modifications of $\phi$ and $K, \bar K$ mesons. The dashed red lines indicate the $\phi$ rapidity distributions with $\phi$ collisional broadening but without in-medium effects for $K, \bar K$ mesons. The solid blue lines depict the results with collisional broadening for $\phi$ mesons and with in-medium modifications of  $K, \bar K$ mesons.
The number of $\phi$ mesons, reconstructed from $K^+K^-$ pairs, is divided  by the branching ratio $Br(\phi\to K^+K^-)$.}
\label{dndym3GeV}
\end{figure*}

\begin{figure*}[th!]
\centerline{
\includegraphics[width=8.6 cm]{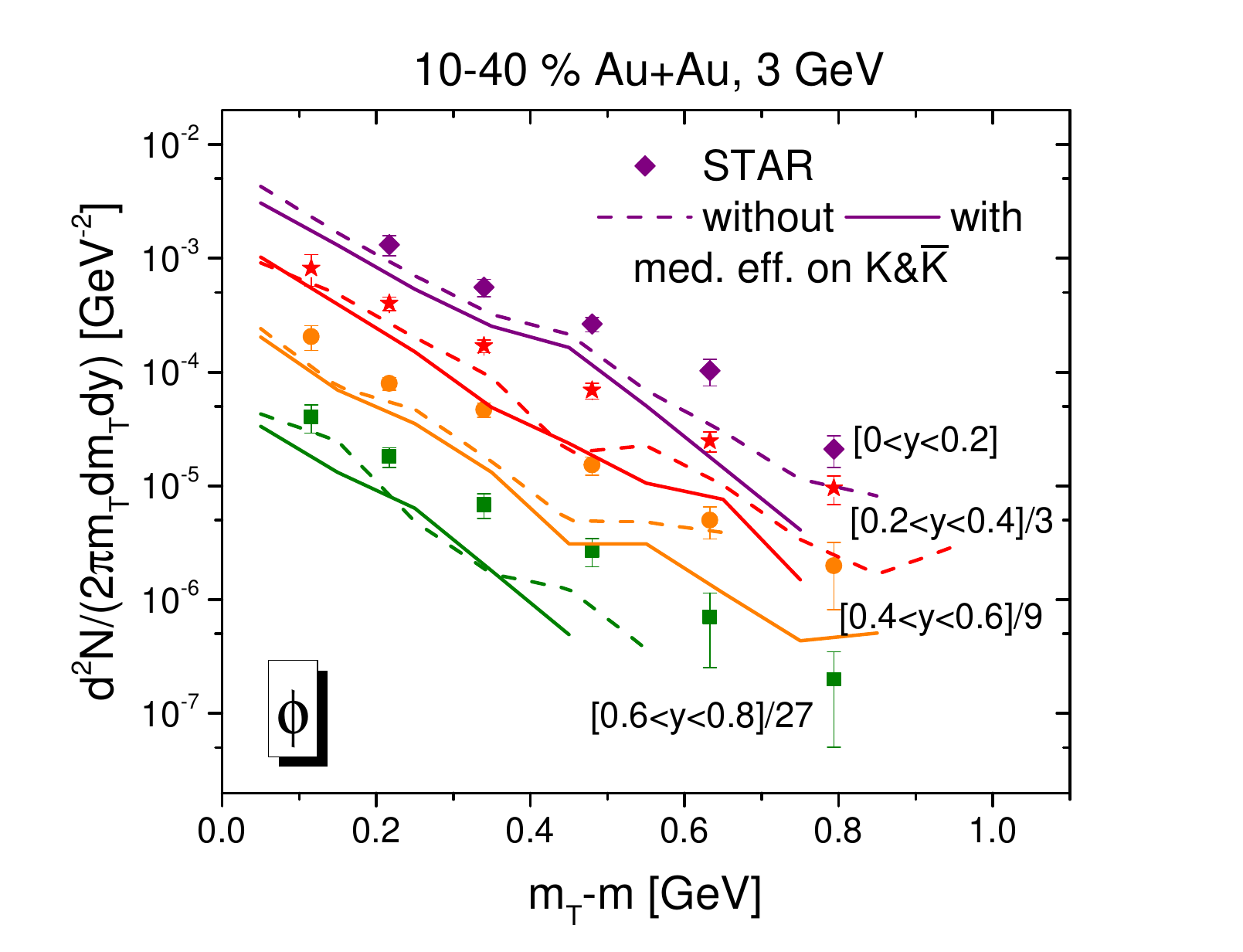} 
\hspace*{-15mm}
\includegraphics[width=8.6 cm]{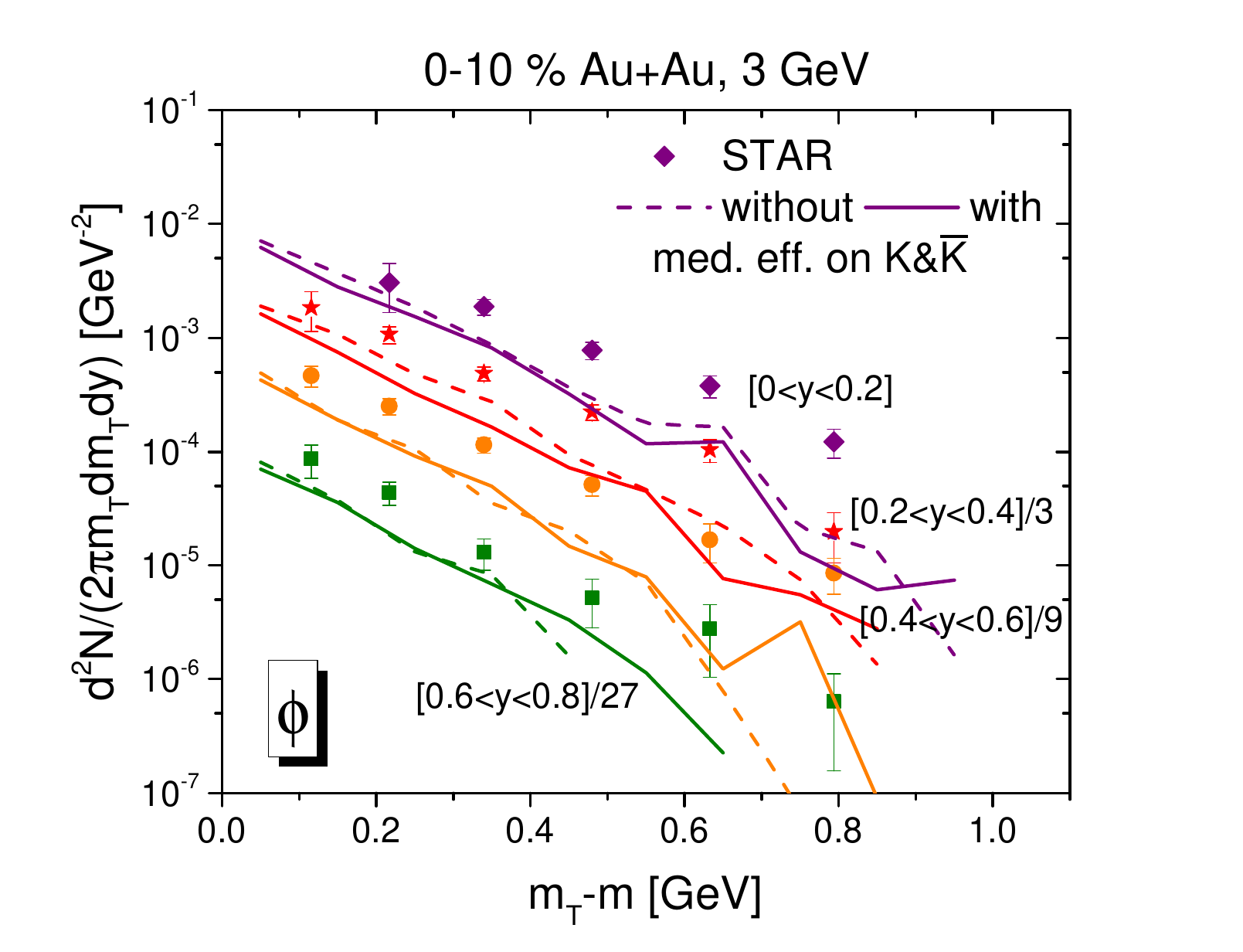} } 
\caption{$m_T$ spectra  of reconstructed $\phi$ mesons in
0-10 \% (left)  and 10-40 \% (right) central Au+Au collisions at $\sqrt{s_{\rm NN}}=$ 3 GeV in comparison with experimental data from the STAR collaboration~\cite{STAR:2021hyx}.
The number of $\phi$ mesons, reconstructed from $K^+K^-$ pairs, is divided  by the branching ratio $Br(\phi\to K^+K^-)$. The calculations are done with the collisional broadening of $\phi$ mesons and  presented for 4 rapidity bins
$0 \le y \le 0.2$, \ $0.2 \le y \le 0.4$ scaled by a factor $1/3$ for better visualization, \ $0.4 \le y \le 0.6$ scaled by a factor $1/9$
and  $0.6 \le y \le 0.8$ scaled by a factor $1/27$. The dashed lines show 
the results without in-medium effects for $K, \bar K$ mesons, while the solid lines that with in-medium modifications of  $K, \bar K$ mesons.}
\label{dndmt3GeV}
\end{figure*}

Figs.~\ref{dndym} and \ref{dndym3GeV} show the PHSD results for the rapidity distribution of reconstructed $\phi$ mesons from the decay into $K^+K^-$ pairs, compared with the experimental data from the HADES and STAR collaborations.
The short dashed orange lines  show the PHSD results  without including of the novel $mB$ channels from the T-matrix approach and without any in-medium modification of $\phi$ and $K, \bar K$ mesons. That was the  "old" setup for $\phi$ production in PHSD, which underestimated substantial the measured experimental data for Au+Au collisions at 1.23 A GeV and 3 GeV. 
All other lines demonstrate the influence of the in-medium effects when the T-matrix $mB$ channels are included:
the dash-dotted lines show the results without in-medium modifications of $\phi$ and $K, \bar K$ mesons. One can see that only including the novel $mB$ channels leads to a substantial enhancement of the $\phi$ yields, which, however, are below the experimental data.
Furthermore, we show the influence if we include the in-medium modification of $K, \bar K$:
the dashed lines indicate the $\phi$ rapidity distributions with $\phi$ collisional broadening, but without in-medium effects for $K, \bar K$ mesons. 
The solid  lines depict the results with collisional broadening  for $\phi$ mesons and with in-medium modifications of  $K, \bar K$ mesons.
Comparing the dashed and solid lines, no significant effect of in-medium modifications of  $K, \bar K$ mesons  on $\phi$ production is observed. The reason is that the main production channel of $\phi$ mesons is not $K+Y$,  but $\eta+N$ and $\rho+\Delta$ scattering. 
On the other hand, the consequences of the in-medium modification of $\phi$ meson by including  collisional broadening are clearly visible. 
Since the broadening of the $\phi$ meson spectral function enhances the $\phi$ production, the dot-dashed lines are always below the solid and dashed lines.

Thus, as follows from Figs.~\ref{dndym} and \ref{dndym3GeV}, including the novel 
$mB$ channels, calculated within the T-matrix of SU(6) coupled channel approach, 
as well as the collisional broadening of the $\phi$ spectral function,
PHSD reproduces the $\phi$ production in Au+Au collisions at  $E_{kin}$ = 1.23 A GeV without introducing the unobserved decay of high mass baryonic resonances to $\phi$ mesons as proposed in other approaches ~\cite{Steinheimer:2015sha,Steinberg:2018jvv}. This shows the importance of the meson-baryon channels for $\phi$ production incorporated in PHSD, 
which are absent or different in strength in the other approaches. We note that $mB$ channels are accessible only theoretically. The T-matrix approach based on the extended SU(6) coupled channel approach provides the presently best framework to take them into account.

Additionally to the rapidity distributions, we show in Fig. \ref{dndmt3GeV} for 4 rapidity bins the
$m_T$ spectra  of reconstructed $\phi$ mesons in
0-10 \% (left)  and 10-40 \% (right) central Au+Au collisions at $\sqrt{s_{\rm NN}}=$ 3 GeV in comparison with  experimental data from the STAR collaboration~\cite{STAR:2021hyx}.
The calculations are done with the collisional brodening of $\phi$ mesons;
the dashed lines show the results without in-medium effects for $K, \bar K$ mesons, while the solid lines that with in-medium effects.
The PHSD calculations approximately reproduce the slope of the $m_T$ spectra
of $\phi$ mesons while the yield is underestimated with collisional broadening scenario as follows from Fig. \ref{dndym3GeV}.
One can see again that the influence of in-medium effects for $K, \bar K$ mesons
is relatively small as explained above.

In spite that the PHSD results are comparable to the HADES data for Au+Au collisions at $E_{kin}$ = 1.23 A GeV, the PHSD overestimates the $\phi$ production in Ag+Ag collisions at $E_{kin}$ = 1.58 A GeV when the collisional broadening of $\phi$'s is included. 
On the other hand, the STAR data on Au+Au collisions at $\sqrt{s_{\rm NN}}=$ 3 GeV are slightly underpredicted for both, for 0-10 \% and 10-40 \% centrality, though the collisional broadening of $\phi$ mesons is taken into account.

The PHSD offers the opportunity to investigate how in-medium effects, acting on $\phi$ and on (anti-)kaon, which will be briefly described in the next section, affect the $\phi$ meson production in heavy-ion collisions.
 

\section{Influence of new mechanisms of $\phi$ meson production on the (anti-)kaon dynamics}\label{influence}

The enhanced $\phi$ production - discussed above - has  consequences for the (anti-)kaon dynamics due to the coupling of the hidden and open strange hadrons. This coupling is more pronounced for the $\bar K$ mesons since at low energies $\bar K$'s are  more rarely produced  compared to $K$, i.e. the "feed down" from $\phi$ decay to $\bar K$ is more visible than that to $K$ mesons. Here we investigate this impact by studying the channel decomposition for $K^-$ production as well as by comparing  the PHSD calculations for different in-medium scenarios with experimental data.

\begin{figure}[!th]
\centerline{
\includegraphics[width=8.6 cm]{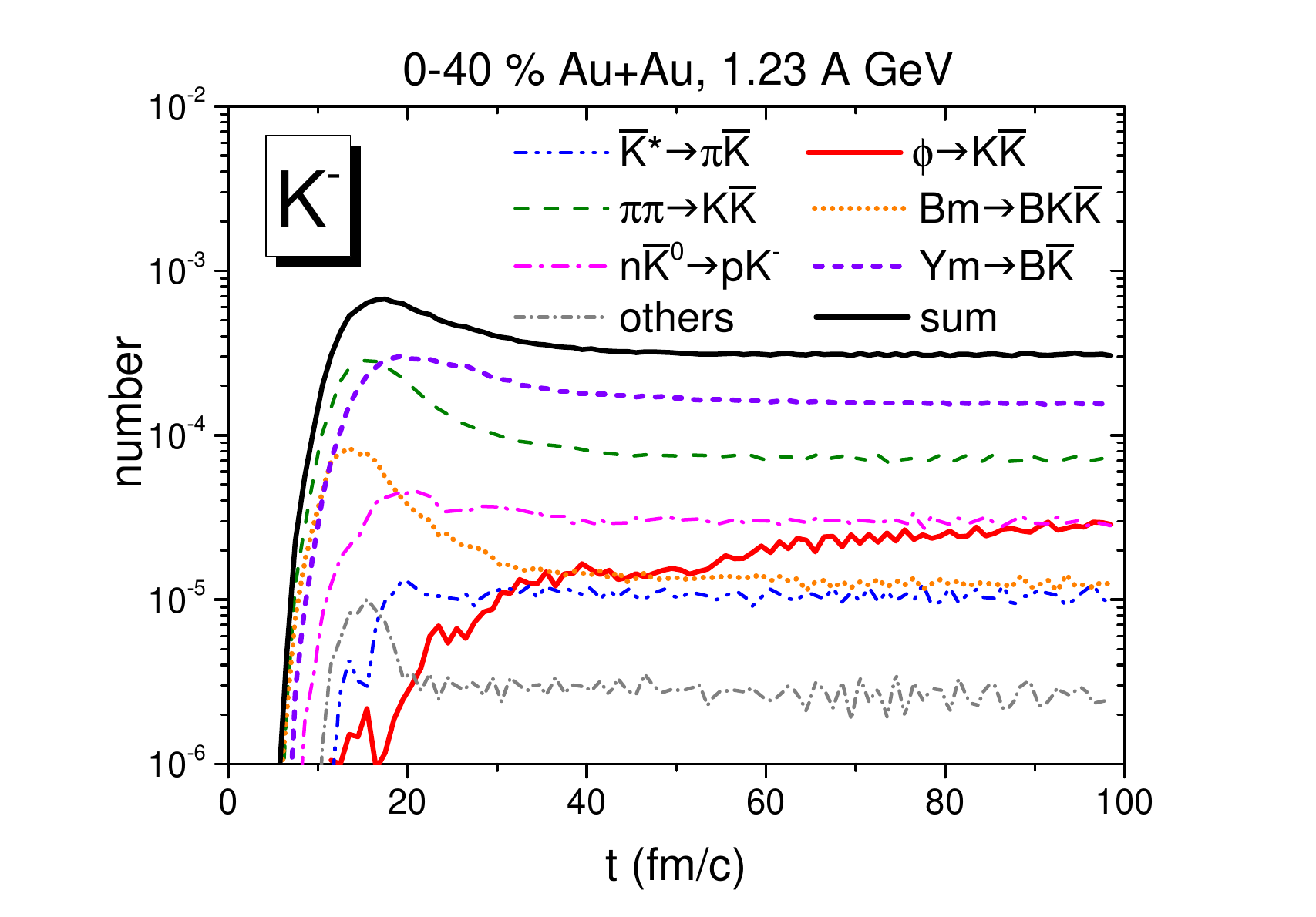}}
\caption{Channel decomposition of the $K^-$ yield for 0-40 \% central Au+Au collisions at $E_{kin}$ = 1.23 A GeV as a function of time. The individual channels are depicted in the legend. The solid red line shows the contribution from the $ \phi \to K^+K^-$ decay. The PHSD calculations are performed including $\phi$ and (anti-)kaon in-medium effects.
}
\label{sources}
\end{figure}

Figure~\ref{sources} shows the channel decomposition of the $K^-$
yield as a function of time in 0-40 \% central Au+Au collisions at $E_{kin}=$ 1.23 A GeV. In-medium effects 
for the $\phi, K, \bar K$ are included in the PHSD calculations. The dominant contributions are hyperon-meson and meson-meson scattering.
The contribution from $\phi$ meson decay (red solid line) is similar to that from the isospin exchange $\bar{K}^0 \to K^-$. 
The contribution from  $\phi \to K^+K^-$ is much larger than in our previous work~\cite{Song:2020clw} since we are now equipped with the T-matrix approach to describe the production of  $\phi$ mesons through meson-baryon scattering.
We mention that only a part of the $\phi$ mesons decay to $K^+$ and $K^-$. Therefore, to obtain the total number of $\phi$ mesons, one has to divide those obtained from reconstructed $K^+K^-$ pairs by the branching ratio $Br(\phi\to K^+K^-)$. Moreover,  about 20 \% of the decays $\phi\rightarrow K^+ + K^-$ take place after 100 fm/c
and are therefore not shown in Fig. ~\ref{sources}.

\begin{figure*}[th!]
\centerline{
\includegraphics[width=8.6 cm]{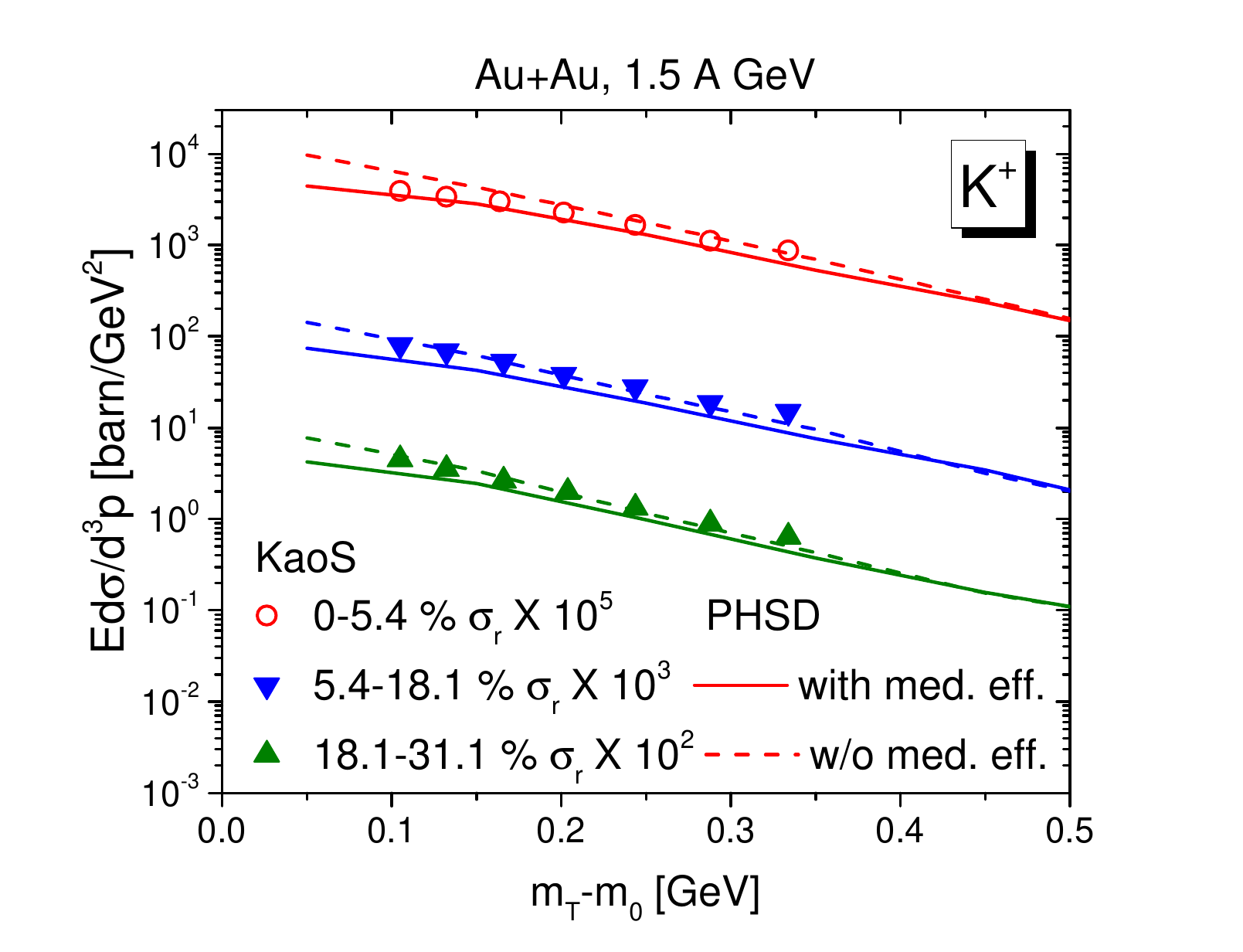}
\includegraphics[width=8.6 cm]{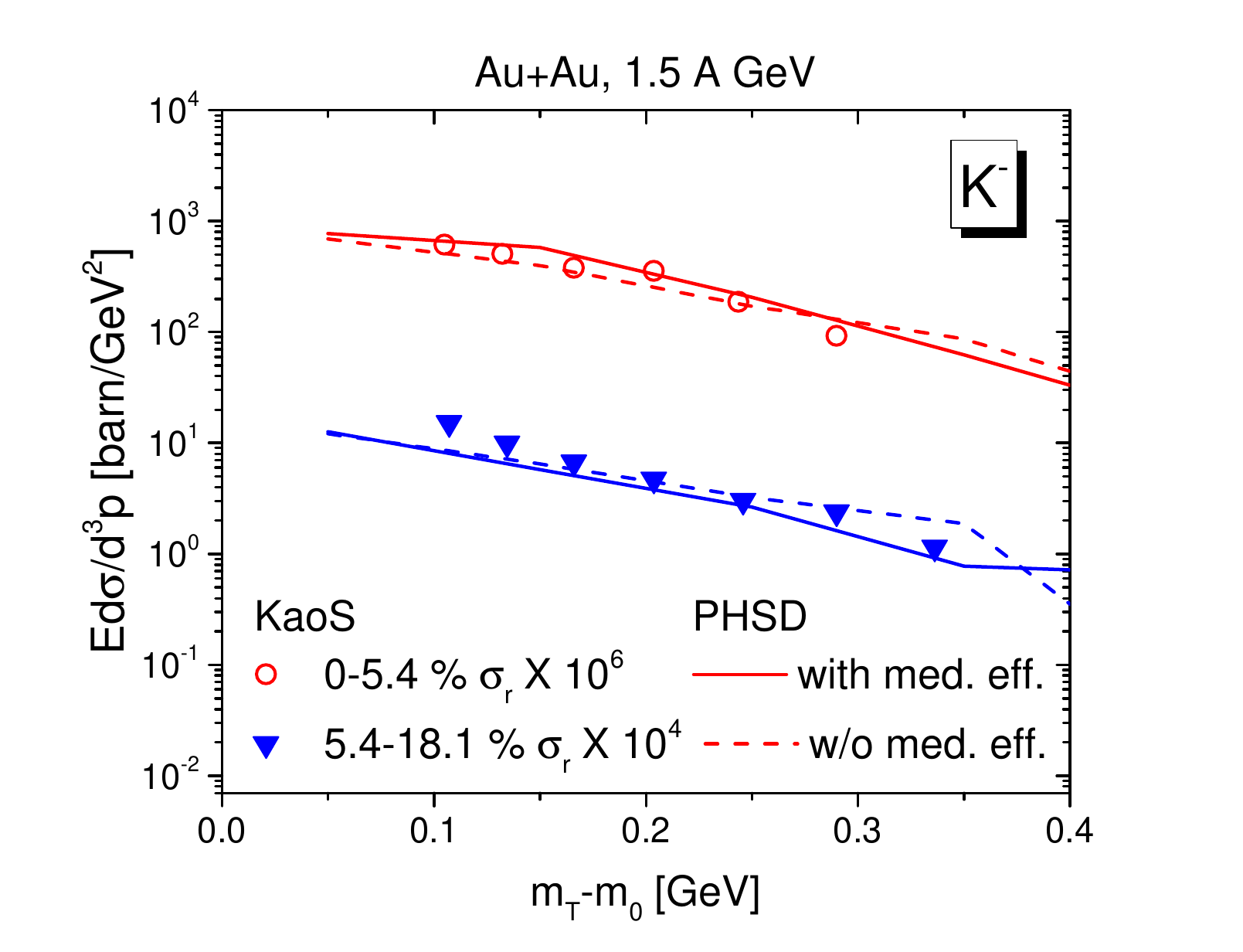}}
\caption{ $m_T$-spectra of $K^+$ (left) and $K^-$ (right) calculated with and without in-medium effects for (anti-)kaon production in Au+Au collisions at $E_{kin}$ = 1.5 A GeV, in comparison with experimental data from the KaoS collaboration~\cite{Forster:2007qk}.}
\label{mt-spectra}
\end{figure*}

Fig.~\ref{mt-spectra} shows the transverse mass spectra of $K^\pm$, calculated without and with in-medium effects for (anti-)kaon production in Au+Au collisions at $E_{kin}$ = 1.5 A GeV at different centralities.
The experimental data from the KaoS collaboration for $K^+$ as well as those for $K^-$ are well reproduced by PHSD. 
Comparing the dashed and solid lines, we see that in-medium effects suppress the $K^+$ production but enhance the $K^-$ production, especially in central collisions where the nuclear density is higher. The in-medium effects also harden the $m_T$ spectra of $K^+$ due to the repulsive potential and soften the $K^-$ spectra.

\begin{figure}[th!]
\centerline{
\includegraphics[width=8.6 cm]{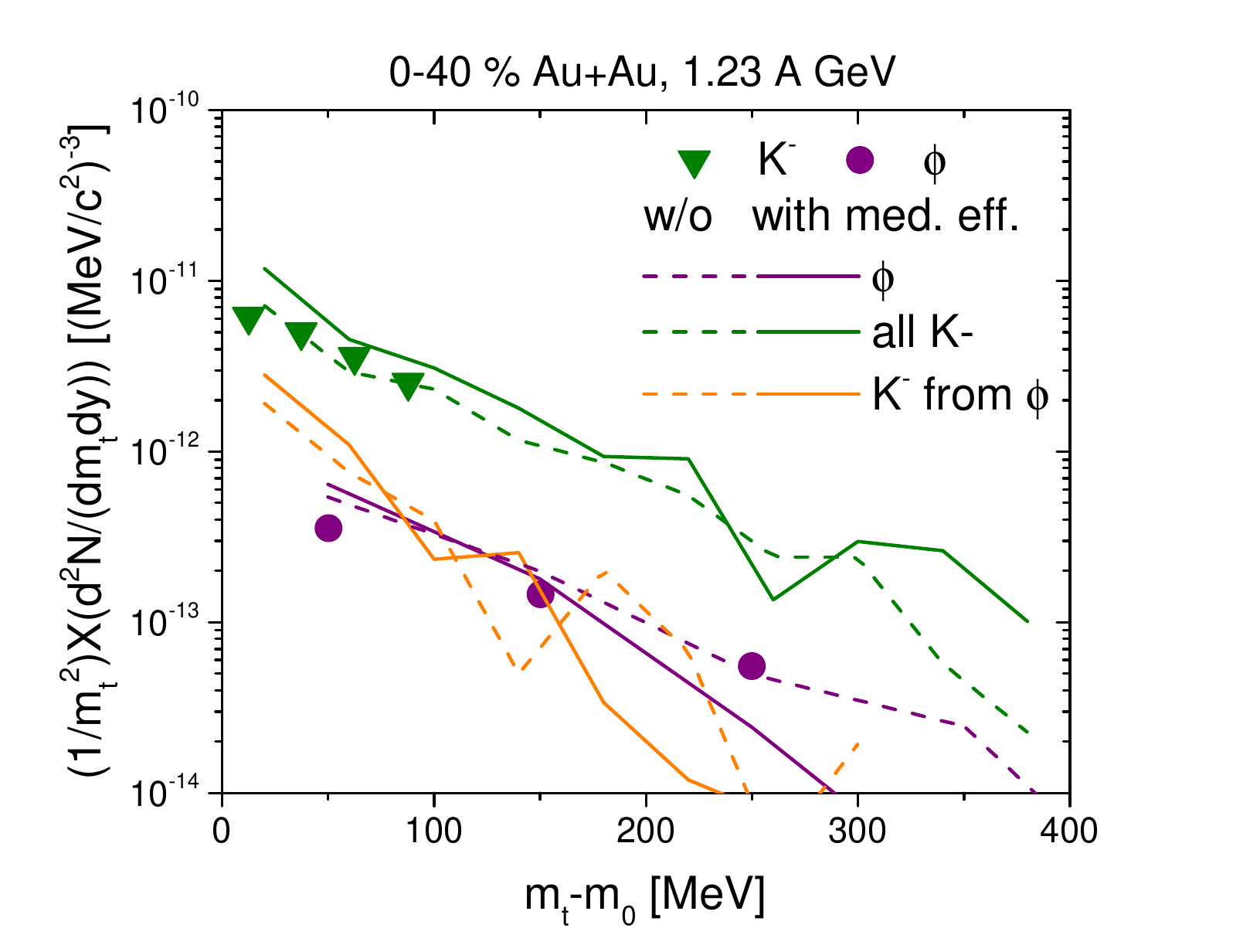}}
\caption{ $m_T$-spectra at midrapidity of final $K^-$ (green lines) and $\phi$ mesons (purple lines) calculated with (solid lines) and without (dashed lines) in-medium modifications for (anti-)kaon production in Au+Au collisions at $E_{kin}$ = 1.23 A GeV,  in comparison to the experimental data from the HADES Collaboration~\cite{HADES:2017jgz}.
The number of $\phi$ mesons is corrected by the branching ratio.}
\label{mt-Kphi}
\end{figure}

In Fig. \ref{mt-Kphi} we show the midrapidity $m_T$-spectra of final $K^-$ (green lines) and $\phi$ mesons (purple lines), calculated with (solid lines) and without (dashed lines) in-medium effects, for (anti-)kaon production in Au+Au collisions at $E_{kin}$ = 1.23 A GeV, in comparison to the experimental data from the HADES Collaboration~\cite{HADES:2017jgz}.
One can see that the PHSD calculations reproduce approximately (within the achieved statistics) the measured $m_T$-spectra of $\phi$ and $K^-$ mesons. We find that the influence of the in-medium effects on the slope of the $K^-$ spectra is rather strong - the antikaon $m_T$-spectra with medium effects are softer than without (cf. solid and dashed green lines). On the other hand, medium effects  have only a small effect on the slope of  $K^-$ stemming  from the $\phi$ decay (orange lines) and consequently on that of the reconstructed $\phi$ mesons (purple lines), since the long living $\phi$ mesons decay dominantly at low density. One sees also a strong difference of the slope of the final $K^-$ spectra coming from all possible channels (as discussed above) compared to the 
slope of the $K^-$ coming from the $\phi$ decay, which is much softer. Thus, the $\phi$ contribution is softening  the final $K^-$ spectrum by enhancing the low $m_T$ part of the spectrum. 
The same tendency has been reported by the HADES Collaboration in Ref. \cite{HADES:2017jgz}, 
where the data have been interpreted by a two-component model: a "feed down" from the $\phi$ mesons - coming on a top of a "direct thermal" component - has been considered as a possible origin for the softening of the antikaon spectrum.
We stress that in our model the influence of the in-medium effects on the final slope of $K^-$ is large. Thus, the final slope of $K^-$ is defined by the in-medium effects together with a contribution from the $\phi$ meson decay (which soften the $K^-$ spectra in the same way for both scenarios - with and without medium effect).
\begin{figure*}[th!]
\centerline{
\includegraphics[width=8.6 cm]{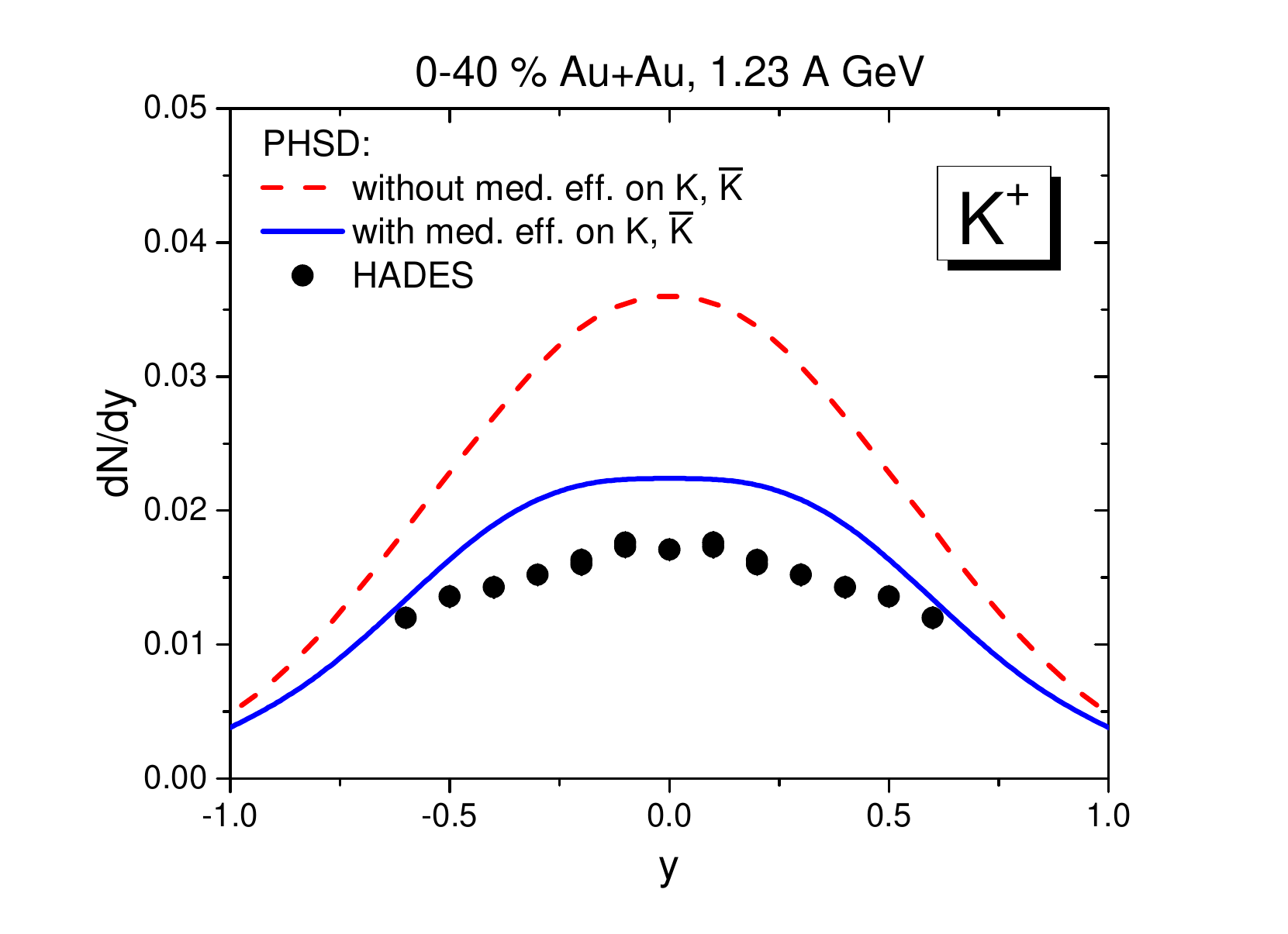}
\includegraphics[width=8.6 cm]{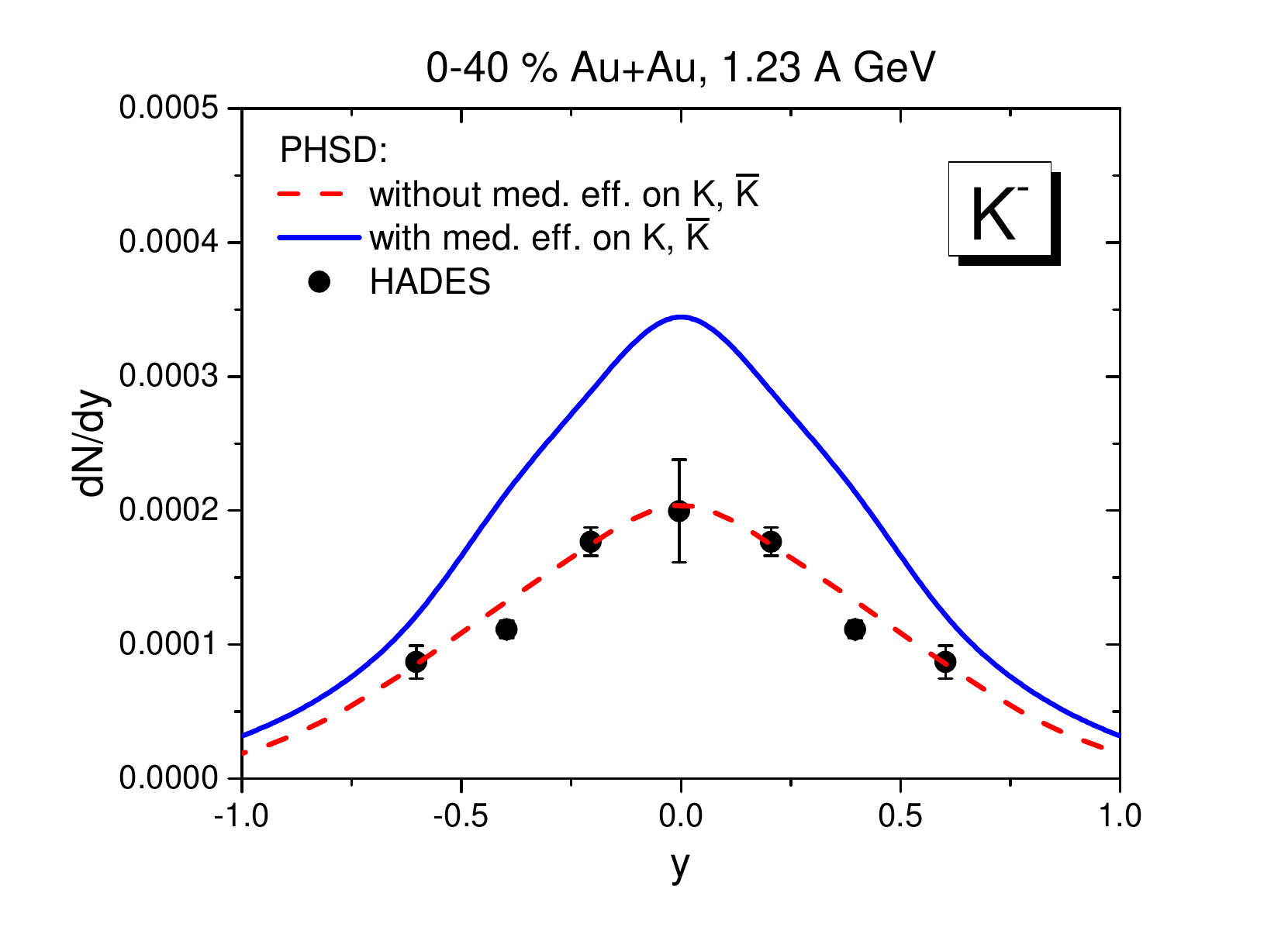}}
\centerline{
\includegraphics[width=8.6 cm]{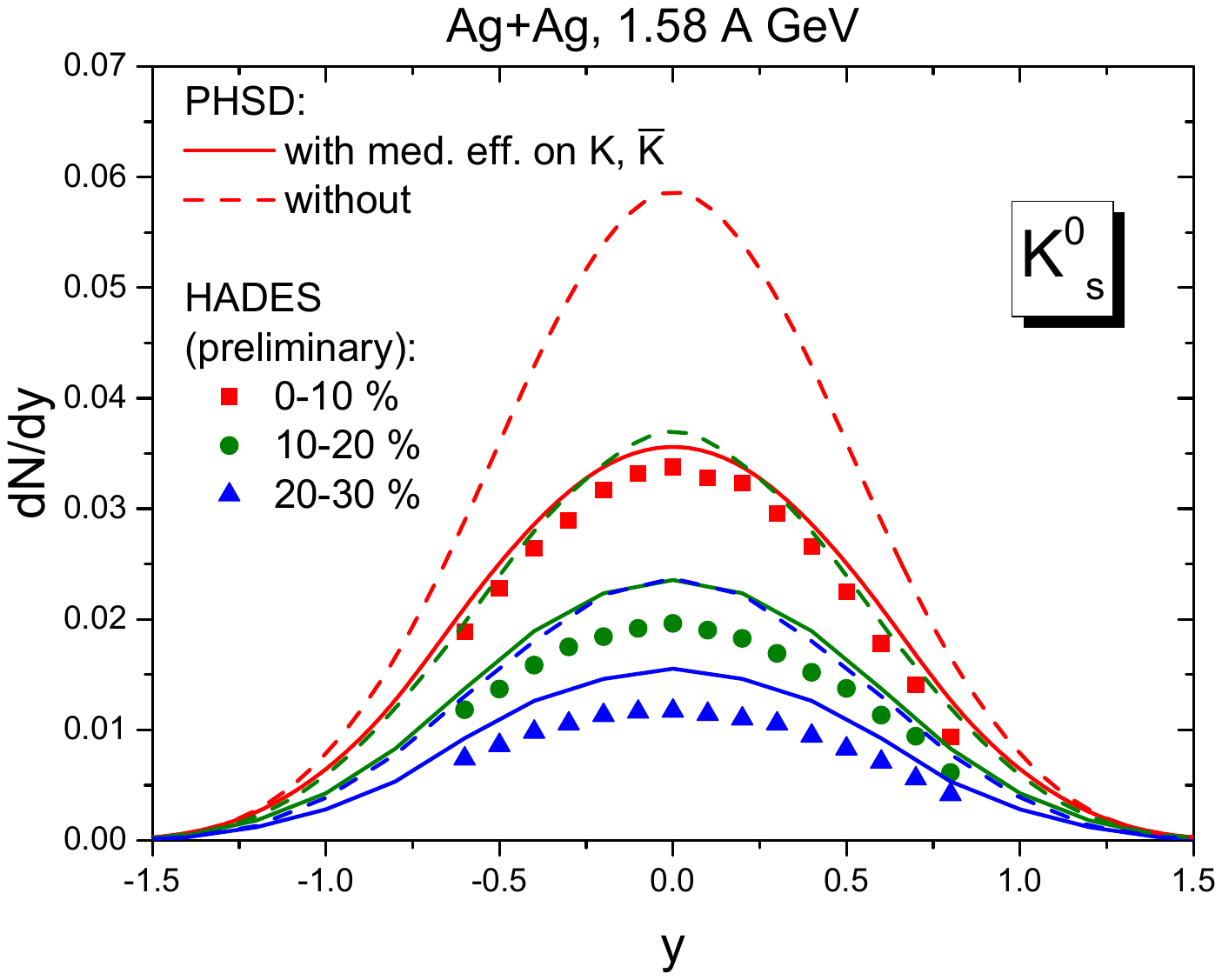}
\includegraphics[width=8.6 cm]{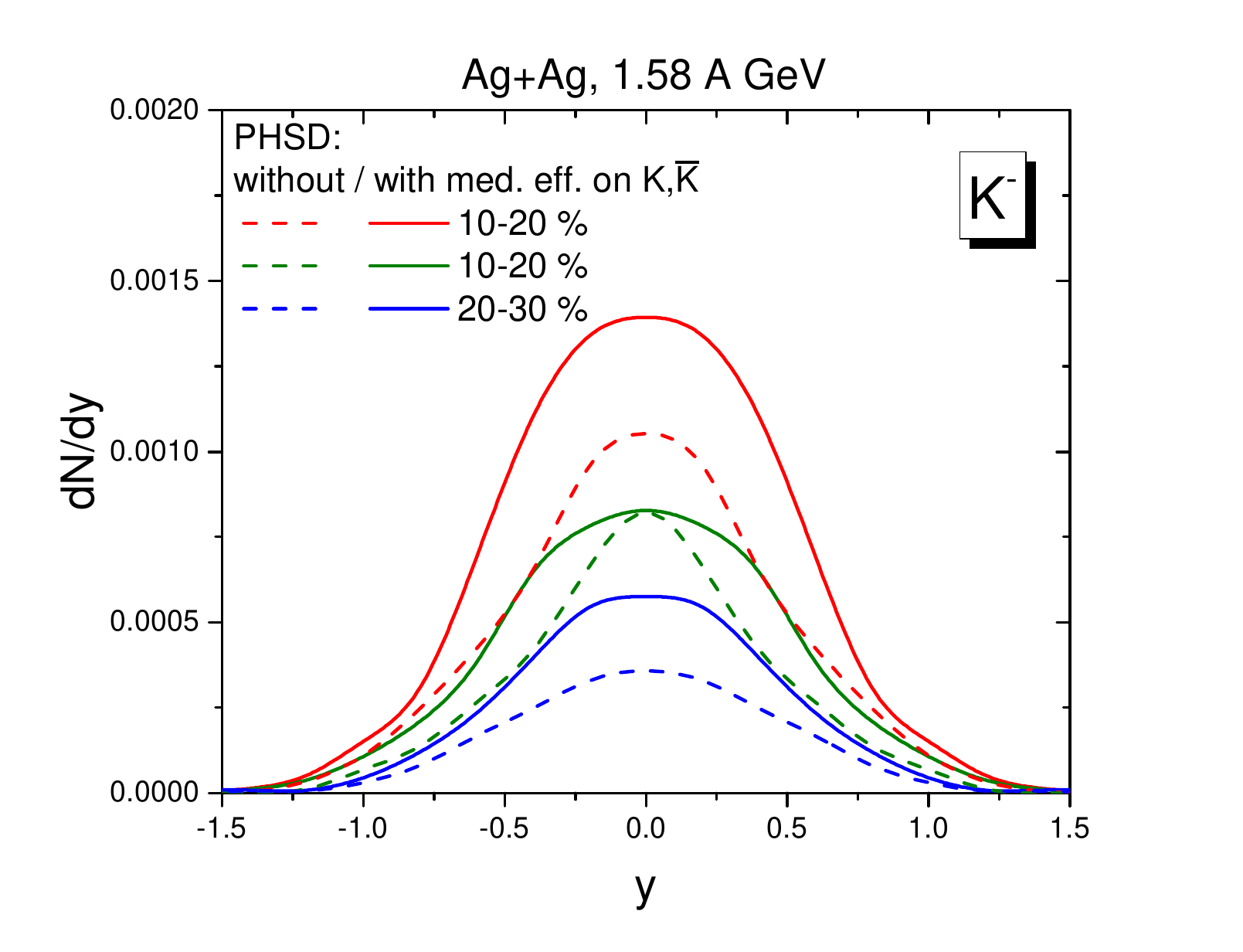}}
\centerline{
\includegraphics[width=8.6 cm]{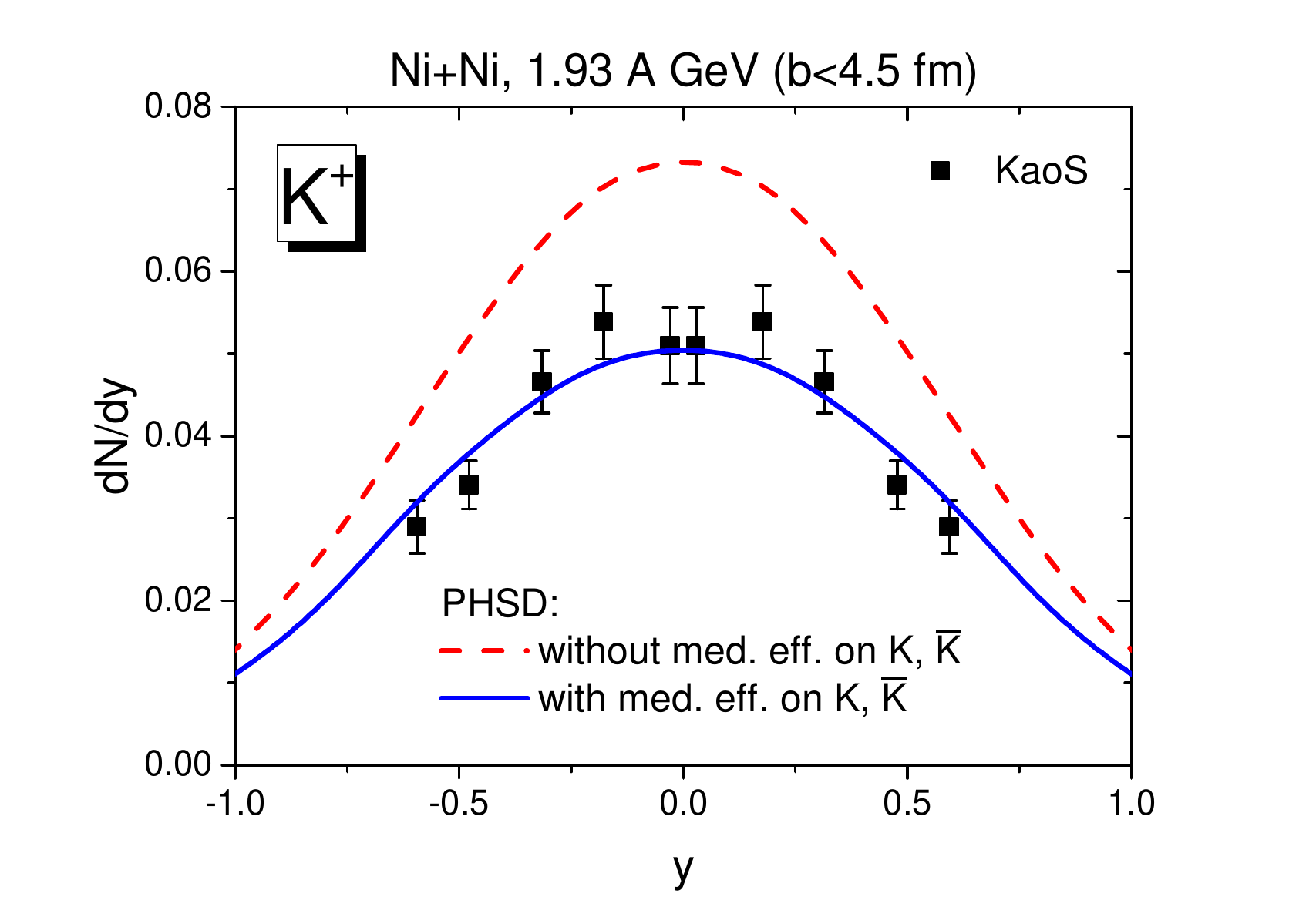}
\includegraphics[width=8.6 cm]{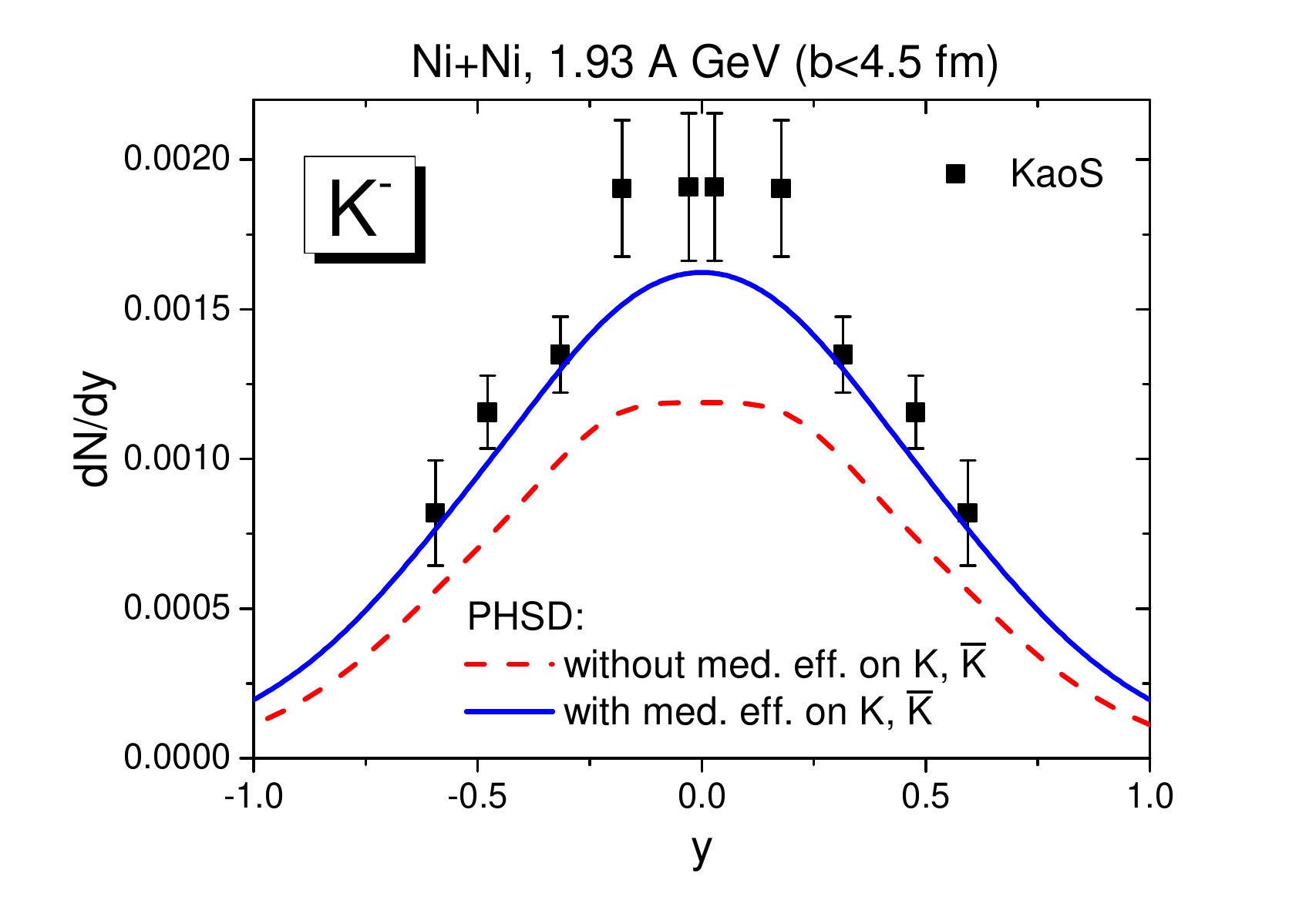}}
\caption{Rapidity distributions of $K^+(K_s^0)$ (left column) and $K^-$ (right column) for 0-40 \% central Au+Au collisions at $E_{kin}$ = 1.23 A GeV (upper row), for 0-10, 10-20 and 20-30 \% central Ag+Ag collisions at $E_{kin}$ = 1.58 A GeV (middle row) and for Ni+Ni collisions at $E_{kin}$ = 1.93 A GeV for $b<4.5 ~{\rm fm}$ (lower row). 
The solid blue and dashed red lines show the PHSD calculations with and without in-medium effects for $K, \bar K$; the  collisional broadening of $\phi$ mesons is accounted for in both cases.
The solid symbols show the experimental data from Refs. ~\cite{HADES:2017jgz,Schuldes:2017arj,KaoS:2000eil} and preliminary Ag+Ag data from HADES~\cite{Spies:2022sfg}.}
\label{kkbar1}
\end{figure*}

In Fig.~\ref{kkbar1} we display the rapidity distributions of $K^+(K_s^0)$ (left column) and $K^-$ (right column) for 0-40 \% central Au+Au collisions at $E_{kin}$ = 1.23 A GeV (upper row), for 0-10, 10-20 and 20-30 \% central Ag+Ag collisions at $E_{kin}$ = 1.58 A GeV (middle row) and for Ni+Ni collisions at $E_{kin}$ = 1.93 A GeV for $b<4.5 ~{\rm fm}$ (lower row).  The PHSD results are compared with the KaoS and HADES data from Refs. ~\cite{HADES:2017jgz,KaoS:2000eil}, though Ag+Ag data from HADES are preliminary~\cite{Spies:2022sfg}. The solid blue and dashed red lines show the calculations with and without in-medium effects for $K, \bar K$; the  collisional broadening of $\phi$ mesons is accounted for in both cases.

Compared to our previous results, Ref.~\cite{Song:2020clw}, these calculations include the (anti-)kaons from the $\phi$ decay, whose production is enhanced due to the novel meson-baryon channels. 
However, the rapidity distributions of $K^+$ and $K^-$ for Au+Au collisions at $E_{kin}$ = 1.23 A GeV (upper row) and for Ni+Ni collisions at $E_{kin}$ =  1.93 A GeV (lower row) are similar to our previous calculations \cite{Song:2020clw} since the "feed down" from $\phi$ mesons to the $K^+$ yield is very small compared to other 
kaon production channels; the $K^-$ production is slightly enhanced from $\phi$ meson decay, however, it is compensated by the enhanced $K^-$ absorption by $\Delta$ baryons. The contribution of the $\phi$ meson decay to $K^-$ production we will discuss in the next Section.
In the middle row we display the results for Ag+Ag at $\sqrt{s_{\rm NN}}=$ 1.58 A GeV for which recently
preliminary experimental data on $K^0_s$ have been made available by the HADES collaboration. Also here the in-medium modifications of the kaon properties are necessary to describe the experimental data. On the right hand side we show the predictions for the $K^-$ rapidity distribution. The upcoming HADES data will be important to investigate the system size dependence of strangeness production: as shown in Ref.~\cite{Song:2020clw} and confirmed now by the updated PHSD calculations, including in-medium effects, PHSD overestimates the HADES data on $K^-$ production in Au+Au collisions at 1.23 A GeV but describes, on the other hand, the KaoS, FOPI and HADES data for lighter systems as Ni+Ni, Ar+KCl and C+C at different SIS energies including in-medium modifications. Moreover, including these in-medium effects, PHSD reproduces the $K^+$ experimental observables for light and heavy systems as well as experimental data on $\Lambda$ production which give solid constraints on the strangeness balance in the calculations.
\begin{figure*}[th!]
\includegraphics[width=7. cm]{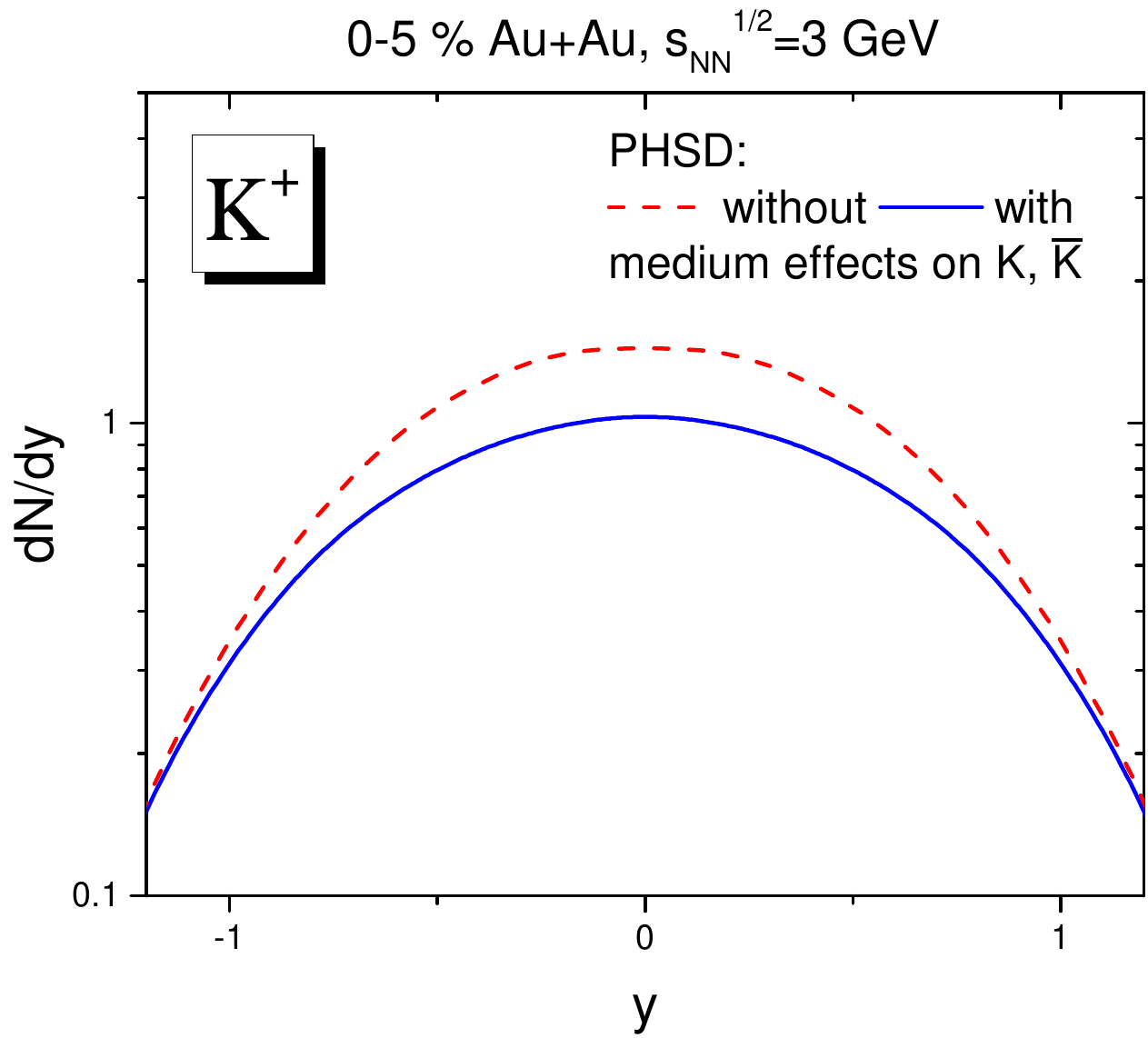} \hspace*{1.5cm}
\includegraphics[width=7. cm]{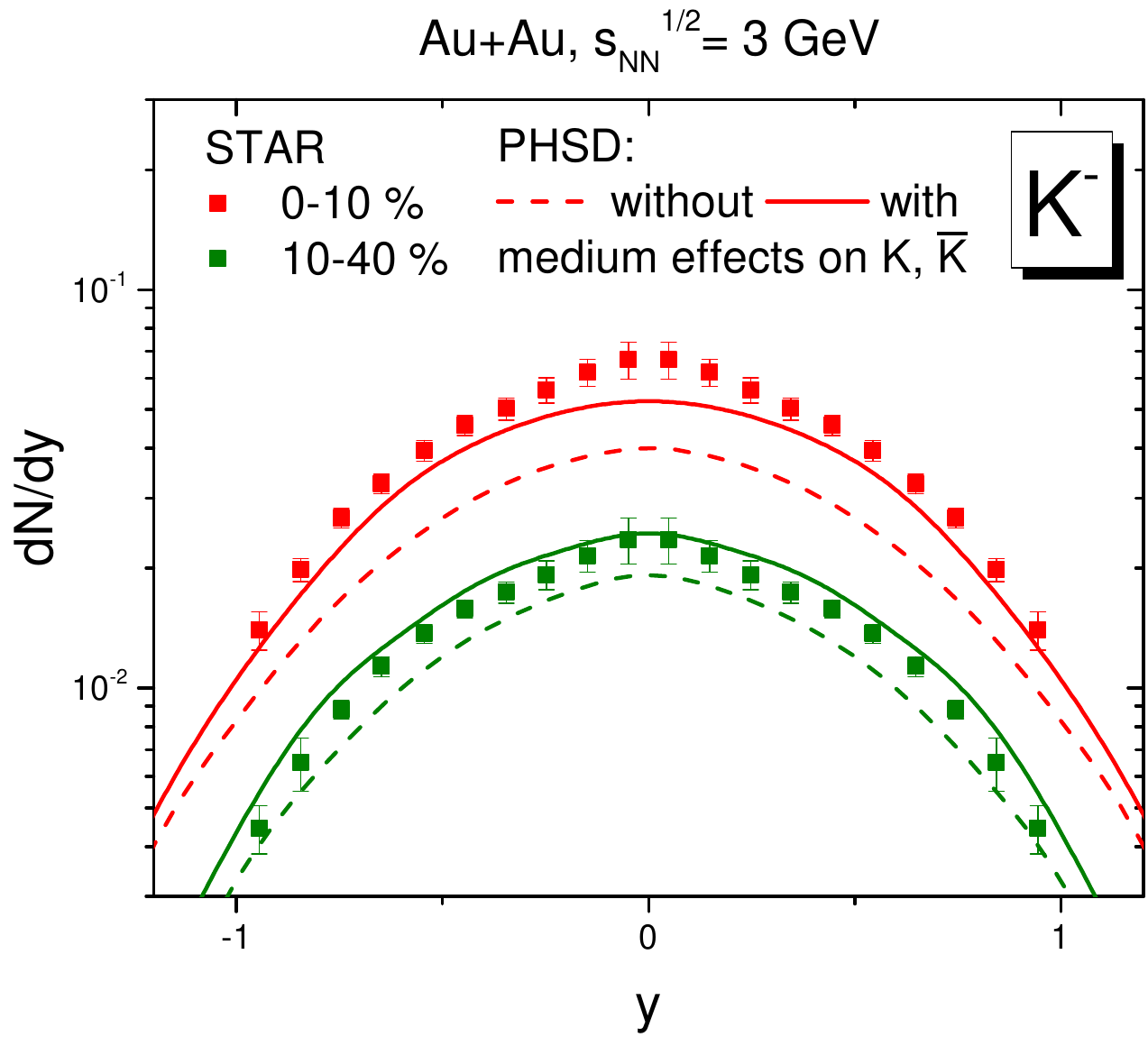}
\includegraphics[width=8.6 cm]{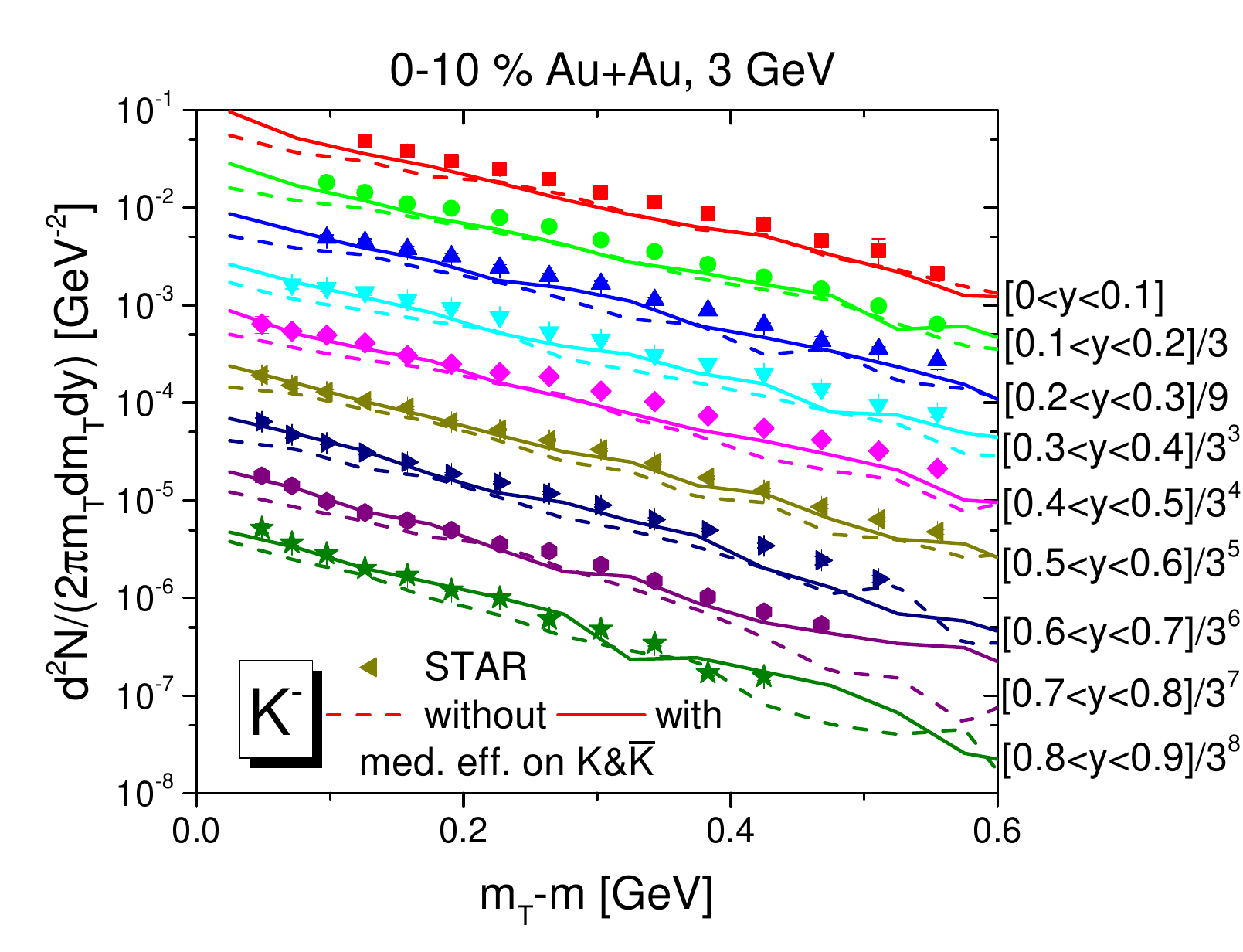} 
\includegraphics[width=8.6 cm]{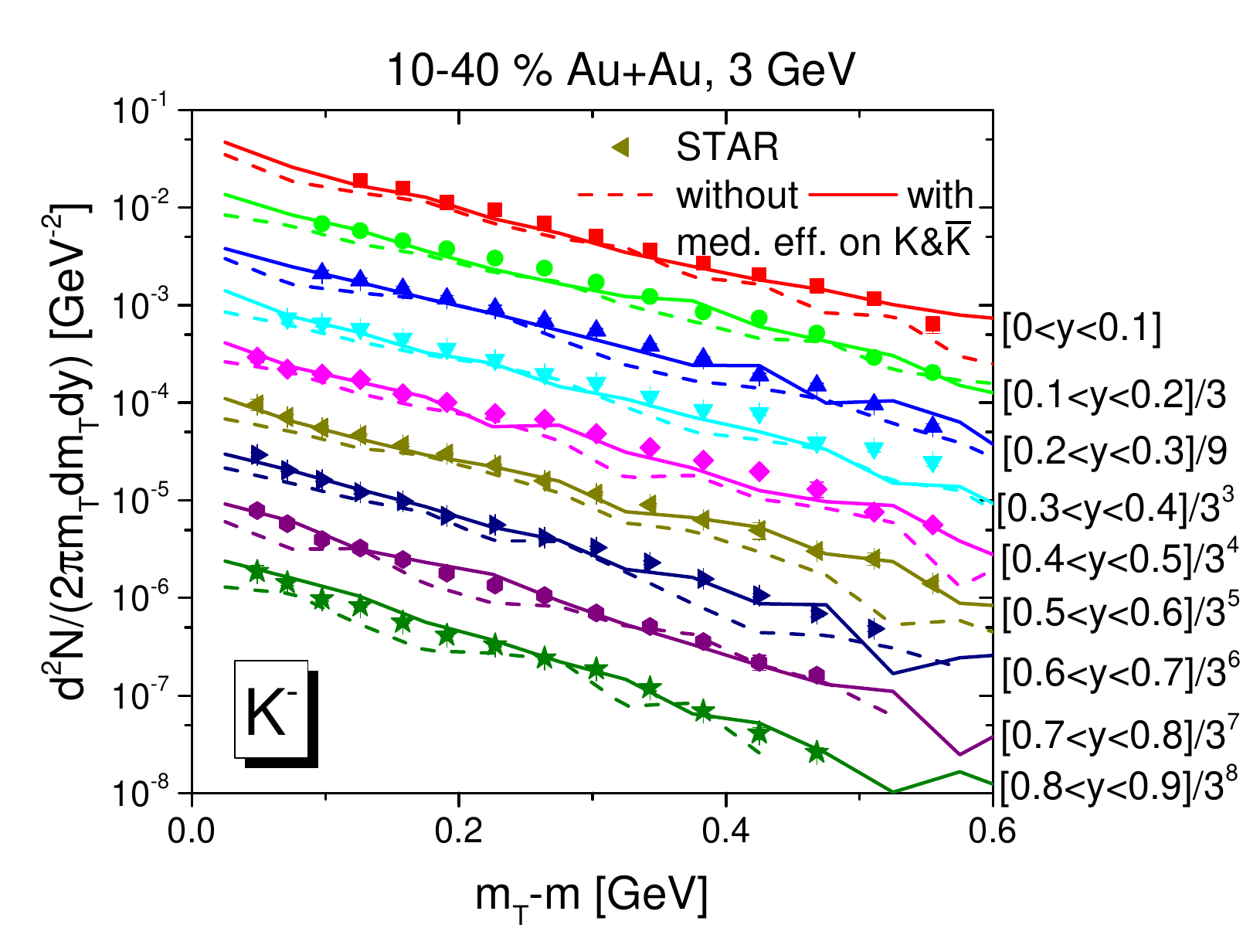}
\caption{Rapidity distributions of $K^+$ (top left) and $K^-$ (top right)  for  Au+Au collisions at $\sqrt{s_{\rm NN}}=$ 3 GeV. 
The PHSD  calculations with (solid lines) and without (dashed lines) in-medium effects on (anti-)kaon production are displayed and compared for the $K^-$ with the preliminary data of the STAR collaboration ~\cite{STAR:2021hyx}.
The $K^+$'s are calculated for 0-5 \% centrality, while $K^-$ mesons are calculated and compared for two centrality classes: for 0-10 \%  (red) and 10-40 \% (green) central Au+Au collisions. 
The $m_T$ spectra of $K^-$ for 0-10 \% central (bottom left) and 10-40 \% central (bottom right) Au+Au collisions at $\sqrt{s_{\rm NN}}=$ 3 GeV.
The calculations are done with the collisional broadening of $\phi$ mesons and  presented for 9 rapidity bins $0 \le y \le 0.1$, \ $0.1 \le y \le 0.2$, ...
\ $0.8 \le y \le 0.9$, scaled by a factor $1/3$ each for better visualization. The dashed lines show  the results without in-medium effects for $K, \bar K$ mesons, while the solid lines that with in-medium modifications of  $K, \bar K$ mesons.
}
\label{kkbar2}
\centerline{
\includegraphics[width=8.6 cm]{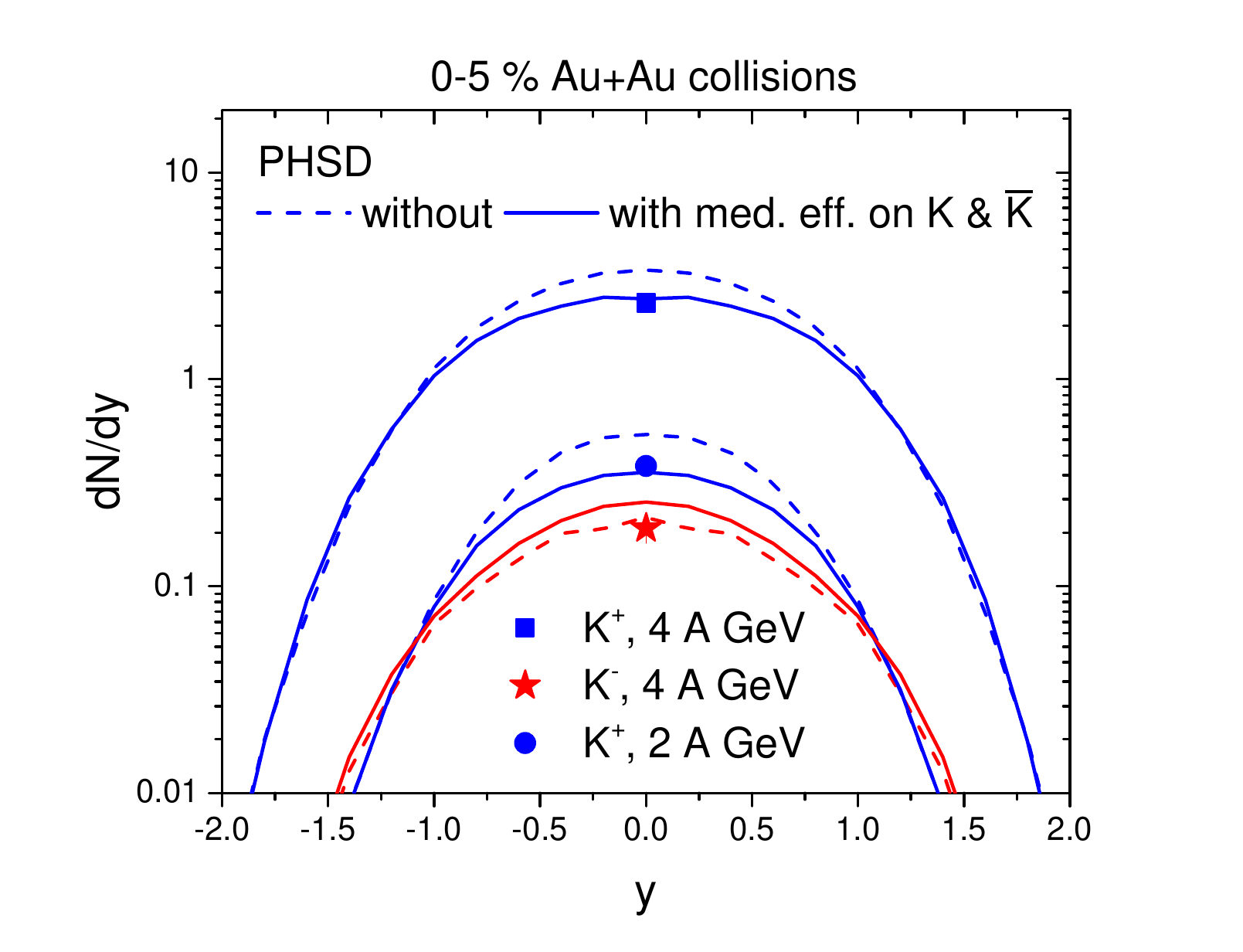}
\includegraphics[width=8.6 cm]{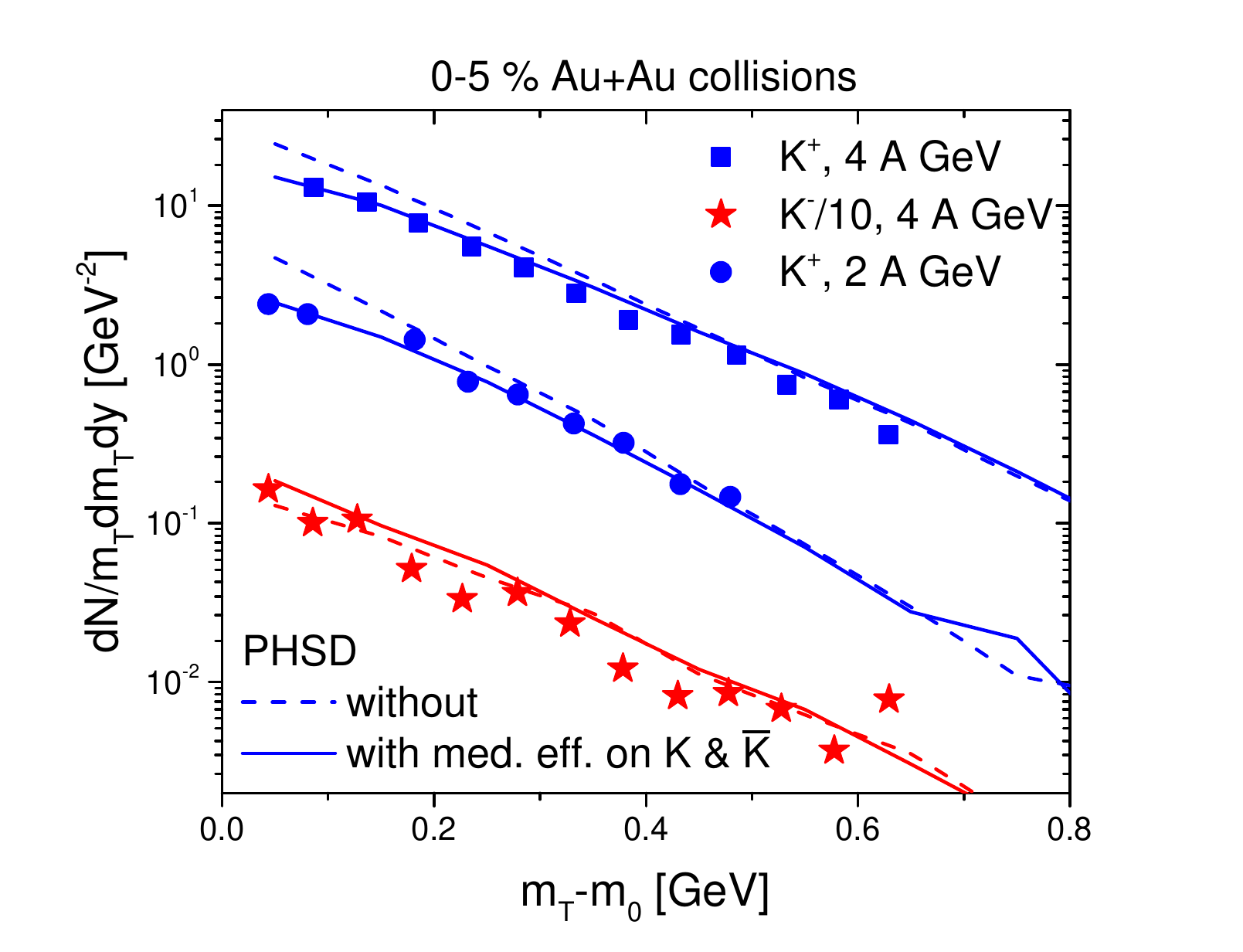}}
\caption{Rapidity distributions (left) and $m_T$-spectra (right) at midrapidity of $K^+$ (blue) and $K^-$ 
(red) in 0-5 \% Au+Au collisions at $E_{kin}$= 2 and 4 A GeV. 
The PHSD  calculations with (solid lines) and without (dashed lines) in-medium effects for (anti-)kaon production are compared to the experimental data of the E866 Collaboration \cite{E866:2000dog}.}
\label{AGS}
\end{figure*}

Also above the kinematical threshold (for pp collisions) the experimental rapidity distribution of $K^-$ mesons agrees well with the experimental data if in-medium effects are included. 
In Fig.~\ref{kkbar2} (right)  we compare our calculations with the  $K^-$ rapidity distributions measured by the STAR collaboration for 0-10 \% and 10-40 \% central Au+Au collisions at $\sqrt{s_{\rm NN}}=$ 3 GeV \cite{STAR:2021hyx}.  As seen, the $K^-$ rapidity distributions
are better described by including in-medium modifications of the antikaon.
On the lower panel of Fig. \ref{kkbar2} we show 
the $m_T$ spectra of $K^-$ for 0-10\% central (left) and 10-40\% central (right) Au+Au collisions at $\sqrt{s_{\rm NN}}=$ 3 GeV.
The calculations are done with the collisional broadening of $\phi$ mesons and  presented for 9 rapidity bins.  The dashed lines show  the results without in-medium effects for $K, \bar K$ mesons, while the solid lines that with in-medium modifications of  $K, \bar K$ mesons. One can see in the top right figure \ref{kkbar2} that PHSD calculations with in-medium effects on $K, \bar K$ mesons reproduce the experimental $m_T$ spectra rather well.

On the left part of Fig. \ref{kkbar2} we show our predictions for the $K^+$ production with (solid blue line)
and without (dashed red line) medium effects for 0-5 \% central  Au+Au collisions at $\sqrt{s_{\rm NN}}=$ 3 GeV, where the new STAR data are expected. One can see the suppression of the $K^+$ yield at midrapidity for the case of in-medium modifications.

Fig.~\ref{AGS} (left) shows the PHSD results - with (solid lines) and without (dashed lines) medium effects - for the $K^+$ (blue) and $K^-$ (red) rapidity distributions in Au+Au collisions at $E_{kin}=$ 2 and 4 AGeV in comparison to the E866 data at  midrapidity \cite{E866:2000dog}. Also here one can see that the experimental data for $K^+$ are better reproduced by calculations with medium effects.
Moreover, the $m_T$-spectra of $K^+$ mesons for central Au+Au collisions at midrapidity (shown in Fig~\ref{AGS}, right), clearly favour the in-medium scenario for $K^+$ for both energies.  
For the $K^-$ mesons the in-medium effects enhance only slightly the yield at midrapidity
and soften the slope of the $m_T$ spectra at low $m_T$. 

\begin{figure*}[th!]
\centerline{
\includegraphics[width=8.6 cm]{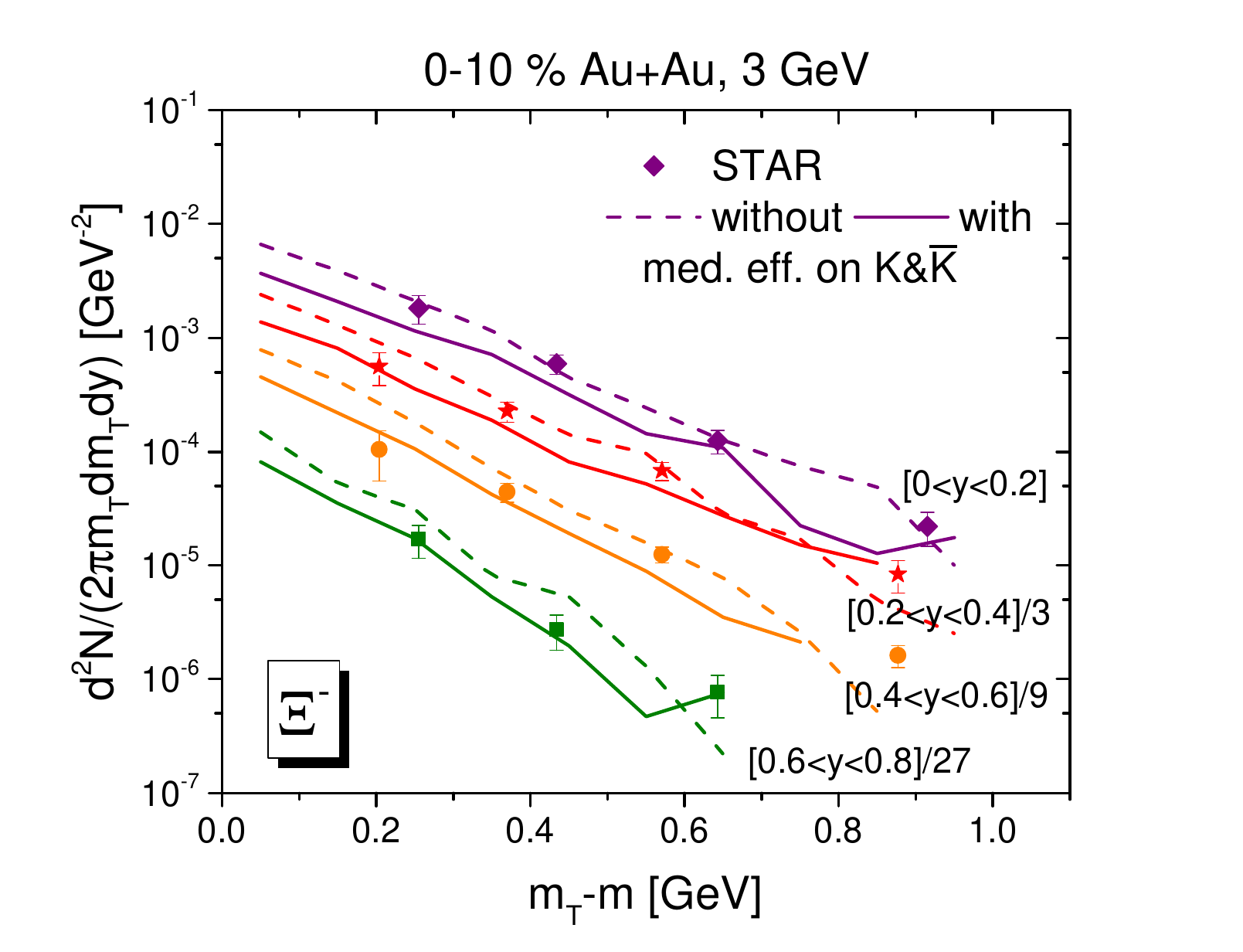}
\includegraphics[width=8.6 cm]{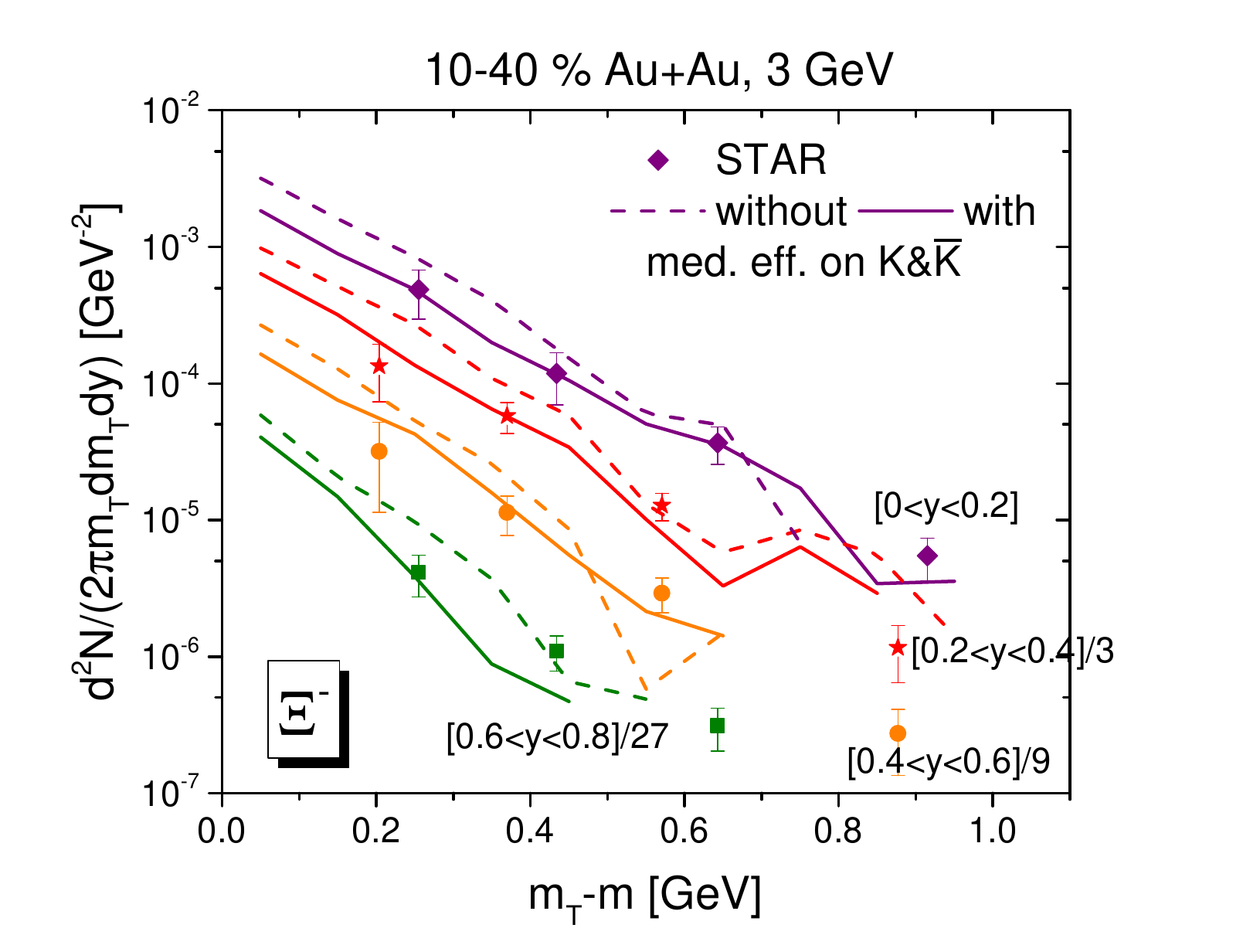}}
\caption{$m_T$ spectra of $\Xi^-$ in 0-10 \% (left) and 10-40 \% central (right) Au+Au collisions at $\rm \sqrt{s_{NN}}=$ 3 GeV, compared with the experimental data from the STAR Collaboration~\cite{STAR:2021hyx}.
The calculations are done with the collisional broadening of $\phi$ mesons and  presented for 4 rapidity bins $0 \le y \le 0.2$, \ $0.2 \le y \le 0.4$ scaled by a factor $1/3$ for better visualization, \ $0.4 \le y \le 0.6$ scaled by a factor $1/9$ and  $0.6 \le y \le 0.8$ scaled by a factor $1/27$. The dashed lines show  the results without in-medium effects for $K, \bar K$ mesons, while the solid lines that with in-medium modifications of  $K, \bar K$ mesons.}

\label{xi-mt}
\end{figure*}

In Fig. \ref{xi-mt} we show the $m_T$ spectra of $\Xi^-$ in 0-10 \% (left) and 10-40 \% central (right) Au+Au collisions at $\rm \sqrt{s_{NN}}=$ 3 GeV, compared with the experimental data from the STAR Collaboration~\cite{STAR:2021hyx}.
At low energies the $\Xi$ is produced in the PHSD through the  elementary reactions $Y+Y\rightarrow \Xi+N$ and $Y+\bar{K}\rightarrow \Xi+\pi$ 
with cross sections calculated within the model of Refs. \cite{Li:2012bga,Chen:2003nm,Li:2002yd}.
The inverse reactions are implemented by detailed balance. At high energies the $\Xi^-$'s are produced via string fragmentation.
The calculations are done with the collisional broadening of $\phi$ mesons 
for 4 rapidity bins. The dashed lines show  the results without in-medium effects for $K, \bar K$ mesons, while the solid lines that with in-medium modifications of  $K, \bar K$ mesons.
One sees that the PHSD spectra are consistent with the experimental data (within the achieved statistics). The in-medium modifications of (anti)kaons 
leads to the reduction of the $\Xi^-$ yield, since hyperons are less produced with the medium effects.

\section{Strange particle ratios}\label{ratios}

In this section we show the PHSD results for the ratios of strange particles $K^-/K^+$,  $\phi/K^-$ and $\phi/\Xi-$ based on the results presented in Sections~\ref{heavy-ion} and \ref{influence}.

\begin{figure}[th!]
\centerline{
\includegraphics[width=8.6 cm]{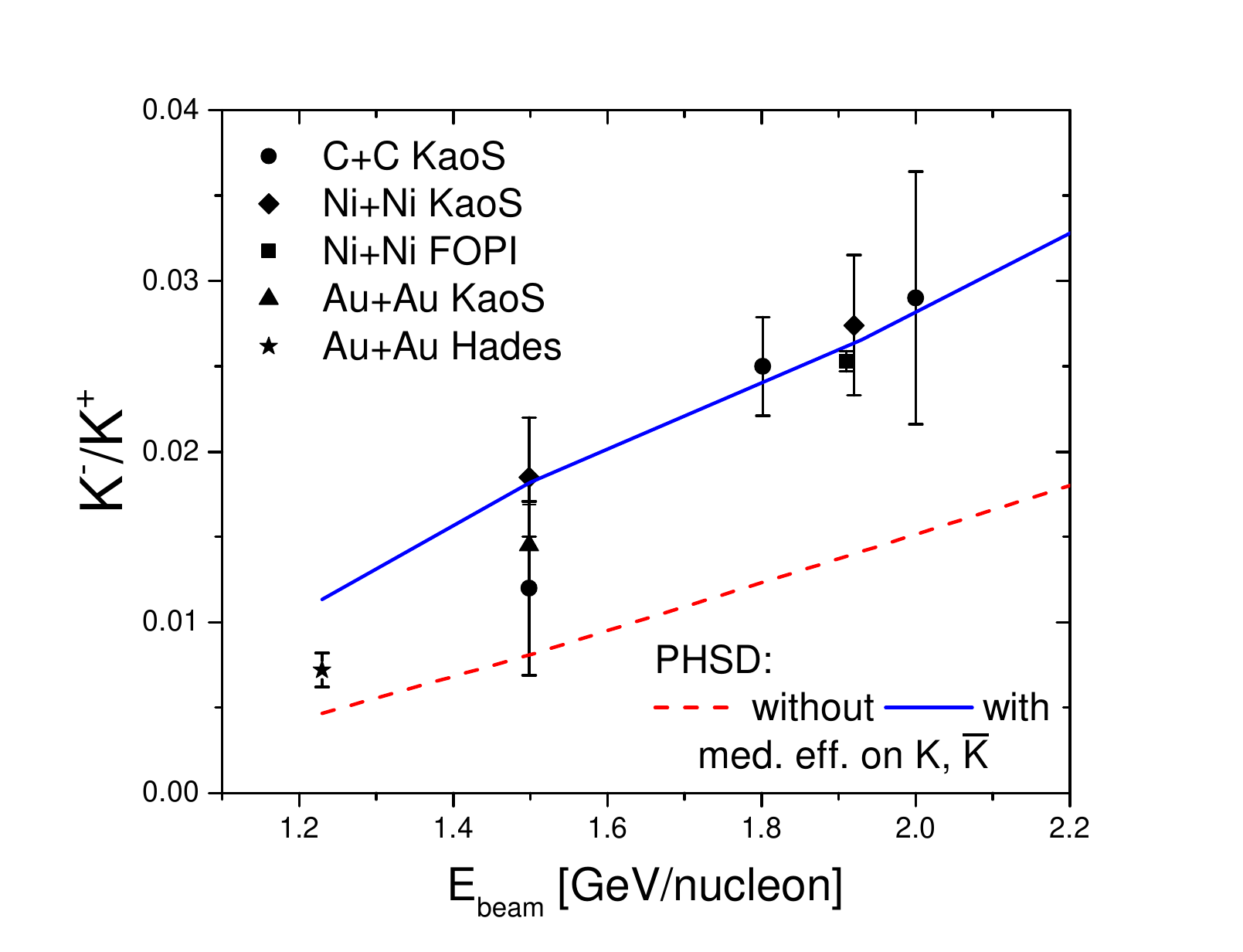}}
\caption{The PHSD results for the ratio $K^-/K^+$ calculated with (solid blue line) and without (dashed red line) in-medium effects for (anti-)kaon production in A+A collisions versus the beam energy $E_{beam}$. 
The symbols represent the experimental data of ref.~\cite{Forster:2007qk,HADES:2017jgz,FOPI:2018siq}.}
\label{RatioKmKp}
\end{figure}

In Fig. \ref{RatioKmKp} we present the PHSD results for the $K^-/K^+$ ratio calculated with (solid blue line) and without (dashed red line) in-medium effects for (anti-)kaon production in A+A collisions versus the beam energy $E_{beam}$ in comparison with  the experimental data of ref.~\cite{Forster:2007qk,HADES:2017jgz}.
We note that the $K^-/K^+$ ratio is rather flat as function of centrality, therefore one can simply look at the ratio in different systems and centrality classes.
As follows from Fig. \ref{RatioKmKp} the experimental data are in favour of the in-medium scenario for the increasing $K^-/K^+$ ratio. Only the HADES data for Au+Au at 1.23 A GeV are overestimated due to the overestimation of the $K^-$ yield, as discussed above.
More $K^\pm$ and $\phi$ experimental data at the low energy and their excitation functions would be helpful to clarify the discrepancies.

\begin{figure}[th!]
\centerline{
\includegraphics[width=8.5 cm]{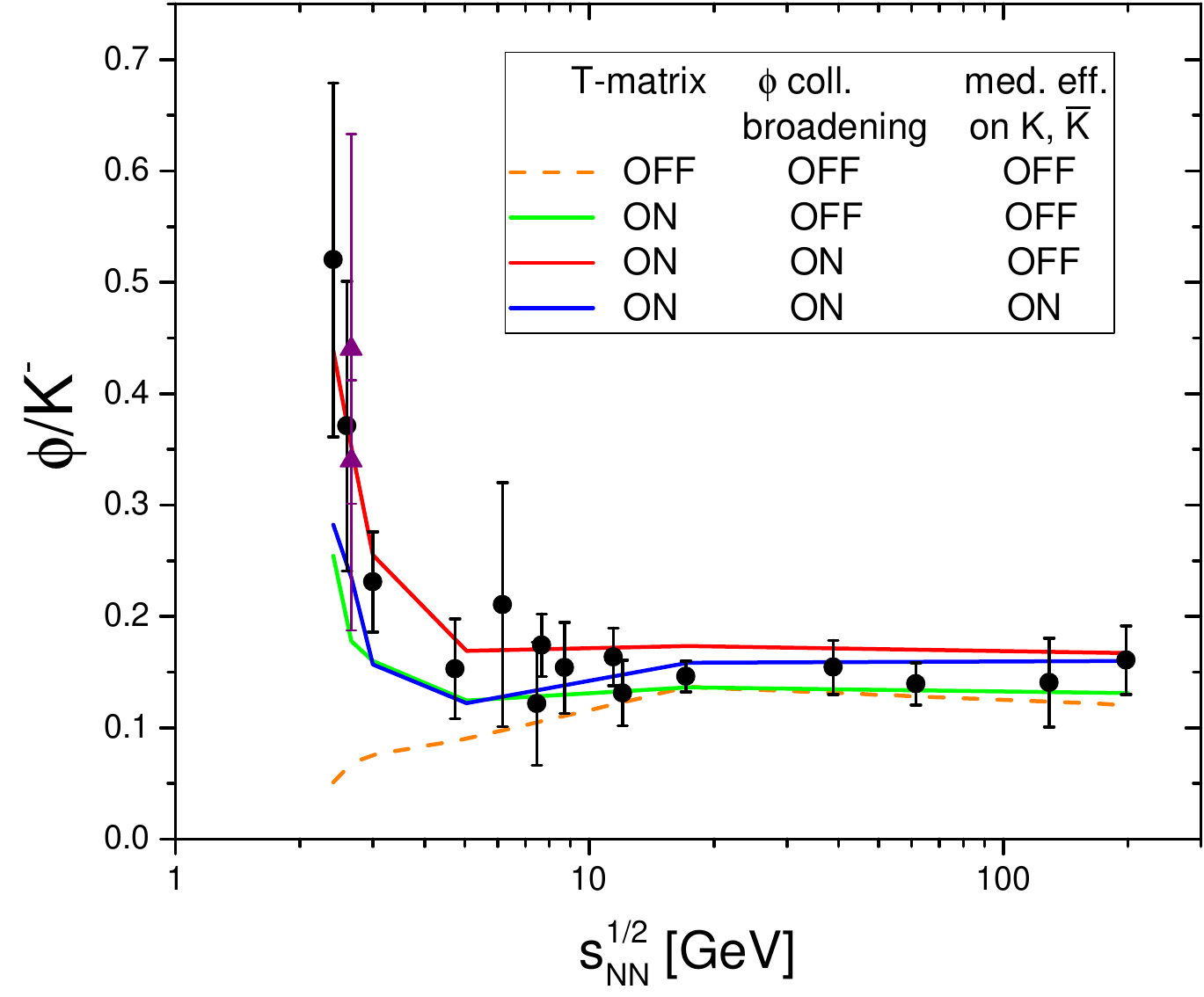}}
\caption{The PHSD results for the ratio $\phi/K^-$ in midrapidity ($|y|\le $0.3) as a function of the collision energy for four different scenarios: with and without  novel $mB$ channels for the $\phi$ meson production from the T-matrix approach and 
with and without the collisional broadening of $\phi$ meson width and in-medium effects on (anti-)kaons (cf. the legend). 
The number of $\phi$ mesons, reconstructed from $K^+K^-$ pairs, is divided  by the branching ratio $Br(\phi\to K^+K^-)$.
The solid symbols show the compilation of the experimental data from Refs.~\cite{E917:2003gec,NA49:2002pzu,NA49:2007stj,NA49:2008goy,STAR:2004yym,STAR:2021hyx,FOPI:2016jgt}.}
\label{ratio1}
\end{figure}
Fig.~\ref{ratio1} shows the $\phi/K^-$ ratio as a function of the collision energy from $E_{kin}$ = 1.23 A GeV to $\sqrt{s_{\rm NN}}$ = 200 GeV. The PHSD results are presented for  four different scenarios: \\
1) the orange dashed line is the ratio without any in-medium effect and also excluding the novel $mB$ production channels of $\phi$'s from the T-matrix approach. This ratio is comparable to the experimental data at high energies ($\sqrt{s_{\rm NN}} > 20~{\rm GeV}$), but underestimates significantly the experimental data at low energies ~\cite{E917:2003gec,NA49:2002pzu,NA49:2007stj,NA49:2008goy,STAR:2004yym,STAR:2021hyx,FOPI:2016jgt}. \\
2) The green line shows the $\phi/K^-$ ratio including the novel $mB$ channels for $\phi$ mesons produced from our T-matrix approach, but without in-medium effects for $K, \bar K$.
These new $mB$ production channels increase the ratio at low collisional energies and the results are consistent with the experimental data down to $\sqrt{s_{\rm NN}}=$ 3 GeV.
At high collisional energies the contribution from meson-baryon scattering is not important because the dominant mechanisms for $\phi$ production at high energies are string fragmentation and hadronization of the quark-gluon plasma, which are dominated by phase space. \\
3) In a baryon environment the width of the $\phi$ mesons broadens, modifying the spectral function. Including this collisional broadening (red line) the $\phi/K^-$ ratio becomes comparable to the experimental data at all energies. \\
4) Finally we display (by the blue line) the results if, additionally to the collisional broadening of the $\phi$ meson width,  the in-medium modifications of the (anti-)kaon properties are taken into account.
As already mentioned in the previous sections, these in-medium effects enhance the $K^-$ production in nuclear matter while the yield of the $\phi$ meson is not strongly affected by the in-medium modification of the (anti-)kaon.  As a result, the ratio decreases due to the enhanced $K^-$ production.

\begin{figure}[th!]
\centerline{
\includegraphics[width=7.4 cm]{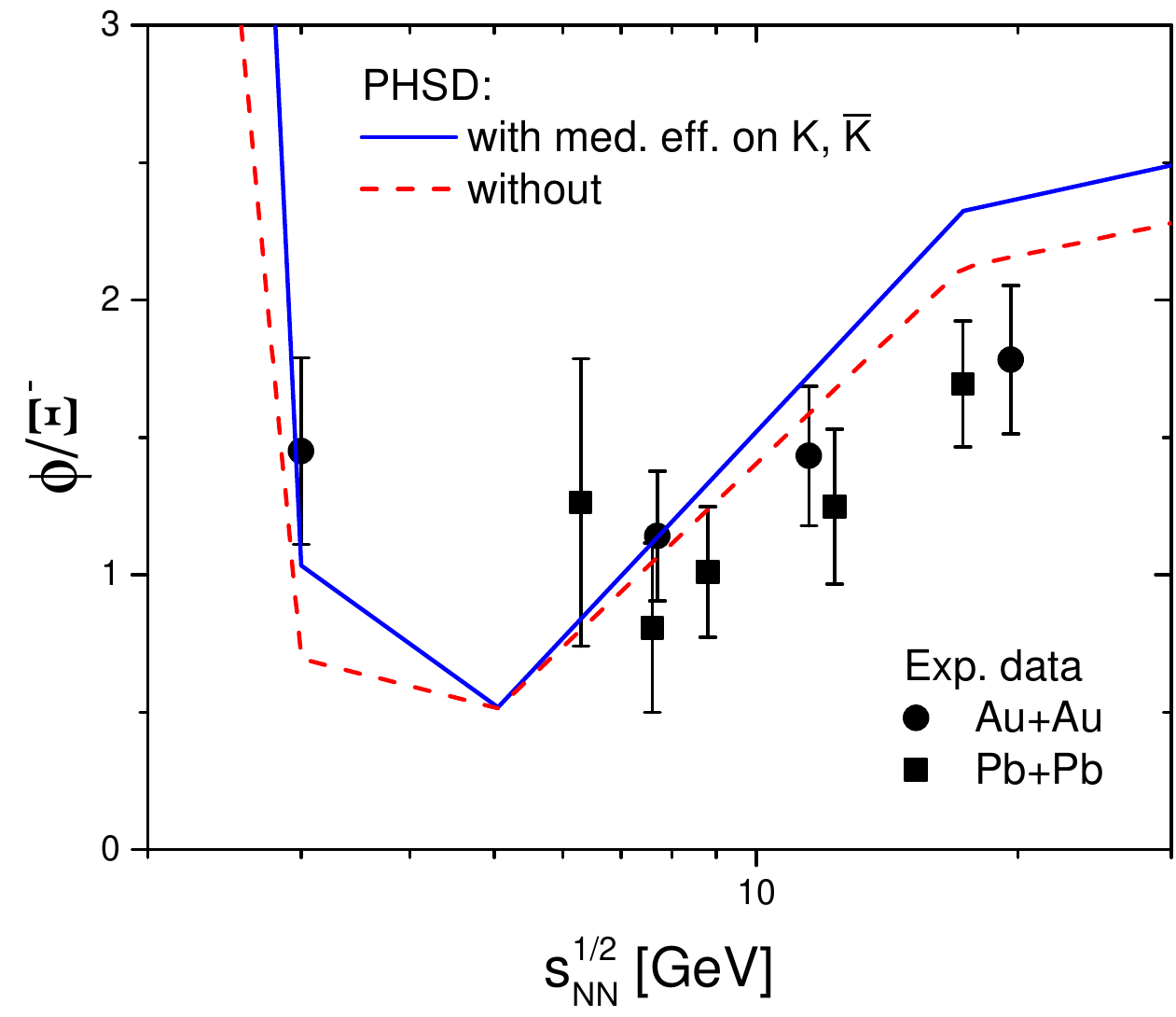}}
\caption{PHSD calculations for the $\phi/\Xi^-$ ratio as a function of the collision energy 
for calculations with (solid blue line) and without (dashed red line) in-medium effects for (anti-)kaons in comparison to the experimental data~\cite{STAR:2021hyx}. The number of $\phi$ mesons, reconstructed from $K^+K^-$ pairs, is divided  by the branching ratio $Br(\phi\to K^+K^-)$.
 }
\label{ratio2}
\end{figure}

We study furthermore the  $\phi/\Xi^-$ ratio. 
Fig.~\ref{ratio2} shows the $\phi/\Xi^-$ ratio as a function of the collision energy with (solid blue line)
and without (dashed red line) in-medium effects for (anti-)kaons. The collisional broadening of the $\phi$ 
meson widths is included in both cases.
Since the $\phi$ production depends little on the in-medium effects of (anti-)kaons, this ratio is 
insensitive to the (anti-)kaon properties.
The ratio is largest at HADES energies and decreases rapidly with increasing collision energy. 
After a minimum it slowly increases as shown in the figure. Our calculations are consistent with the experimental data~\cite{STAR:2021hyx}.

\begin{figure}[th!]
\centerline{
\includegraphics[width=8.6 cm]{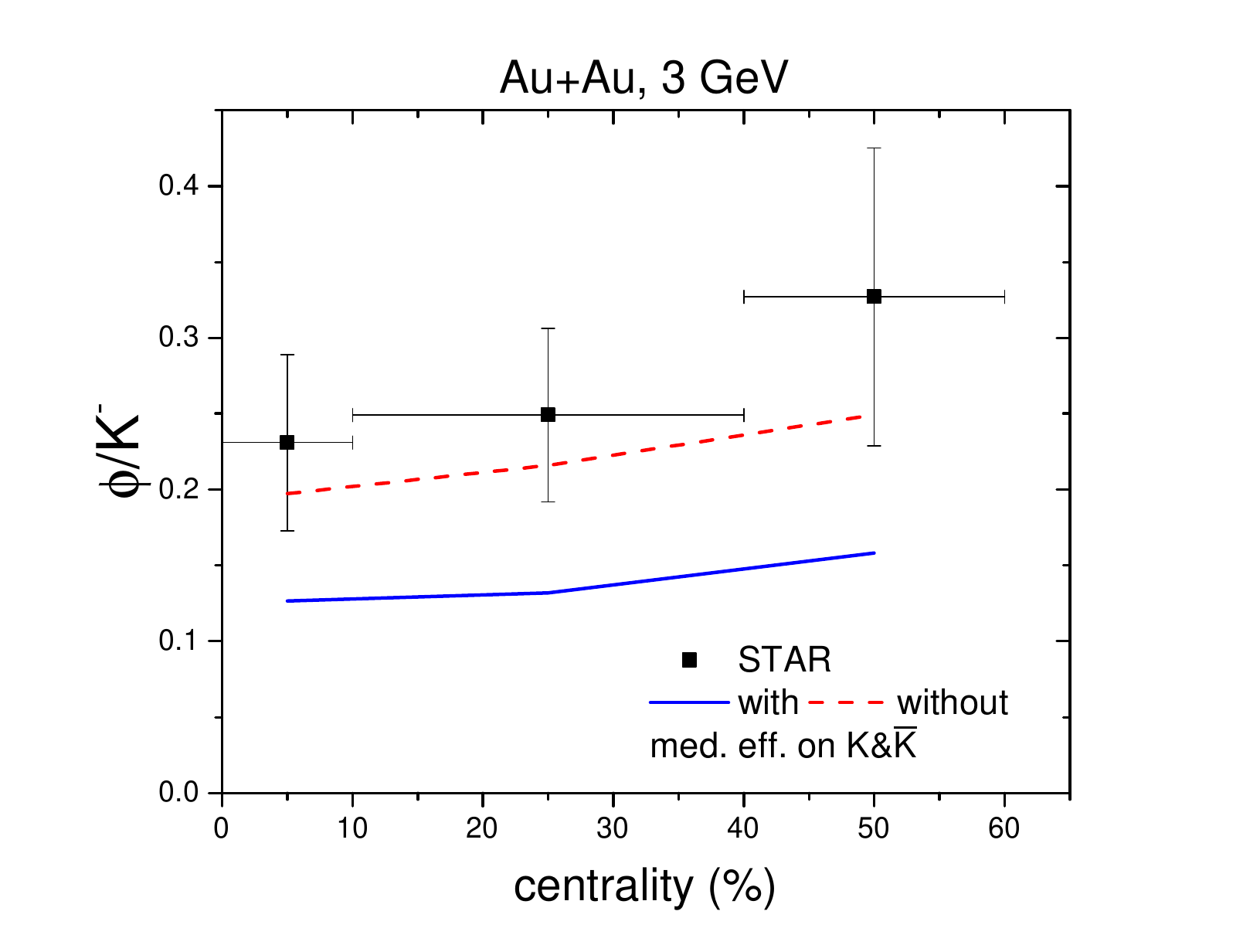}}
\centerline{
\includegraphics[width=8.6 cm]{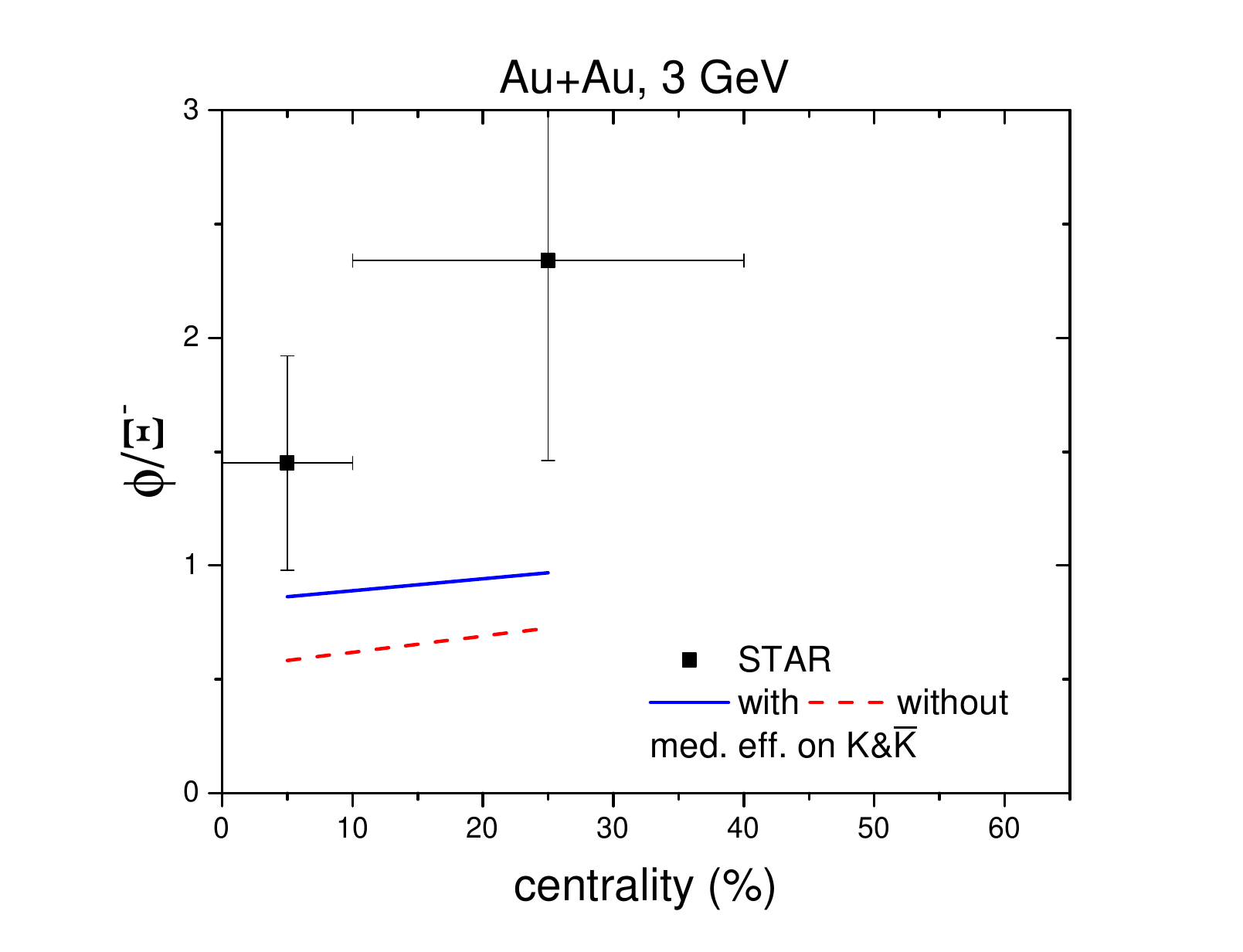}}
\caption{Ratios $\phi/K^-$ (upper) and $\phi/\Xi^-$ (lower) as a function of centrality with (solid blue line) and without (dashed red line)  medium effects on (anti)kaon in Au+Au collisions at $\sqrt{\rm s_{NN}}=$3 GeV in comparison with experimental data from the STAR Collaboration~\cite{STAR:2021hyx}.
 }
\label{ratio-star}
\end{figure}

Finally, in Fig. \ref{ratio-star} we show the ratios $\phi/K^-$ (upper) and $\phi/\Xi^-$ (lower) as a function of centrality with (solid blue line) and without (dashed red line) medium effects on (anti)kaon in Au+Au collisions at $\sqrt{\rm s_{NN}}=$3 GeV in comparison with experimental data from the STAR Collaboration~\cite{STAR:2021hyx}. 
One gets an impression that the experimental $\phi/K^-$ ratio is better described by the scenario without medium effect. However, looking at the $\phi$ and $K^-$ spectra separately in Figs. \ref{dndym} and \ref{kkbar2}, one can see that with in-medium effects we better describe the $K^-$ rapidity distributions, while without in-medium effects it is substantially underestimated. On the other hand, the $\phi$ yield is less sensitive to the in-medium effects on $K, \bar K$. Thus, the ratio $\phi/K^-$
is larger for the scenario without in-medium effects due to the underestimation of $K^-$ yield.
For the $\phi/\Xi^-$ ratio the situation is opposite since the $\Xi$ yield
is smaller with in-medium effects for (anti)kaons as follows from Fig. \ref{xi-mt}, thus the $\phi/\Xi^-$ ratio is larger for the in-medium scenario for (anti)kaons.

\section{Summary}\label{summary}

In this study we have investigated the hidden ($\phi$ meson) and open (kaon and antikaon) strangeness production and their mutual dependence in heavy-ion collisions from subthreshold to relativistic energies within the microscopic off-shell PHSD transport approach.
Motivated by the fact that conventional transport models, which propagate on-shell particles, cannot explain the $\phi$ meson yield in heavy-ion reactions at low beam energies without
introducing a not-observed $\phi$ decay of heavy baryonic resonances, we studied systematically the properties of hidden and open strange mesons in the dense and hot medium created in heavy-ion collisions and the reaction channels which contribute to the production of $\phi$ mesons in a nuclear environment. This includes

i) The in-medium modifications of $\phi$ mesons. \\
Collisional broadening leads to an increase of the width of $\phi$ mesons, proportional to the baryon density, and hence to a broadening of the $\phi$ meson spectral function.
Consequently, the $\phi$ meson production threshold in all hadronic reactions is lowered if the $\phi$ is in a strongly interacting environment. This enhances the $\phi$ meson production at subthreshold energies and leads to a modification of the (elastic and inelastic) cross sections with hadrons. 

Based on the off-shell Kadanoff-Baym equations, PHSD is able to change dynamically the width of the $\phi$ while travelling through the dense baryonic medium. This allows for a consistent transport of the $\phi$ meson in the dense matter which is created in heavy ion reactions.

ii) New relevant $mB \to \phi$ production channels.\\
We used the SU(6) chiral effective Lagrangian in the T-matrix approach to calculate the different meson-baryon cross sections which contribute to $\phi$ production. Taking into account the five sets of total quantum numbers $(I,S,J)=(1/2,0,1/2)$, $(3/2,0,1/2)$, $(1/2,0,3/2)$, $(3/2,0,2/2)$, $(3/2,0,5/2)$, 10 scattering channels to produce $\phi+N$ and 7 channels to produce $\phi+\Delta$ are calculated. By this we methodically extended the usually considered $m+B$ channels involving pions, nucleons and $\Delta$'s by accounting for s-wave scattering $m+B \to \phi + B$ reactions with $m=\eta, K, \rho, K^*$ mesons and $B=N, \Delta, \Lambda, \Sigma, \Sigma^*$ baryons for all possible $I=1/2$ and 3/2 channels. We have found that among them the $\eta N \to \phi N$ channel is the most important. Although the cross section is not the largest, the $\eta$ abundance, larger than that for all other mesons (besides the pion), makes this channel important at SIS energies. 
The reverse reactions $\phi B \to m B$ have been realized by detailed balance. These $mB$ channels have been neglected up to now in transport approaches.

iii)  The in-medium modifications of $K$ and $\bar K$ mesons. \\
Following our previous study \cite{Song:2020clw}, we have accounted for the in-medium modifications of strange mesons, too:
a repulsive  potential for kaons  which increases linearly with the baryon density and which leads 
to the increase of the kaon mass with increasing density while the width remains negligibly small;
the  density and temperature dependent complex self-energy obtained by a G-matrix approach for antikaons
which leads to a substantial broadening of the spectral functions and to the decrease of the pole mass.

Our findings can be summarized as follows:
\begin{itemize}
\item In-medium effects, as a collisional broadening of the $\phi$ meson spectral function, lead to an enhancement of $\phi$ meson production, especially at subthreshold energies, as demonstrated in Figs. \ref{dndym}, \ref{dndym3GeV}.

\item  The novel $mB\to \phi B$ channels from the SU(6) chiral Lagrangian in the T-matrix approach enhance considerably the $\phi$ production in heavy-ion collisions, see Figs. \ref{dndym}, \ref{dndym3GeV}, \ref{ratio1}.

\item Due to rescattering of the $K^+$ or $K^-$ in the medium, in Au+Au reactions at energies between $E_{kin}=$ 1.23 A GeV and $\sqrt{s_{\rm NN}}=$ 3 GeV  only  60-70\% of the decaying $\phi$'s can be reconstructed by the invariant mass method.

\item The in-medium effects for open strange mesons - a  repulsive potential for the kaons and density and temperature dependent complex self-energies from the G-matrix approach for antikaons -
lead to a sizeable modification of the yield and spectra of (anti-)kaons  as already investigated in our previous study \cite{Song:2020clw}. However, they also have an  impact
on the final $\phi$ meson observables:  by reducing the reconstructed $\phi$ mesons since
the number of $K^+$ decreases and the rescattering of $K^-$ is getting stronger with the in-medium effects.

\item The measured $K, \bar K$ rapidity and $m_T$ distributions for different systems at SIS energies as well as the $K^-/K^+$ ratio favour the in-medium scenario for the (anti-)kaons,
except of the Au+Au data at 1.23 A GeV where we overestimate the $K^-$ yield.

\item The experimental $\phi/K^-$  ratio can be reproduced by PHSD calculations, incorporating the collisional broadening for the $\phi$ meson spectral function and including novel $mB$
channels for the $\phi$ production. This leads to a strong enhancement of $\phi$ production close to threshold.

On the other hand, an inclusion of the in-medium effects for $K, \bar K$, which leads to a strong enhancement of the $K^-$ yield and overestimation of Au+Au data at 1.23 A GeV
leads to a reduction of the $\phi/K^-$ ratio and an underestimation of this experimental data point while for the other energies and systems the experimental $\phi/K^-$ ratio is still within the low experimental errorbars for the full in-medium scenario for $\phi, K, \bar K$.

\item The ratio $\phi/\Xi^-$ is also found to be reproduced within the PHSD approach including the full in-medium scenario for $\phi, K, \bar K$ and novel $mB$ channels for $\phi$ production.
\end{itemize}

\section*{Acknowledgements}

We are grateful to Daniel Cabrera for providing us the T- and G-matrix calculations.
E.B. is grateful to Herbert Str\"obele  for stimulating discussions on an importance of $\eta N$ reactions for $\phi$ production.
We are also thankful for inspiring discussions with Marcus Bleicher, Wolfgang Cassing,  Philipp Gubler, Benjamin Kimelman, 
Yvonne Leifels, Manuel Lorenz, Laura Tolos and Iori Vassiliev.  
Furthermore, we acknowledge support by the Deutsche Forschungsgemeinschaft 
(DFG, German Research Foundation): grant BR 4000/7-1
and   by the GSI-IN2P3 agreement under contract number 13-70.
This project has, furthermore, received funding from the European Union’s Horizon 2020 research and innovation program under grant agreement No 824093 (STRONG-2020). 
Also we thank the COST Action THOR, CA15213.
The computational resources have been provided by the LOEWE-Center for Scientific Computing and the "Green Cube" at GSI, Darmstadt.

\bibliography{references}

\end{document}